\documentclass[11pt]{article}

\usepackage{amssymb}
\usepackage{amsmath}
\usepackage{multirow}
\usepackage{soul}
\usepackage{comment}
\usepackage{graphicx}
 \usepackage{lineno}
\usepackage[T1]{fontenc}
\usepackage[utf8]{inputenc}
\usepackage{authblk}

\begin{document}

\title{
The modellers' encounter with ecological theory. Or, what is this thing  called `growth rate'?
}

\author[1]{Michael Deveau}
\author[1]{ Richard Karsten}
\author[1,2]{Holger Teismann\thanks{hteisman@acadiau.ca}}
\affil[1]{\footnotesize{Department of Mathematics and Statistics\\
  Acadia University, Wolfville, 
  Canada}}
\affil[2]{\footnotesize{Basque Center for Applied Mathematics, Bilbao, Spain}}

\maketitle 

\begin{abstract}
The attempt to determine the population growth rate from 
field data reveals several ambiguities in its definition(s), which seem to throw  
into question the very concept itself. 
However, an alternative point of view is proposed that not only preserves the identity of the concept, but also helps  discriminate between competing models for capturing the data. 

\end{abstract}




\section*{Introduction}
Imagine a group of quantitative scientists, such as statisticians or mathematicians, with only a textbook knowledge of ecology,
who are charged  with analyzing a dataset of population counts.  Let us call these scientists the ``Modellers''. 
The Modellers feel that if their analysis is to be relevant to ecologists, it should relate to what they understand to be the  most important notion 
in population ecology, the \textit{population growth rate}.   This note is written from the perspective of the Modellers and 
describes the difficulties they experience in the process, as well as their attempts at overcoming them. 
\section*{What is the growth rate?}
  We give just one piece of evidence of the fundamental importance of the growth 
rate by quoting Berryman (2003)  \cite{berrymanLaws} who makes it the unsung hero of a fundamental law of ecology: 
\begin{quote} {\sl  {\bf The first principle (geometric growth)}  \\
Ecologists seem to agree, in general, that geometric (exponential) growth is a good candidate for a 
law of population ecology.
 {[\ldots ]} since geometric growth is a fundamental and self-evident property of all populations living under a certain set of conditions (unlimited resources), I prefer to think of it as the first founding principle of population dynamics {[\ldots ]} }
 \end{quote}  
We have no intention here to enter the fray of the extensive debate among ecologists about whether Berryman's law (a.k.a. the Malthusian Law) is in fact a Law of Nature and/or whether 
ecology has any laws at all\footnote{In addition to \cite{berrymanLaws}, interested readers might find e.g. \cite{anarchist,ginzburg2007,lockwood2008,raerinne2013} useful as potential entry points into the pertinent literature,
which also contain older and widely discussed contributions such as \cite{Levins1966} and \cite{turchin2001}.};
although we admit that the title of  O'Hara's spirited article \cite{anarchist} on the subject provided some inspiration 
for the title of the present paper.
 
 Rather, we imagine the Modellers taking Berryman's law to be the\linebreak  \textit{definition} of  the growth rate, according to which it is simply the number $r$ in an exponential-growth expression of 
the form $N \sim e^{rt}$ (or the discrete-time version thereof\footnote{In this paper we use continuous-time models throughout.  
However, the discussion could equally be applied to discrete--time models (Leslie matrix models).}), where $N=N(t)$ denotes the total number of individuals at time $t$.   As a result, the existence of a well-defined growth rate is contingent on populations under the condition of  ``unlimited resources" following an exponential 
growth law.

\begin{figure}[h]
\includegraphics[height=9cm,width=13cm]{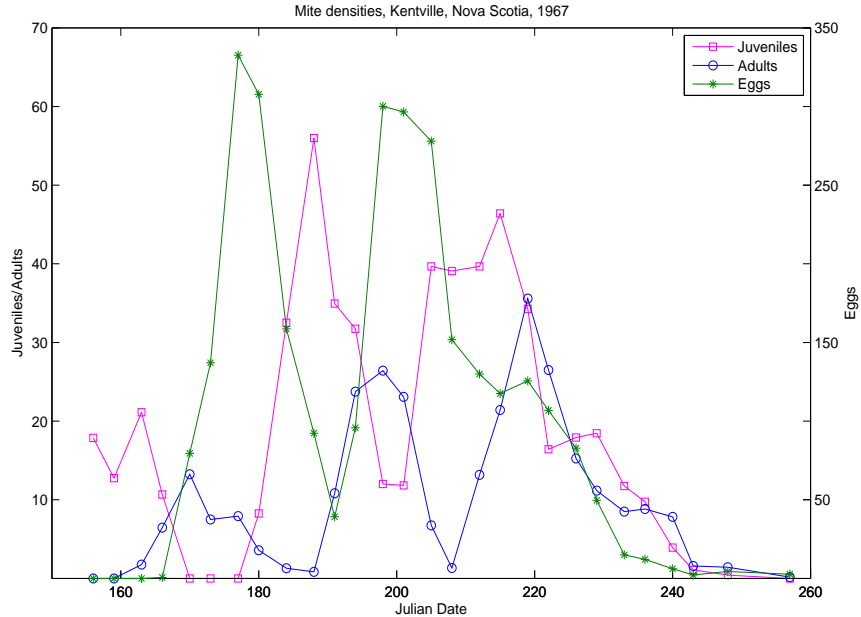} \vspace{-5ex}
\caption{Mite-population field data (note the different scales for juveniles/adults and eggs).} \label{Fig1}
\end{figure}

 But then, what about the Modellers' data, which are shown in Fig.~\ref{Fig1} below?\footnote{This data set is remarkable for its quality and detail, and it has therefore recently attracted renewed interest. While various aspects of the data have been described in the literature \cite{herbertsanford1969,herbert1970,hardman1985,hardman1991}, some of the raw
 data have apparently never been analyzed.}
This is a plot of field data of mites on apple trees, which, at
least at the beginning of the season, do not experience any significant resource limitation. So what are the Modellers to make of the 
large oscillations? This does not look like a simple  exponential $e^{rt}$ at all --- at least not as long as $r$ is a real 
number; and which ecologist would put up with a complex growth rate?
\section*{One, two, three  growth rates?}
The Modellers may start over and look for the standard definition, which,
in the words of e.g. Sibly and Hone (2002) \cite{sibly2002}, reads 
\begin{quote}
{\sl Population growth rate describes the per capita rate of growth of a population, \st{either} as the factor by which population size increases per year, conventionally given the symbol $\lambda (=N_{t+1}/N_t)$, or as $r=\log \lambda $.} 
\end{quote} 

\begin{figure}[h]
\includegraphics[height=9.5cm,width=13cm]{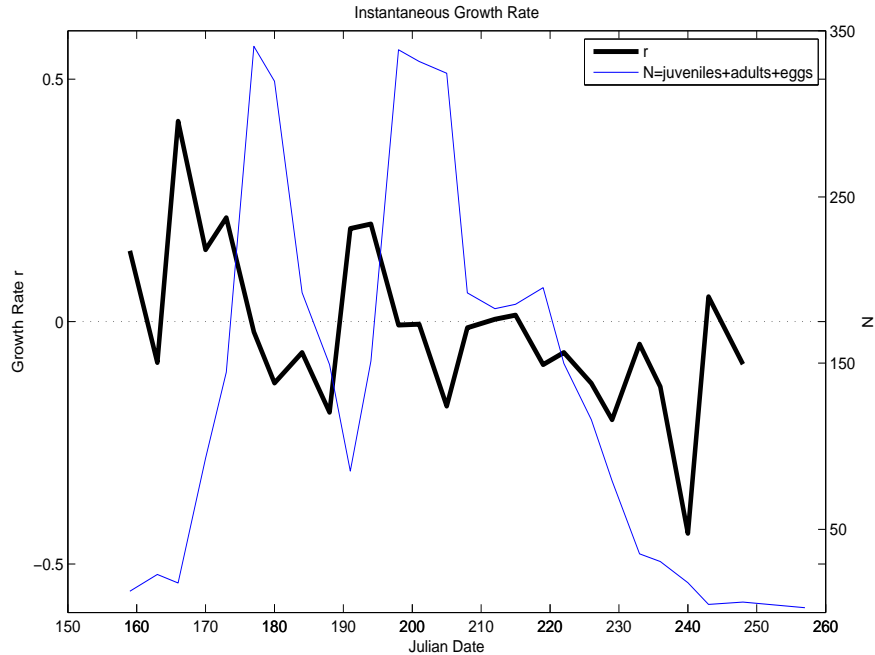} \vspace{-5ex}
\caption{Instantaneous growth rate $r=\log\left(N(t+\Delta t)/N(t)\right)$, where $N(t)=E(t)+J(t)+A(t)$ is the
total number of individuals.} \label{fig2}

\end{figure}

But when this definition is applied to the data shown in Fig. \ref{Fig1} (mutatis mutandis; the relevant time step is obviously not a year)  the growth rate ``suddenly" starts fluctuating wildly between positive and negative values (Fig. \ref{fig2}).  
 Does this mean that a constant growth rate does not exist after all?\footnote{We agree with Chester \cite{chester2012}
 who argues that, if the growth rate is allowed to depend on time, it loses its meaning and utility.} Even if the environmental conditions are 
constant (as they are -- at least approximately -- during the beginning of the mite season)?  

Sibly and Hone continue as one would expect from textbook ecology. 
\begin{quote}
{\sl In the simplest population model all individuals in the population are assumed equivalent, with the same death rates and birth rates, and there is no migration in or out of the population, so exponential growth occurs; in this model, population growth rate = $r$ = instantaneous birth rate -- instantaneous death rate.} 
\end{quote}  

What interests us the most in this quote is the appearance of the term ``model" in the description of the growth rate, but more on this later.  For now, we read on
a little further:
\begin{quote}
{\sl Population growth rate is typically estimated using either census data over time or from demographic (fecundity and survival) data. Census data are analysed by the linear regression of the natural logarithms of abundance over time, and demographic data using the Euler--Lotka equation (Caughley 1977) and population projection matrices (Caswell 2001).} 
\end{quote} 
So here it is: according to this, there are at least two kinds of growth rates: one based
on ``demographic data", and one based on ``census data".  Moreover, the former seems to draw in other 
quantities (``fecundity and survival"), which seem to have to be known independently/beforehand. 

But, if  the growth rate is an inherent property of a population,
 should it not manifest itself in a measured population curve? Should it not be possible to determine it 
 from measured data without reference to outside information?  This appears to discredit the demographic 
 variant of the concept -- or maybe the demographic calculation as described gives an
 ecological quantity that is different from the growth rate.  
 
What is more, Sibly and Hone also give two ways of ``analysing" census data. In the first quote above, they present the 
simple formula $r=\log \left(N_{t+1}/N_t\right) $; i.e.,  the formula for the \textit{``instantaneous"} growth rate \cite{walthall1997}. 
Now they suggest  to use ``linear regression of the natural logarithms of abundance over time'', which will result 
in an averaged or smoothed growth quantity.  But will the numerical value(s) depend on the time interval(s)? If the Modellers are asked to take averages, how are they going to decide 
which time interval to use?

 In any case, the supposedly fundamental (and elementary) concept of growth rate certainly seems to have lost some of its simplicity and clarity. 
  
\section*{Asymptotic vs. transient growth rate (population structure)}  
Experienced ecologists will readily identify the large oscillations in the data of Fig.~\ref{Fig1}  as generational waves, and they will argue that in determining the growth rate one must account for  the (st)age structure of the population. 

The Modellers, obediently,  
consult  e.g. Tenhumberg (2010) \cite{ten2010}  who 
 uses the (instantaneous) census-data definition of the growth rate, complete with its fluctuations during the 
 early season (``transient dynamics"),  and only offers them the   
  piece of mind of a constant-value growth rate asymptotically.  
   \begin{quote}
{\sl If nothing else changes, the population eventually reaches the stable stage distribution and the speed at which the population is growing approaches a constant rate (the asymptotic population growth rate).}
\end{quote}    
This, of course, refers to the 
  mathematical fact, discovered by Lotka \cite{lotka1922}, that solutions to the appropriate demographic model for structured populations eventually settle on the 
  \textrm{stable age distribution} and grow exponentially with a rate that can be computed from the demographic  data 
 (\textit{``Lotka's r"} or \textit{``intrinsic rate of increase"} \cite{walthall1997}). 
The census-data and 
 demographic growth rates of Sibly and Hone are in this sense actually identical!
 
 However, this is still unsatisfactory --  at least for anyone ready to accept Berryman's law: 
 it is, in fact, precisely during the early season that resource  limitations are the least likely to occur;
 so during \textit{this} period the exponential-growth law should work particularly well.  Having to wait until 
 the stable age distribution is assumed seems counter-intuitive --- as well as unrealistic and impractical, as many 
 species will experience diminished growth due to limited resources before they can even approach 
 the asymptotic state and/or will change their characteristics 
  altogether due to seasonal behaviour etc.\footnote{Taylor \cite{taylor1979} used life-table data of various 
 insect and mite species to estimate the time for the populations to get within 5\% of the stable age distribution (SAD). 
 According to those estimates, it is conceivable for \textit{some} species to approach the SAD within a season.}  The Modellers' field data  exhibit evidence of both phenomena:
 the growth of the population slows down and comes to a halt after what appears to be 20-50 days; and during the final 
 part of the season, the population crashes, as the mites switch to laying next--season eggs that will  
 not hatch during the season they are laid (the egg numbers shown in Fig.~\ref{Fig1} are for same-season eggs). 
 
 It is an interesting fact, which probably deserves to be better known, that classical demography 
 itself offers a resolution of this conundrum. Properly ``re-weighting'' of the (st)age groups; i.e., considering\footnote{In keeping with the the notation used by Lotka and Fisher, we adopt the continuous-time-continuous-age framework, in which $\rho (t,a)$ represents the number of 
 age--$a$ organisms at time $t$ and the time and age variables $t$ and $a$ are allowed to vary continuously in $(0,\infty )$.}    
 {\small \[ V(t) := \int _0^\infty v(a) \rho (t,a) da,\] }
does indeed result in an exponential-growth law of the form  $V(t) \sim e^{rt}$, where $r$ is the ``asymptotic growth rate" in the parlance of Tenhumberg (i.e. Lotka's $r$). 
The important point here is that the exponential--growth formula for the  aggregate quantity $V(t)$ 
actually holds \textit{for all $t$}, not just for large $t$, as Tenhumberg's terminology suggests.    
Demographers know the function $v(a)$ to be Fisher's \cite{fisher1927}
  \textit{age-dependent reproductive value}\footnote{Although not considered in this note, we mention that in the discrete-time-discrete-age framework of Leslie matrix models, 
$v(a)$ is given by the dominant left eigenvector of the transition matrix. (The dominant right eigenvector 
corresponds to the stable age distribution; the dominant eigenvalue to Lotka's $r$.)} and $V(t)$ to be the  
\textit{total} reproductive value. The resolution of the ``early-season-vs.-asymptotic" conundrum, therefore,  lies
in the realization that only a very \textit{specific} aggregate population-size quantity (namely $V(t)$) obeys the exponential-growth 
law stipulated by Berryman.  Other quantities, such as the 
total number of individuals $N(t) = \int \rho (t,a) da$,  will generally not grow according to a simple exponential.   

\begin{figure}[h]
\begin{center}
\includegraphics[height=9cm,width=13cm]{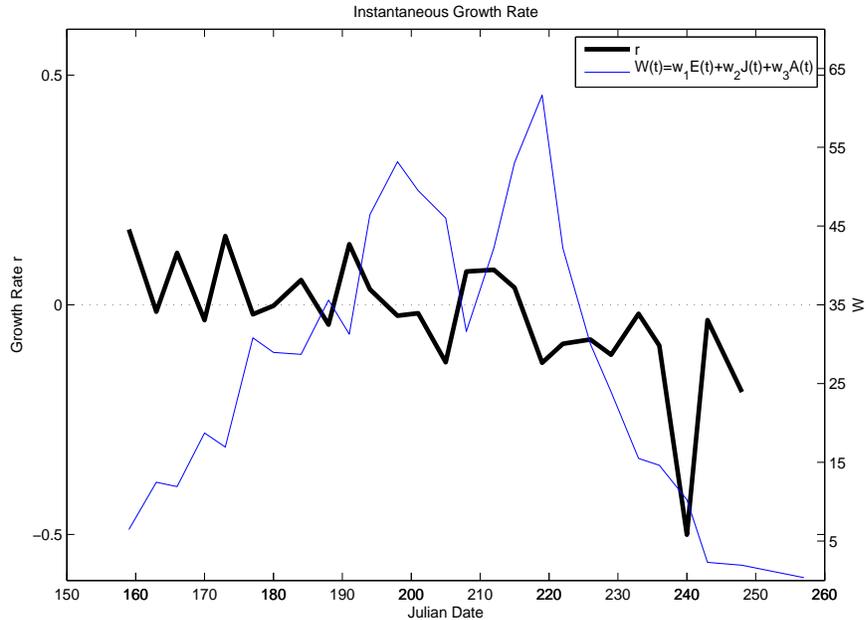} 
 \end{center} 
 \vspace{-5ex}
 \caption{Instantaneous growth rate derived from the heuristically re-weighted aggregate population $W(t) = w_1E(t)+w_2J(t)+w_3A(t)$ (see text). Compared with Fig.~\ref{fig2}, seasonal fluctuations during the growing phase of the season are reduced 
 significantly.}\label{12w}  
 \end{figure}

As an aside, it is amusing to see that applying 
a very simple, purely \textit{heuristic} re-weighting formula of the form 
$W(t) = w_1E(t)+w_2J(t)+w_3A(t) $ 
can reduce the generational fluctuations in the data significantly, as demonstrated in Fig.~\ref{12w}
(as well as Figures \ref{fitsW} and  \ref{figdistW} in \ref{Weight}).
Here the weights $w_j$ are computed by taking ratios of life-stage averages (see eq. (\ref{weights})) and the  step function \\[-1ex]
 {\small \[ w (a) := \left\{ \begin{array}{cl} w_1, & \textrm{$a=$ egg age} \\
  w_2, & \textrm{$a=$ juvenile age} \\
  w_3, & \textrm{$a= $ adult age} \end{array} \right.\] }
serves as a surrogate for $v(a)$.\footnote{Even if the demographic data (fecundity and mortality) as functions of age are piecewise constant, $v(a)$ is not, but piecewise exponential.} 
\begin{table}[h]
\mbox{} \hspace{-0.55cm} 
\begin{tabular}{|c|c|c|c|c|}
\hline 
 & Model  & parameters & seasonal & formula  (SM)\\
 \hline \hline
I & simple exponential &2  & no  & (\ref{simple}), $K=\infty $ \\
\hline
II & simple logistic  & 3 & no  & (\ref{simple})  \\
   \hline \hline 
 III &  linear phenomenological & 11 &yes  & (\ref{cycle})--(\ref{F1A}), $K= \infty $\\
   \hline
 IV & nonlinear phenomenological & 12 & yes & (\ref{cycle})--(\ref{F1A}) \\
   \hline
  V & linear demographic & 10 & yes & (\ref{dE})--(\ref{dF}), $\nu =0 $\\
   \hline
  VI & nonlinear demographic & 11 &yes  &  (\ref{dE})--(\ref{dF}) \\
   \hline
\end{tabular} \vspace{-1ex}
\caption{Models used. For details see \ref{Models} of the Supplementary Material (SM).} \label{tab0} 
\end{table}

This observation provides a non-theoretical illustration of the basic ``re-weighting" rationale behind the definition 
of $V(t)$, which is a reflection of the fact that in structured populations not all individuals can be ``assumed equivalent", as is done 
``in the simplest population model".
So here is this ominous term again -- \textit{model} -- and  it is finally time for us to 
state our thesis (as well as to abandon the, admittedly somewhat contrived, distinction between the imaginary Modellers and the authors). 
\section*{(How) Many growth rates?}
\begin{quote}
The concept of growth rate is \textit{model-dependent};  for any given population, we may have \textit{as many} reasonable 
answers to the question ``what is the population's growth rate?" \textit{as} we have reasonable models for its  dynamics. 
\end{quote} 
Before we offer some consolation to readers who despair at our move from the enlightened monotheism of 
one growth rate to the heathen polytheism of many growth rates, 
we are going to use the data shown above to illustrate our thesis.  
We used six different models to capture the data\footnote{All models considered in this 
paper are continuous in the time variable. We also tested discrete-time (Leslie matrix) models, but they did not provide any advantage; neither in terms of modelling, nor in terms of fitting the data or determining the growth rate.}; see Table~\ref{tab0}. 
The number of parameters varied between 2 for the simple exponential model 
$N_0e^{rt}$ and 12 for the most complex  seasonal phenomenological model.
The modelling of the data was achieved by standard parameter-fitting methods.
The details of the models and the fitting process are not important for this discussion  
and are therefore omitted.
Interested readers may consult the supplementary material available online, which contains 
some basic information about the models and results. Here it suffices  to say that the four seasonal models considered (III-VI in 
Table \ref{tab0}) are virtually 
indistinguishable in term of capturing the data (see Fig.~\ref{fitsDemo} of \ref{Fit}). By contrast, the two simple non-seasonal growth models (I \& II) 
are rather crude models of the data (see Fig.~\ref{fitsSimple}), as expected from the discussion above.

The first  four models (I--IV) of the table explicitly contain the growth rate as a parameter.    
The remaining two are demographic models for which the growth rate is computed from 
the model parameters according to the Euler--Lotka equation (see eq. (\ref{Euler-Lotka}) in \ref{DDE}).  The results are tabulated below. 
\begin{table}[h]
\begin{center}\begin{tabular}{|c||c|c|c||c|}
\hline 
     &        \multicolumn{3}{c}{ Time Window} &  \\ \hline 
  Model  & 30 days & 50 days & 70 days & full season   \\
 \hline \hline
I  & 0.1160  &  0.0686   & 0.0329 & \\
\hline
II &   0.4615  &  0.4636  &  0.4789 &  \\
   \hline \hline 
 III &   & 0.0371  &  0.0552  &  0.0419\\
   \hline
 IV &  &0.1030  &   0.0620 &   0.1255\\
   \hline
  V & & 0.0311 &   0.0210 &   0.0165\\
   \hline
  VI &  & 0.0498  &  0.0445 &   0.0352 \\
   \hline
\end{tabular} \end{center} \vspace{-0.8ex}
\caption{Growth-rate values derived from various models/time windows.} \label{tab} 
\end{table}

 Looking at Table~\ref{tab}, the conclusion seems inescapable: that one and only  growth rate 
 we believed we knew is dead. Using 6 models on 3 time intervals each, results in 18 
 growth rates! (Although some of the values are fairly close and could be interpreted as representing 
 the same quantity.)   

\section*{Back to square one?}
Maybe we overstated our case when we interpreted the second Sibly-and-Hone quote 
above as saying that ``there are at least two kinds of growth rates".  
A more adept reading of the quote might be that there are two 
(or three) 
\textit{ways of determining} the growth rate --- the implication being 
that the quantity itself still has a unique ``identity".    Similarly, we may conclude that what we interpreted as model-dependence of the concept above,  may also be interpreted as multiple ways of computing the same ecological quantity,
the growth rate. This point of view offers, in fact, an intriguing possibility of ``turning the table" on the problem: if we were to assume that the one and only growth rate does exist after all, could we use this to discriminate between the 
models? 

\begin{figure}[h]
\begin{center}
\begin{tabular}{c}
 \includegraphics[height=12.5cm,width=9cm]{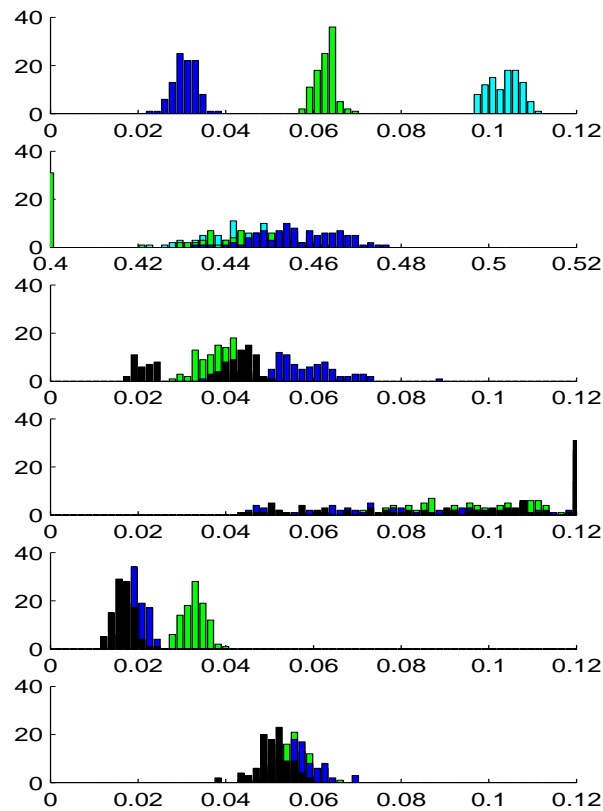} 
 \end{tabular} 
 \end{center} 
 \vspace{-8ex}
 \caption{Growth-rate distributions for models I-VI (rows 1-6) using three different time windows for each model.
 Colour coding of time windows: turquoise - 30 days; green - 50 days; 
 blue - 70 days; black - full season.  (Note the different scale in row 2.) 
} \label{figdist}
 \end{figure}

To apply this idea to our field data, we need to mention that the data actually consist of 24 replicates
(see Fig.~\ref{data} in \ref{Data}); what we showed
above were averaged data.  To increase the sample size
(which we arbitrarily chose to be 100), we used 
a simple partial averaging technique.
For this larger data set (see Fig.~\ref{random}), we repeated the exercise described above; that is, we 
fitted the 6 models to each of the 100 data sets and determined the corresponding growth-rate values. 
Fig.~\ref{figdist} shows the distributions of those values for the various models.   
\vspace{-2ex}

So, which model would you trust? Or, maybe we should say, \textit{en}trust with determining the ``true" growth rate
(assuming you believe it exists)? 

 Obviously, the width/narrowness of the distribution would be 
a factor in making the decision. Given that the data set was computed from \textit{replicates} 
of measurements of the \textit{same} population, one would expect that the different data sets essentially 
contain information about the \textit{same} growth rate (up to some degree of noise), so one would expect the distribution of $r$-values to
be fairly narrow as well. This basically eliminates models II and IV, and probably also 
model III. Perhaps surprisingly, the simplest model (I) has fairly tight distributions, which seems 
to give it an edge. However, it has a very strong dependence on the time window, over which the 
fitting is performed. This introduces an unacceptable arbitrariness (how would we tell which time window is 
the right one?), which leads us to discard this one, too. 
The best compromise between narrowness of the distribution and independence of the time window seem to be 
offered by model VI. So we may be inclined to declare this the ``winner".  It is worth noting that the models that 
perform best in terms of determining the growth rate $r$ (V \& VI)  are the ones that do not contain $r$ explicitly.

Of course, there are other factors  (likely many)  to be considered in selecting models in a particular 
modelling exercise, such as goodness of fit, complexity, derivability from first principles, purpose etc. (see e.g. 
\cite{johst2013} and the literature therein).  
More importantly, the point here is not to actually find the best model for the particular data set shown above. 
Rather, what we want to point out is that stipulating the existence (or ``reality'') of an ecological parameter such 
as the growth rate can potentially provide an additional robustness criterion. This turns the apparent 
\textit{model dependence} of that parameter into a tool for \textit{model selection}.  Ecologists might want 
try to identify other ecological quantities  that could be  utilized in a similar manner.  
\section*{Conclusion}
In this paper we offered the reader a choice: abandon the idea of a unique growth rate inherent of a given population 
and accept that this notion depends on the model used to determine it -- or retain the idea of a single growth rate and 
reject models that are unable to give robust values for it. Is it a matter of mere belief --- a mode of mind 
that many scientists probably feel has no place in scientific inquiry\footnote{Philosophers, even philosophers of science,
may feel differently about this, however: we note with interest that Nancy Cartwright uses the word ``believe" 15 times
 in the introduction to \cite{cartwright1983} alone. } 
--- on which side we come down? Do we even have to decide?

Our final argument is that the two alternatives described above may be used in an iterative 
manner, much like physicists view the genesis of physical theories. That is, we may take the position that an ecological quantity, such as the growth rate, can only reasonably be assumed to have reality/currency if it can be determined robustly; i.e., if at least one model can be found from which consistent and robust values of that quantity can be derived. 
If the quantity \textit{has} gained this kind of credibility, it may be utilized as a criterion for model selection as described above. 
If one model emerges as particularly suitable in a modelling exercise, it can be checked again for robustness in providing 
values for the same and/or other credible and established quantities of interest. 
 
As mentioned at the beginning of this story, we feel no urge (and possess no expertise) to enter the philosophical debate surrounding
the interrelations of reality (or realism), laws of nature, scientific models  etc. and/or  the similarities and differences of ecology and the physical sciences.    Nor do we make claims about 
novelty and originality of the ideas described above.  
For instance, the critical role of models (``nomological machines") and the idea of model robustness
have been discussed by Cartwright \cite{cartwright1999} and Rearinne \cite{raerinne2012} (following Levins \cite{Levins1966}), respectively, as well as
others. 

All we venture to do here is to advocate a pragmatic view, rooted in the simple (simplistic?) observation that scientific practice often proceeds (well) without explicit reference  to abstract 
foundational thinking\footnote{For a better--founded account of the utility of philosophy in biology, see 
\cite{orzack2012}.}.  We, the modellers, view the concept of the recursive definition of ecological quantities based on the constructibility and
robustness of suitable ``nomological machines"  as such a pragmatic approach.

    \section*{Acknowledgements}
 This research was supported by a  
Discovery Grant of the \textit{Natural Sciences and 
Engineering Research Council of Canada (NSERC)}; MD also acknowledges support by NSERC through an undergraduate scholarship.  HT would like to thank  the \textit{Basque Center for Applied Mathematics (BCAM)} for its hospitality and financial support.

\newpage
\appendix
\mbox{} \\[-4cm]
\begin{center}
{\huge \bf Appendix: Supplementary Material}
\end{center}
\section{Data}
\label{Data}
\vspace{-4ex}
\begin{figure}[h]
\hspace{-12.5ex}
\begin{tabular}{cccc}
\includegraphics[height=2.15cm,width=4cm]{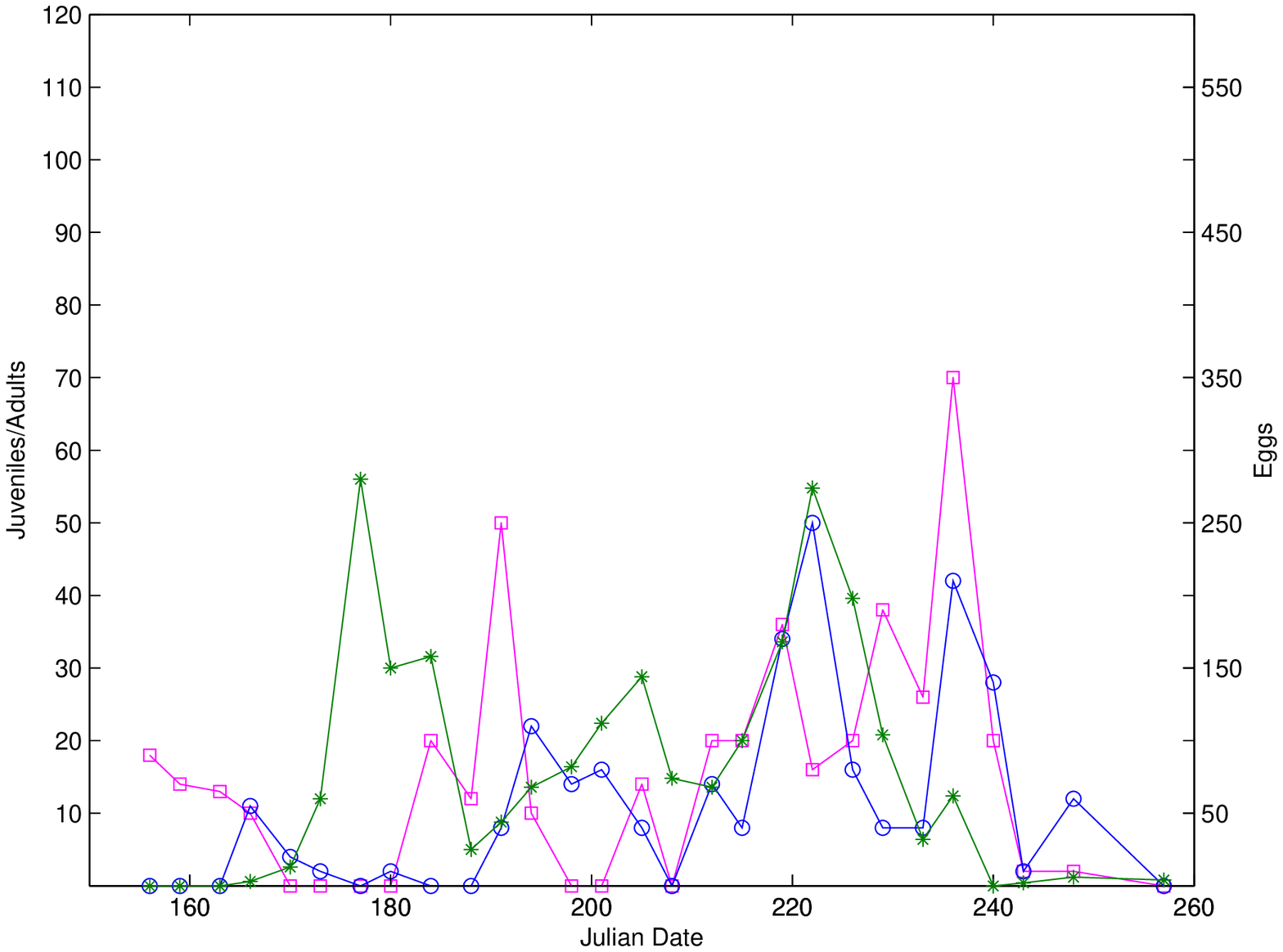} &
 \includegraphics[height=2.15cm,width=4cm]{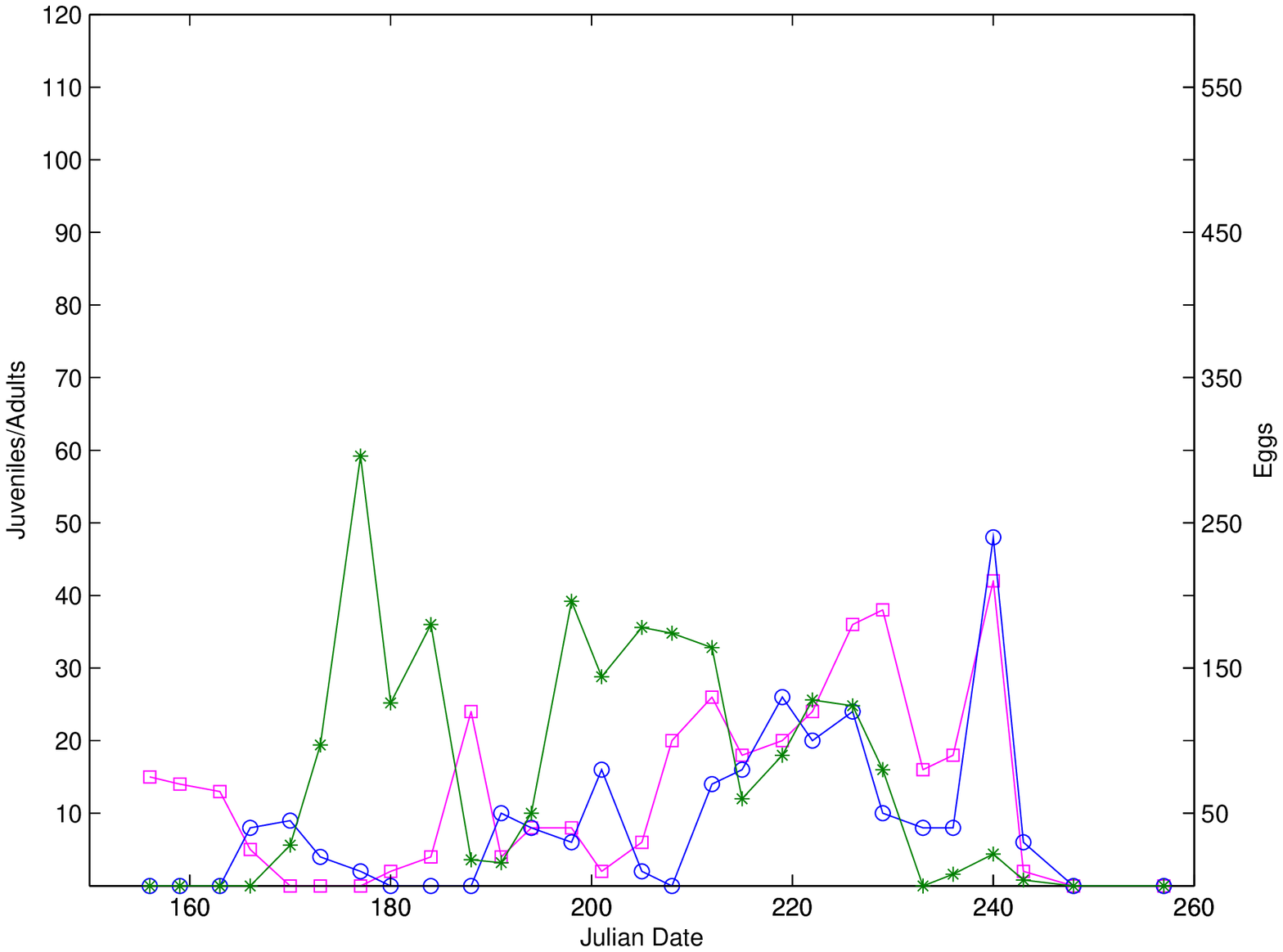} &
 \includegraphics[height=2.15cm,width=4cm]{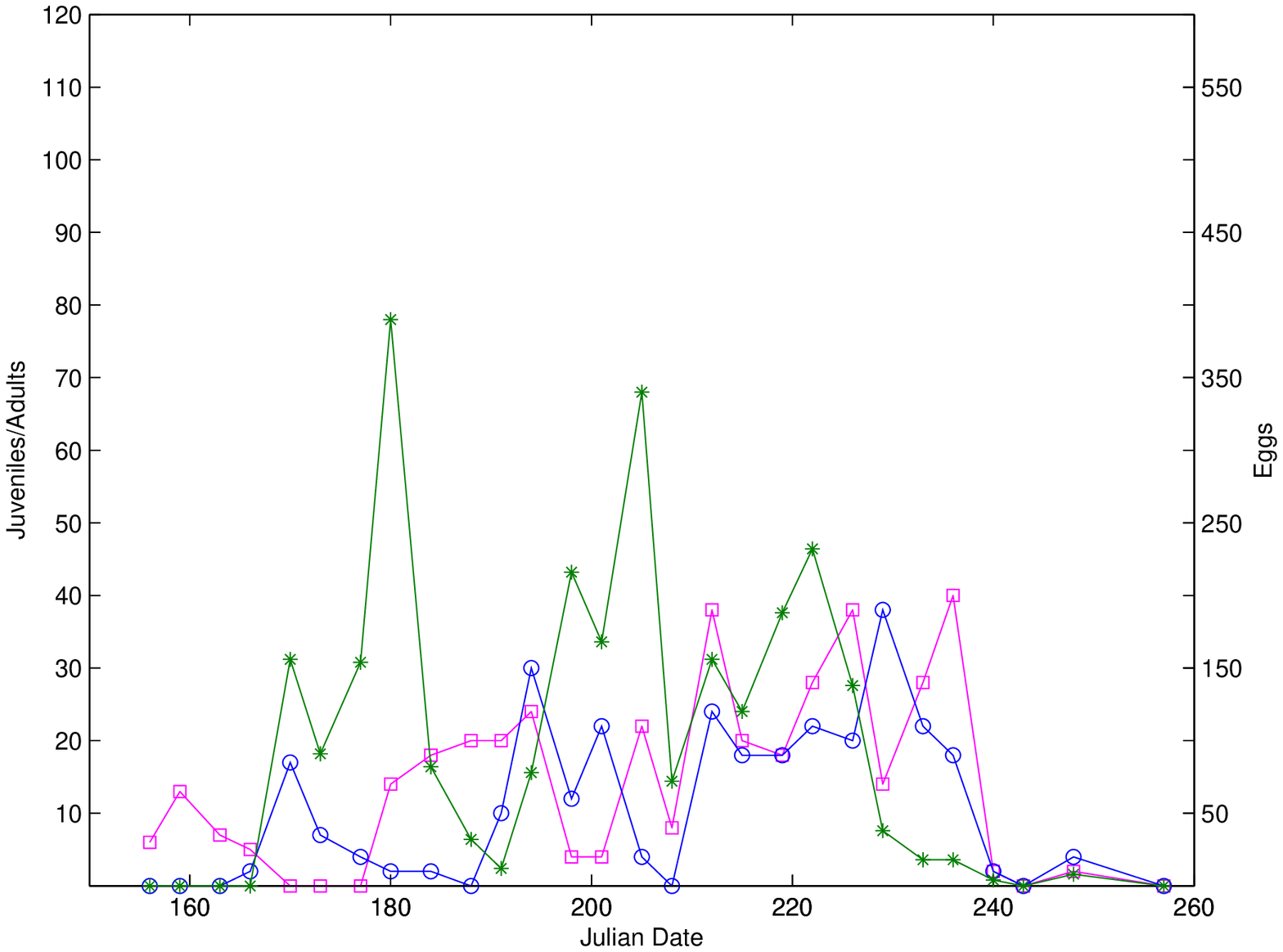}  &
 \includegraphics[height=2.15cm,width=4cm]{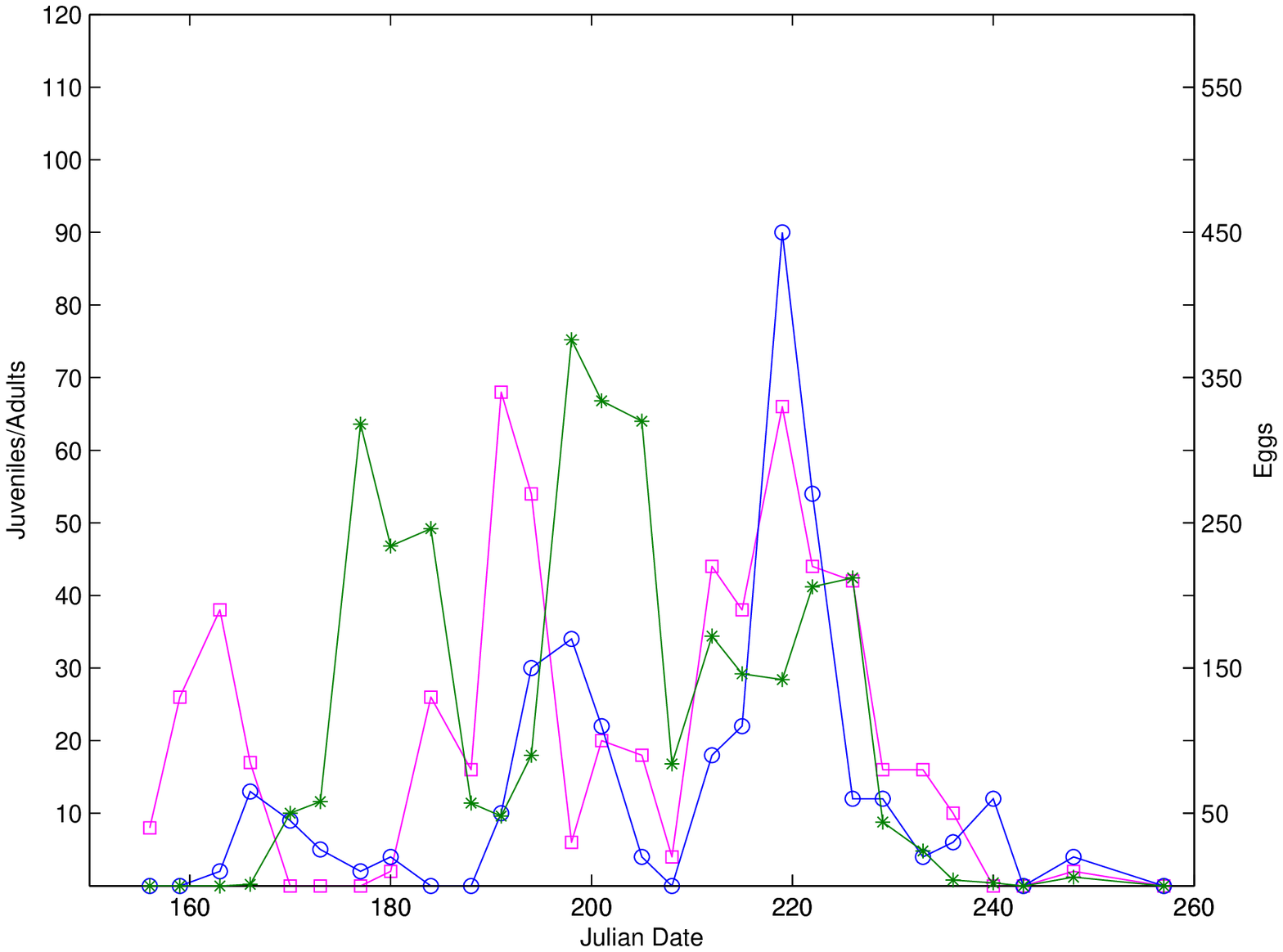} \\
 \includegraphics[height=2.15cm,width=4cm]{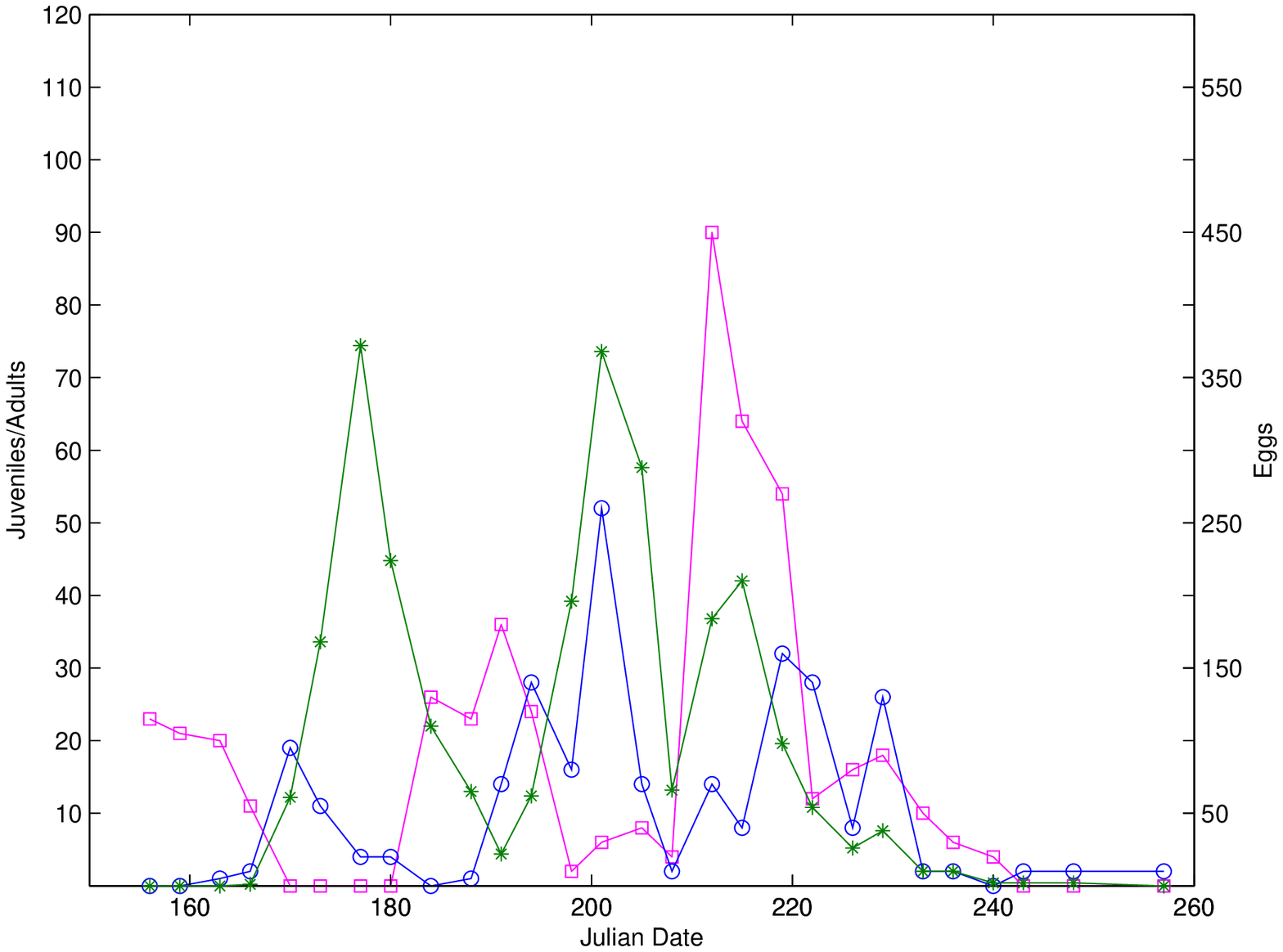} &
 \includegraphics[height=2.15cm,width=4cm]{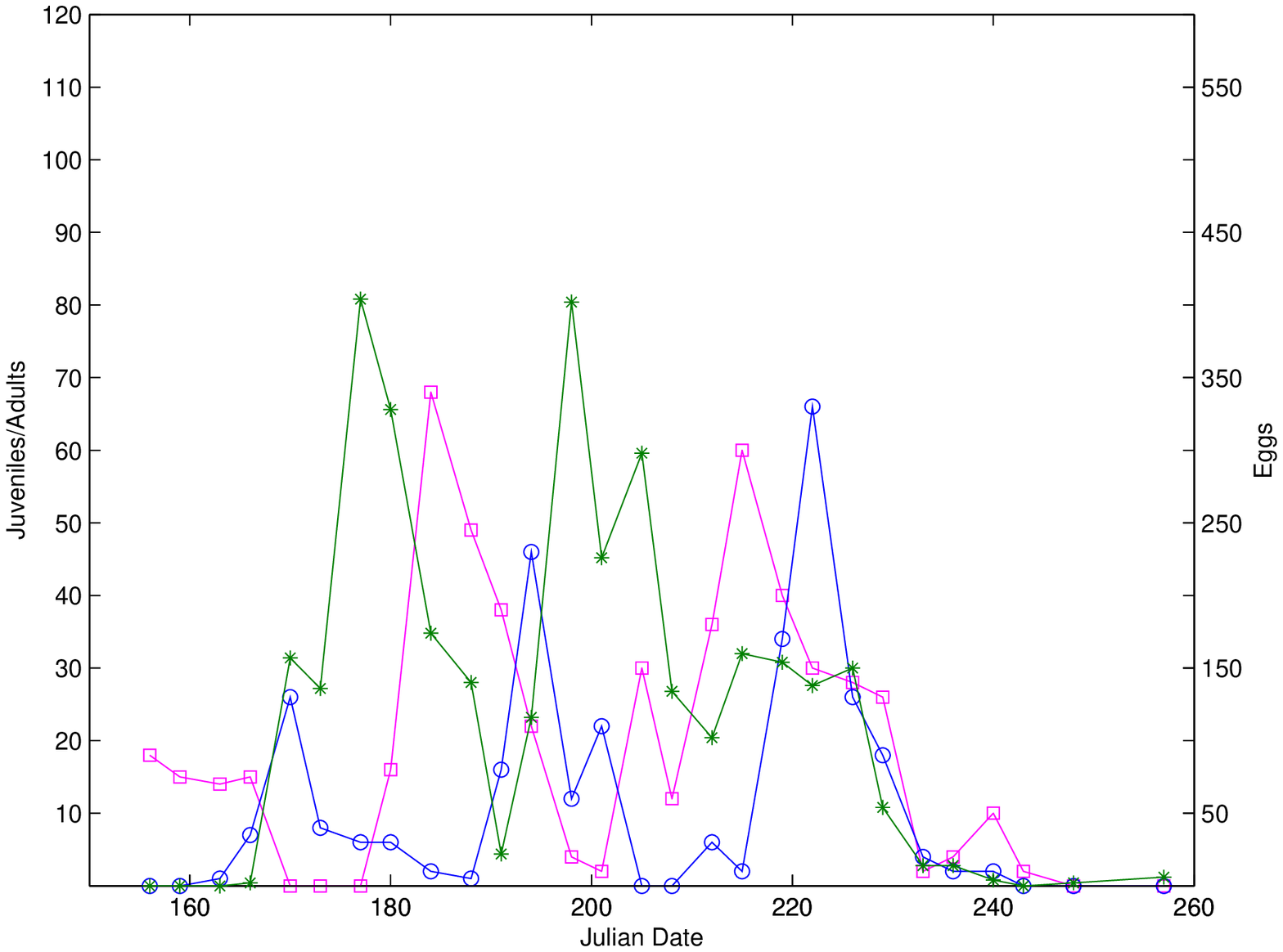}   &
 \includegraphics[height=2.15cm,width=4cm]{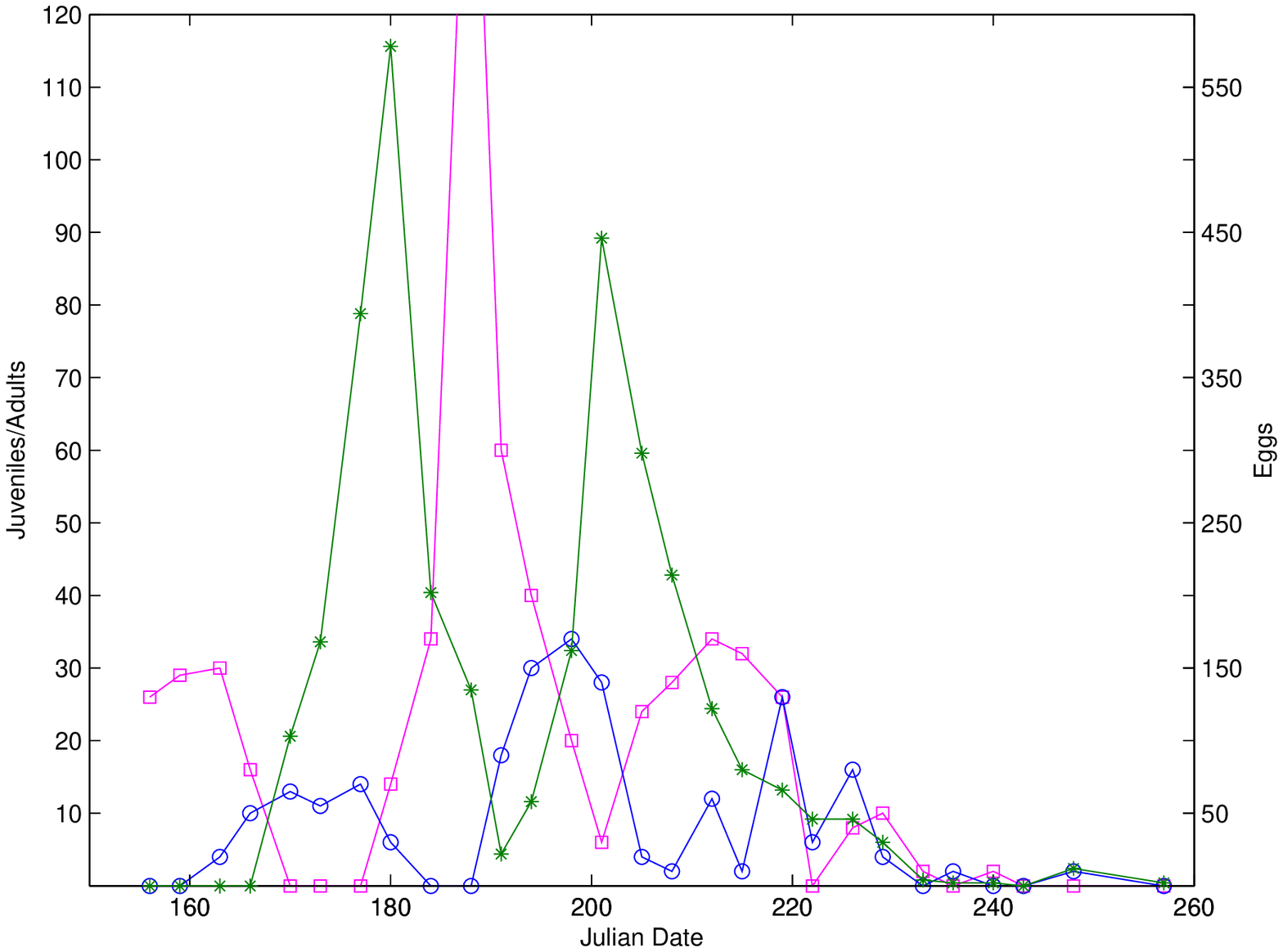} &
 \includegraphics[height=2.15cm,width=4cm]{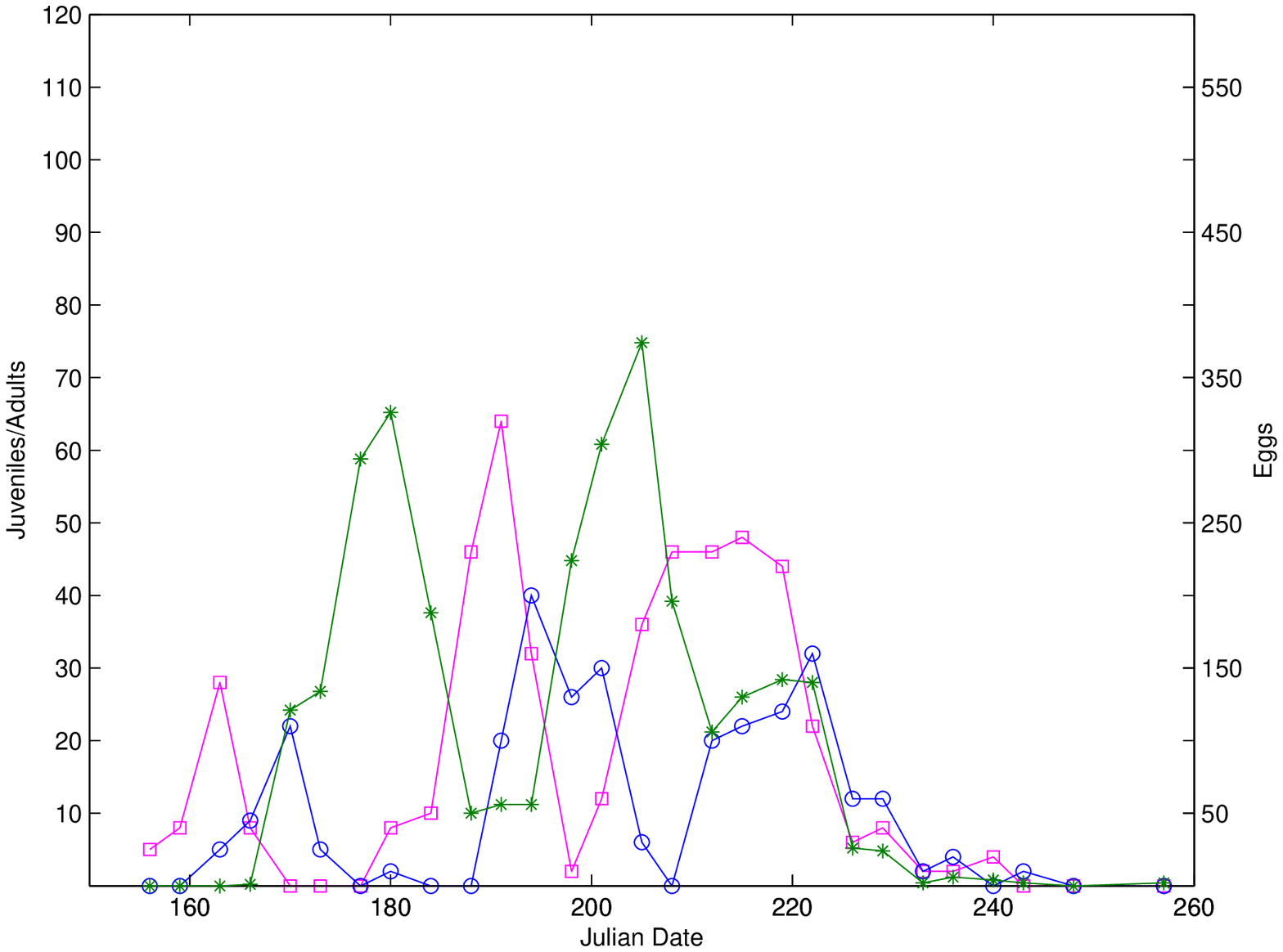} \\
 \includegraphics[height=2.15cm,width=4cm]{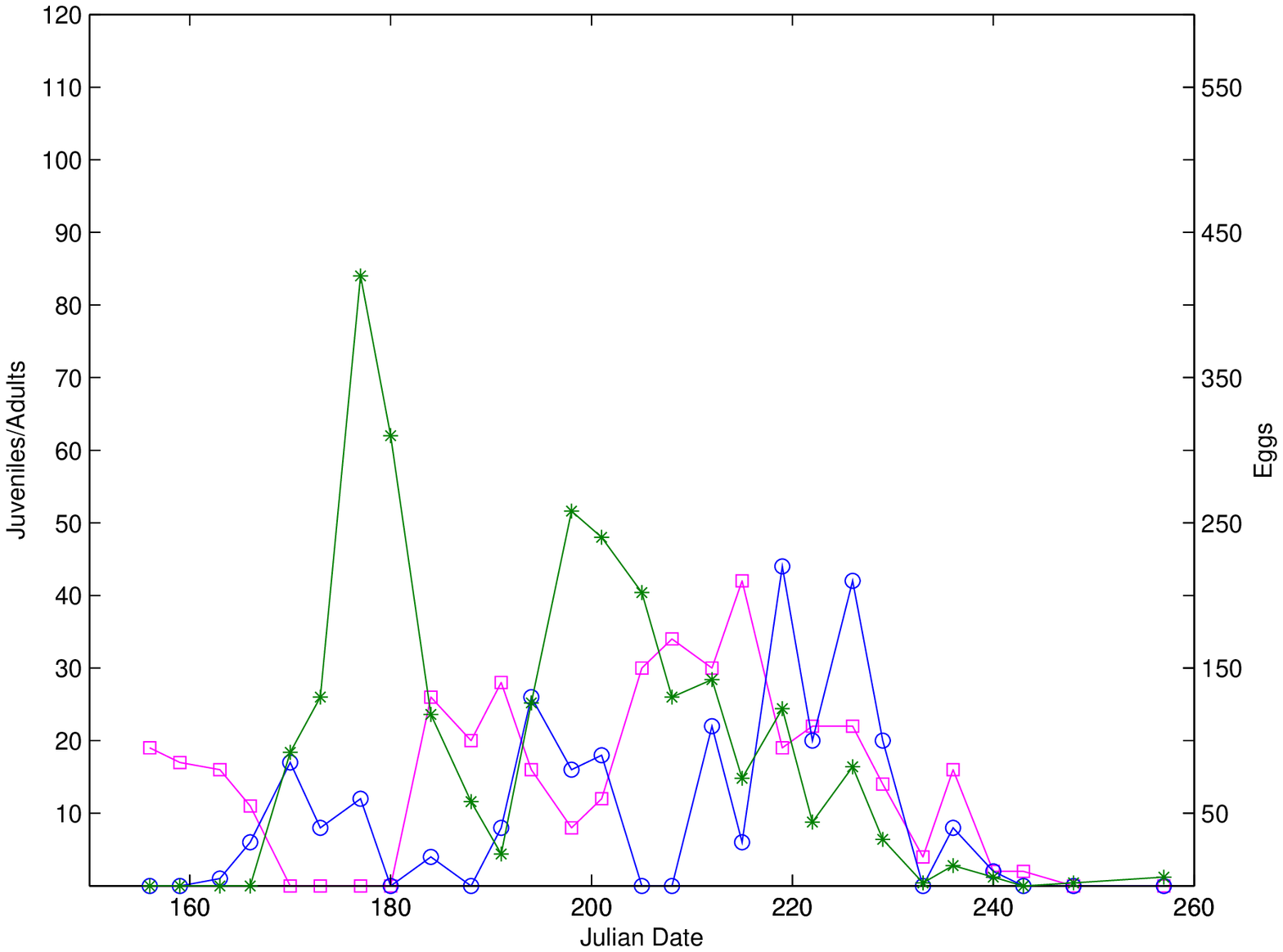}  &
 \includegraphics[height=2.15cm,width=4cm]{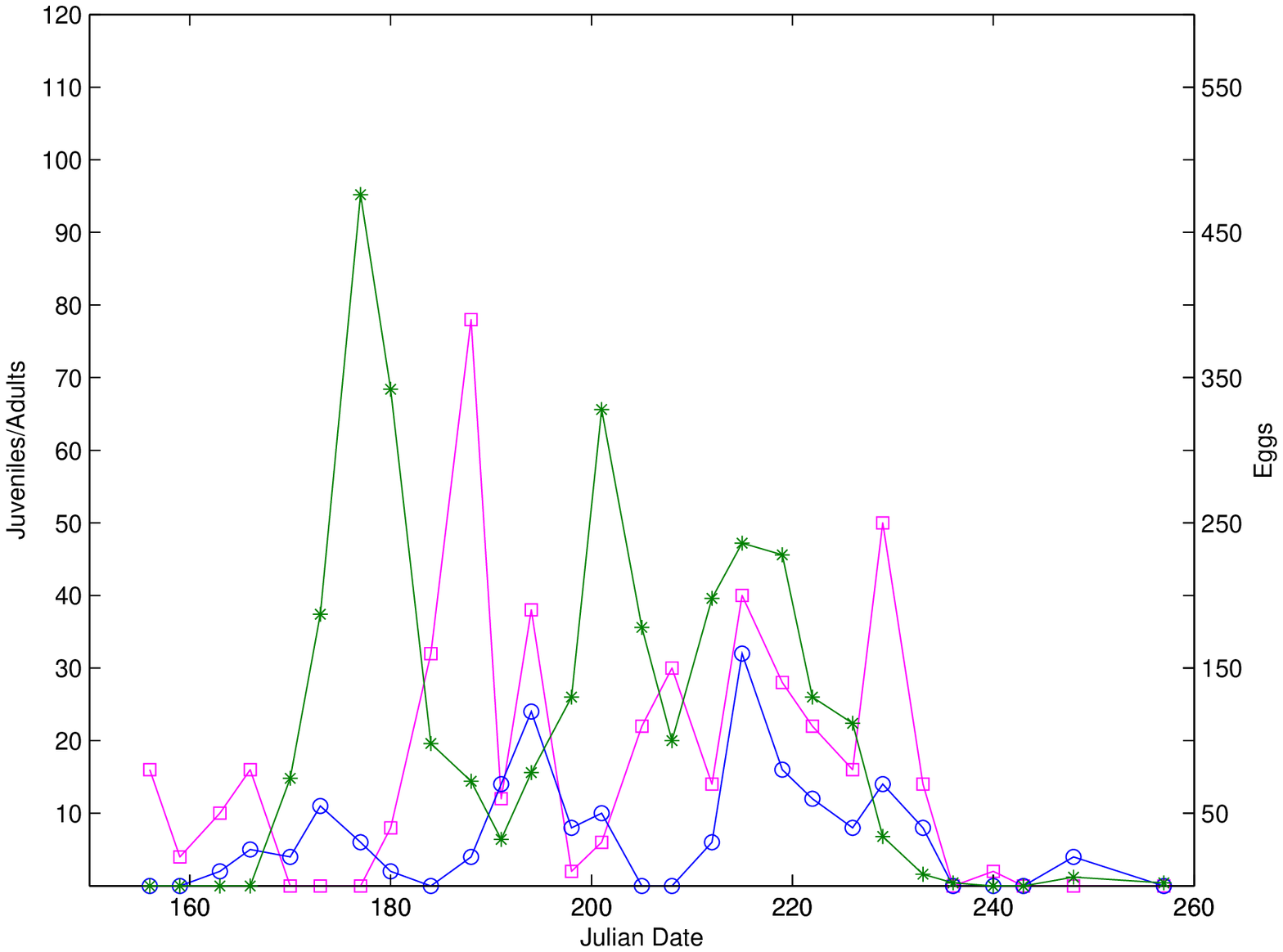} &
 \includegraphics[height=2.15cm,width=4cm]{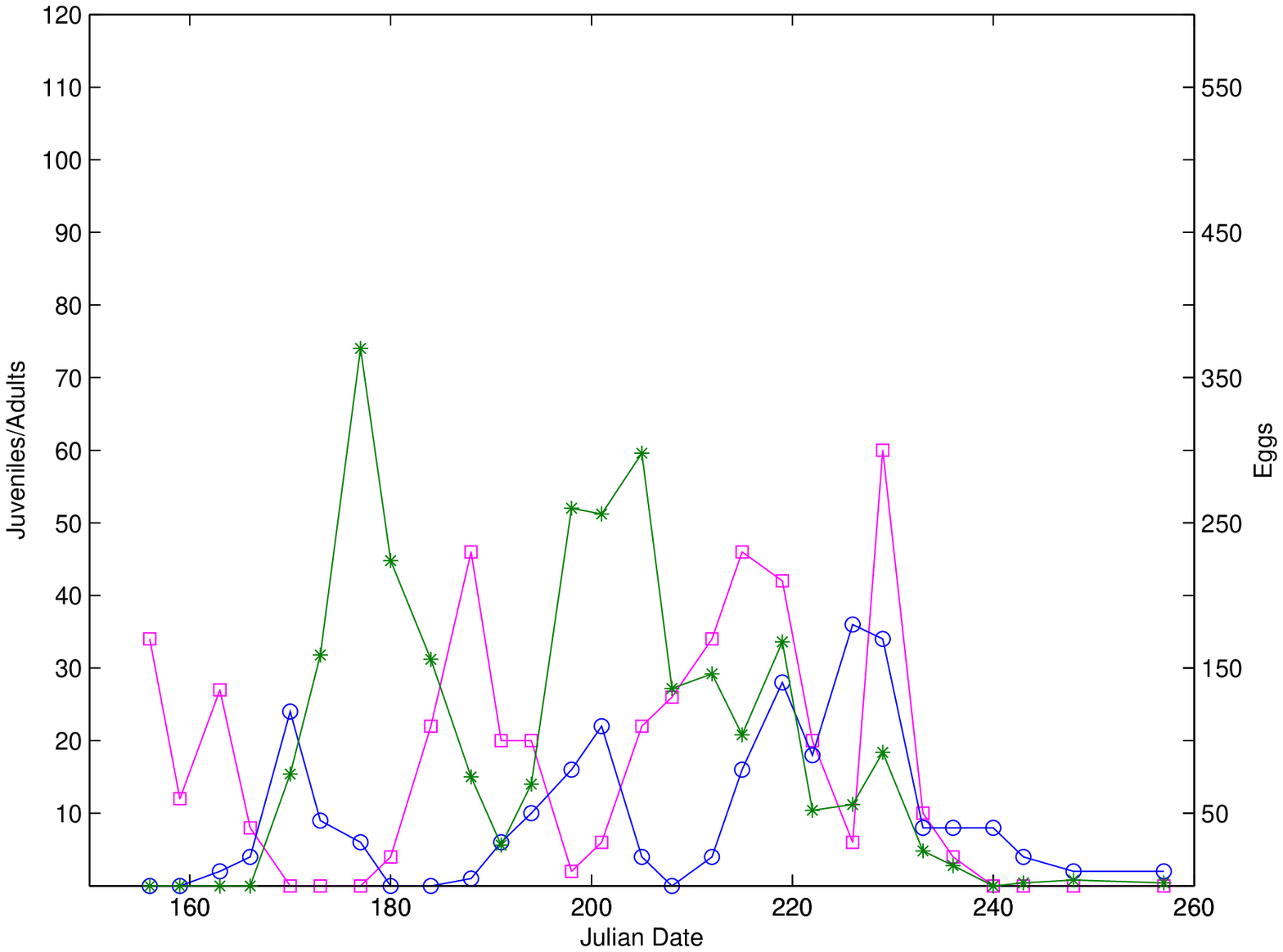} &
 \includegraphics[height=2.15cm,width=4cm]{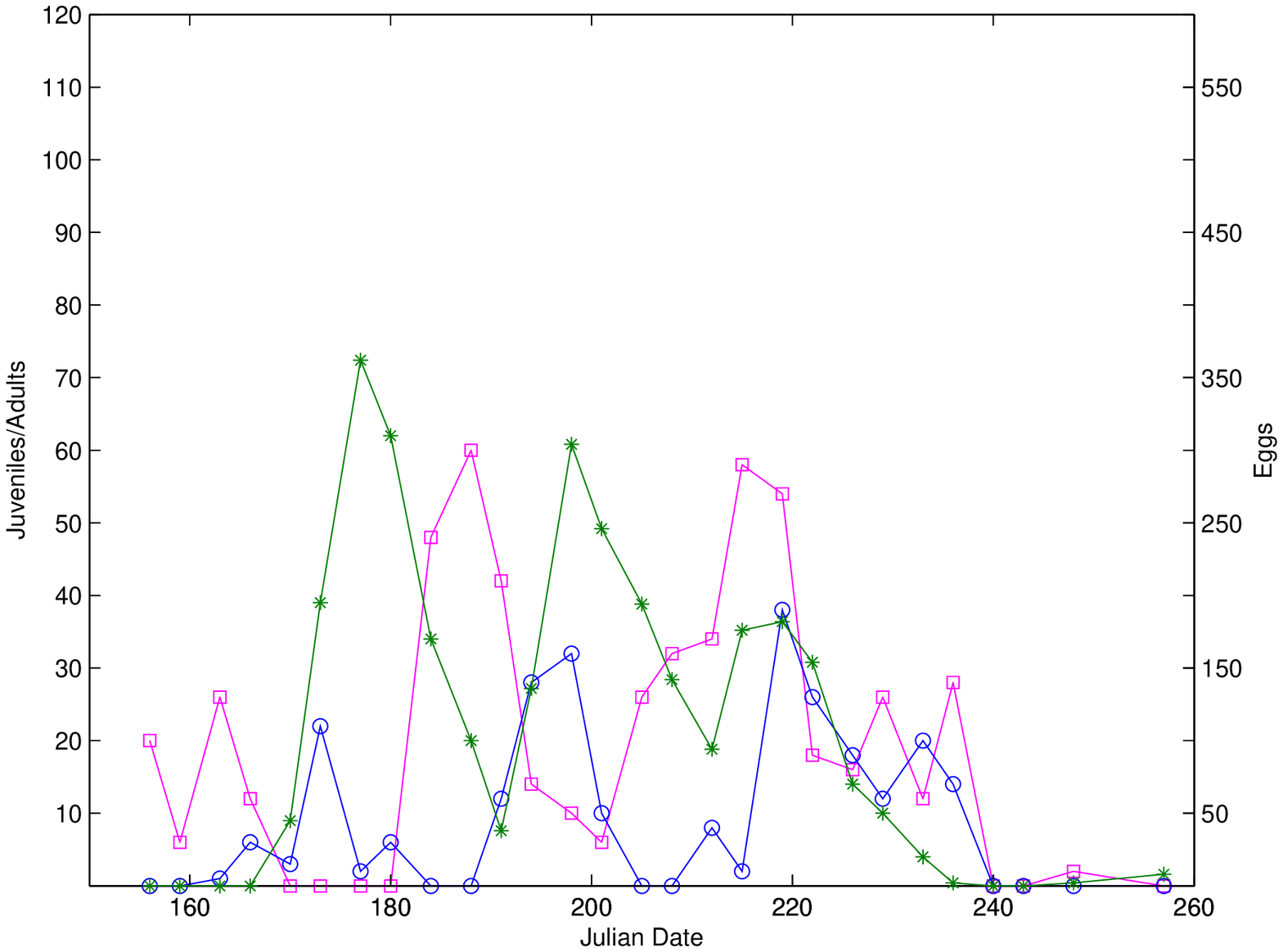}   \\
 \includegraphics[height=2.15cm,width=4cm]{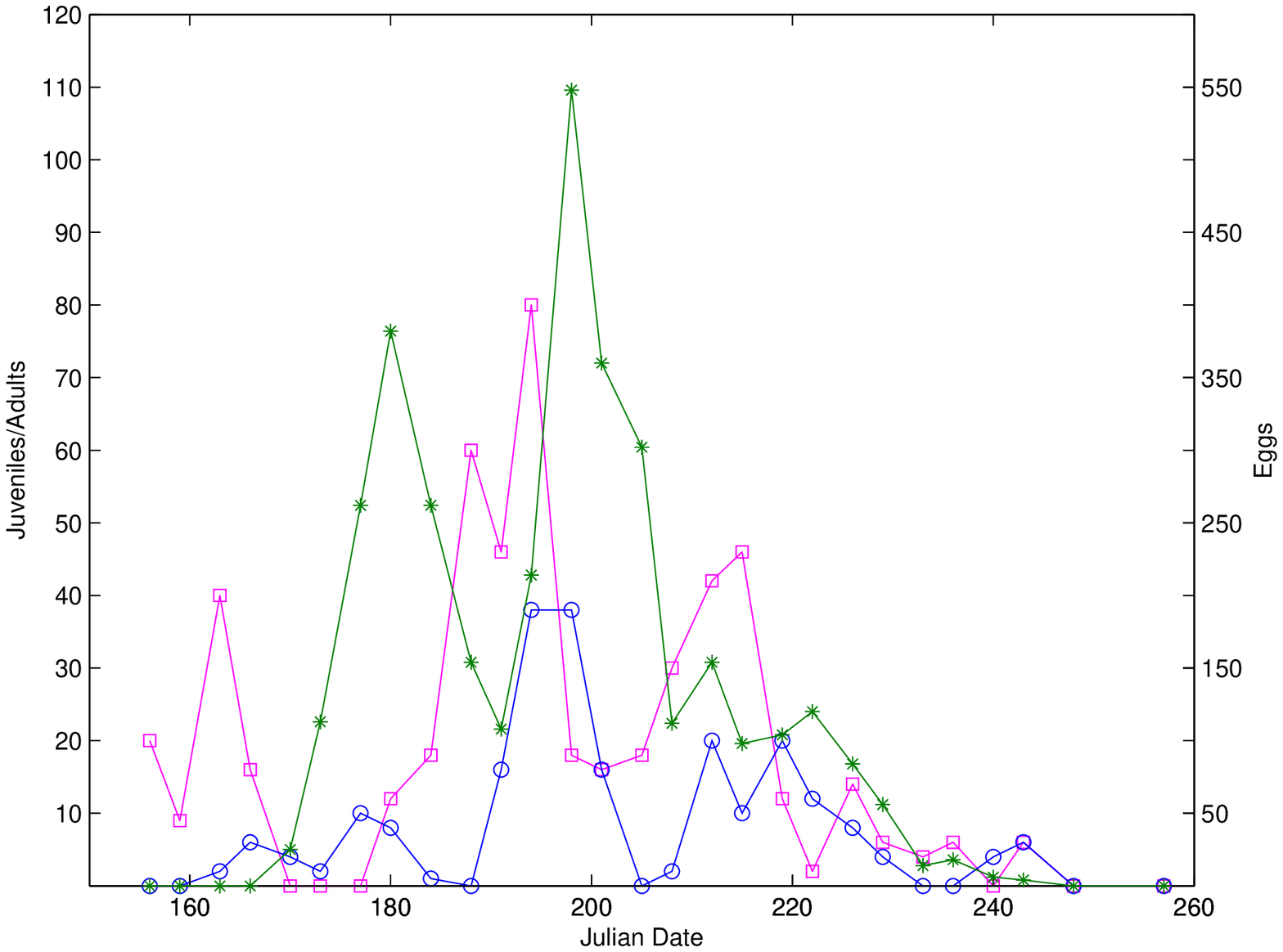} &
 \includegraphics[height=2.15cm,width=4cm]{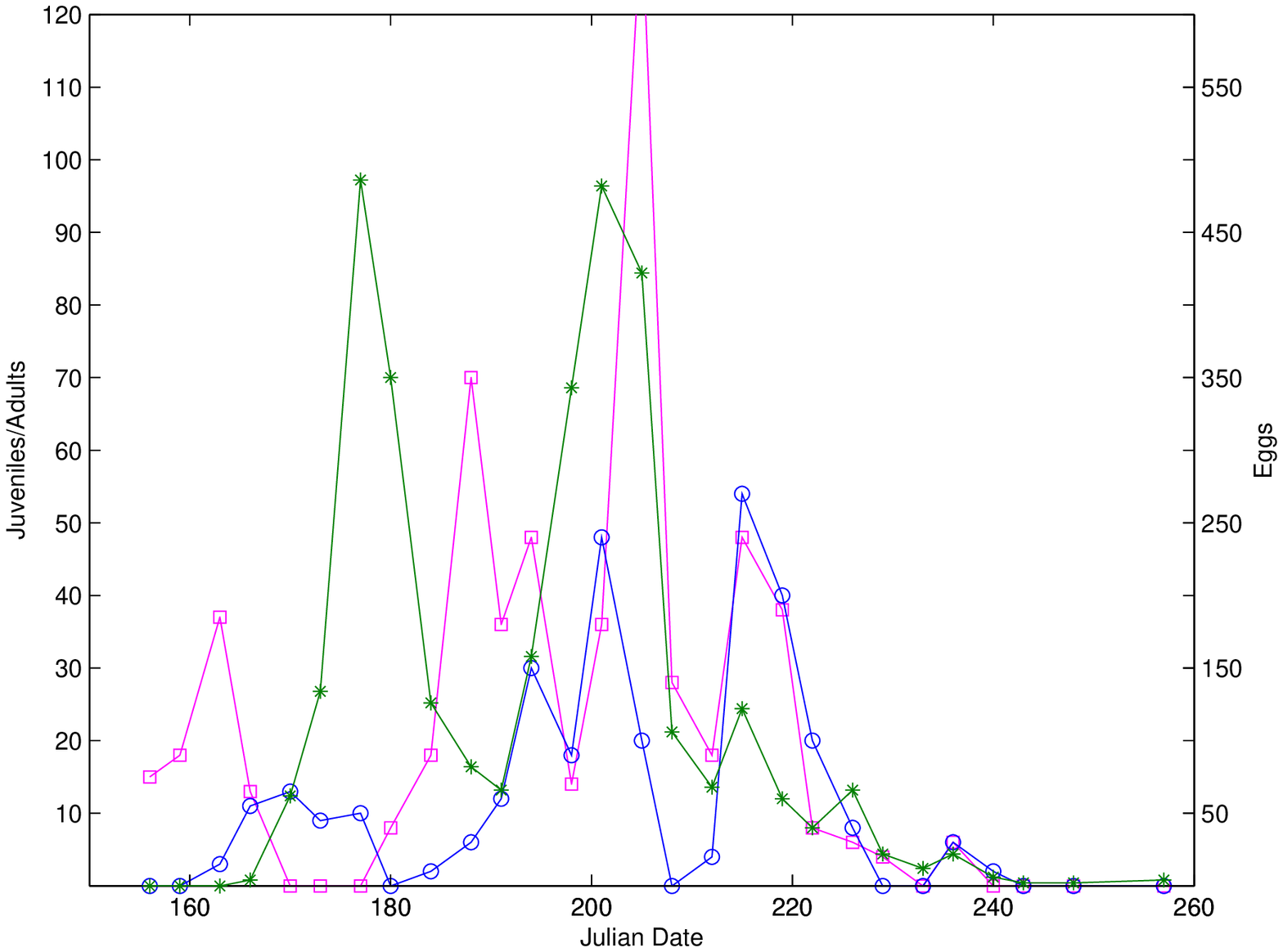} &
 \includegraphics[height=2.15cm,width=4cm]{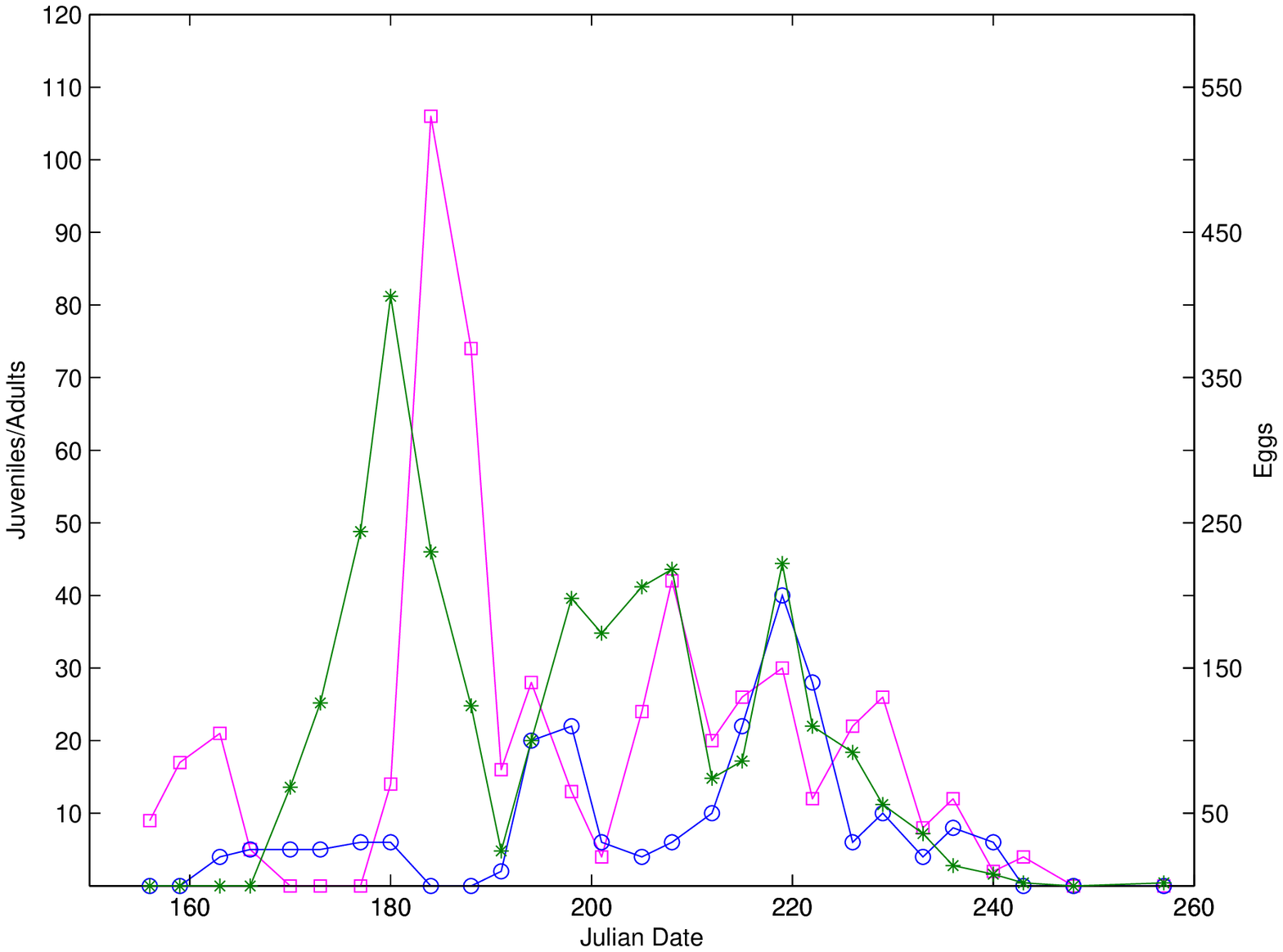}  &
 \includegraphics[height=2.15cm,width=4cm]{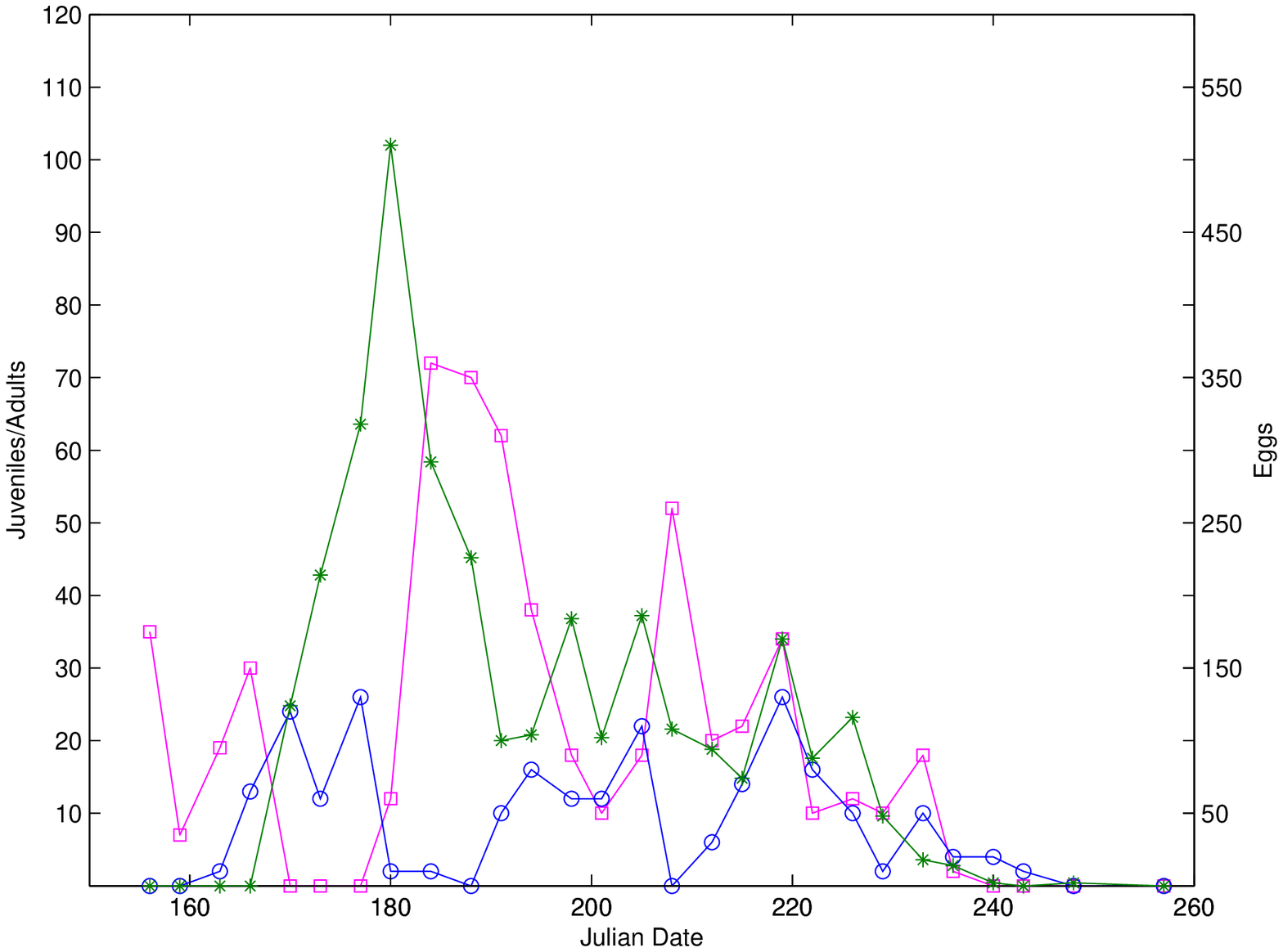} \\
 \includegraphics[height=2.15cm,width=4cm]{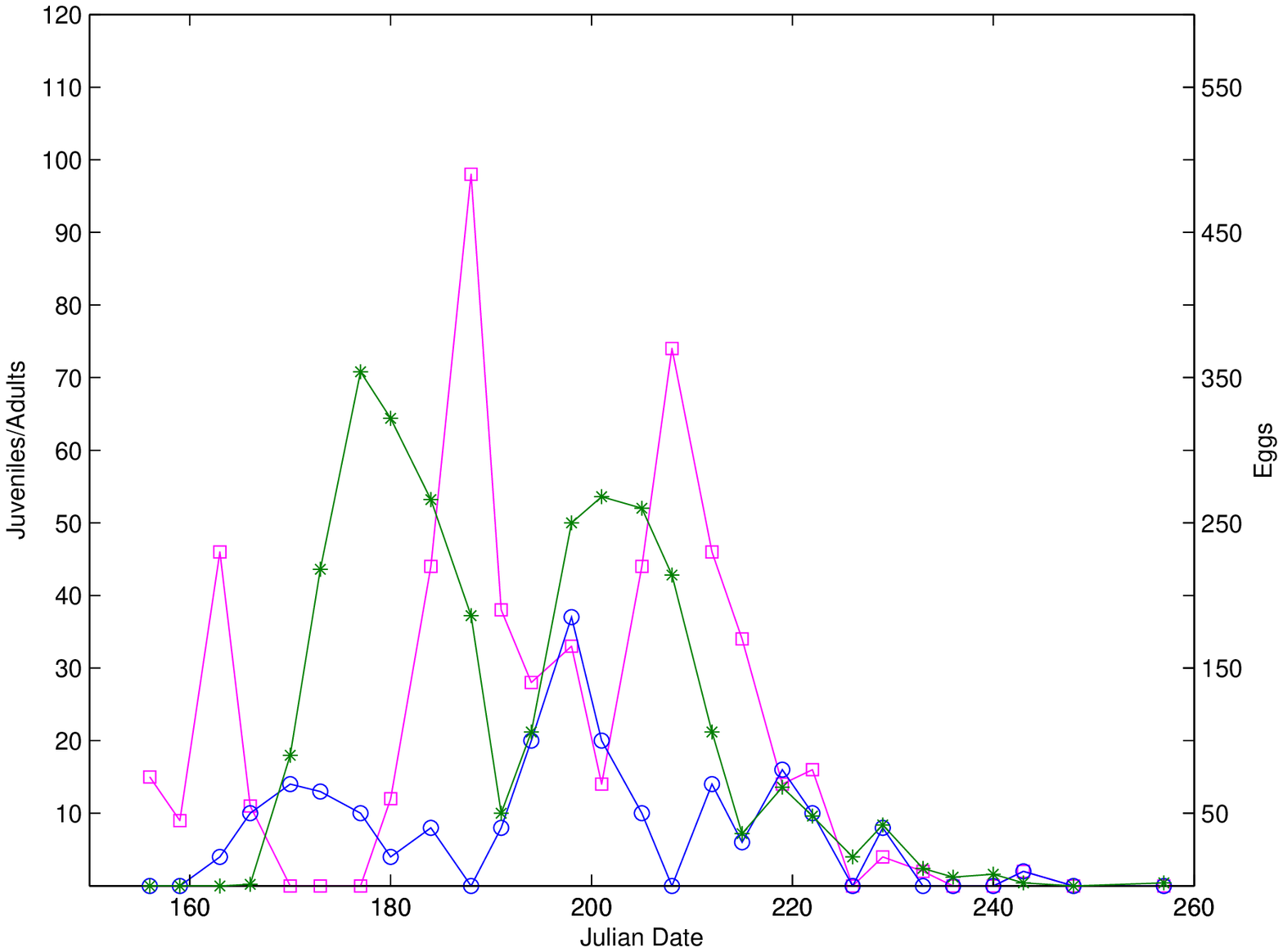} &
 \includegraphics[height=2.15cm,width=4cm]{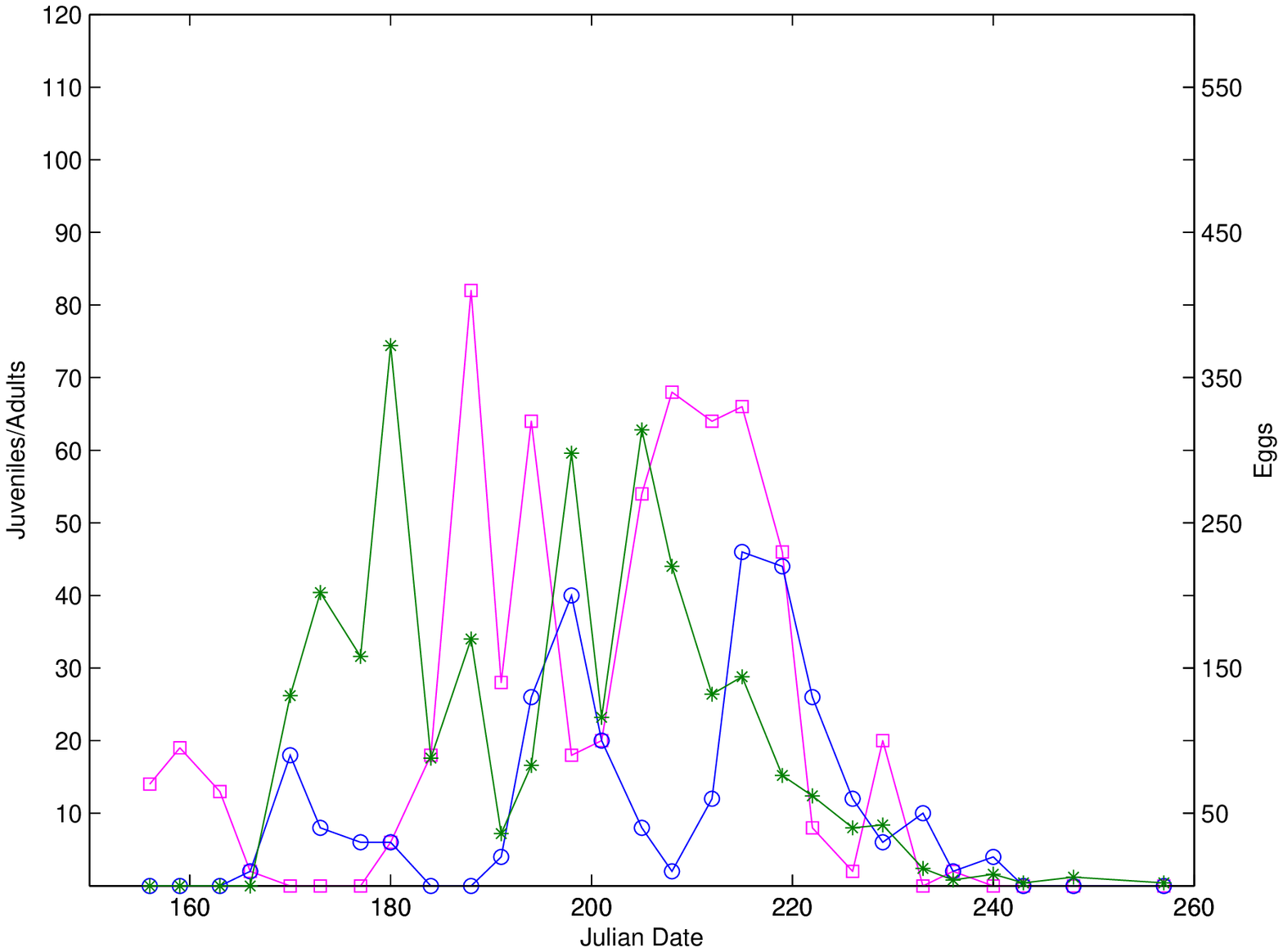}   &
 \includegraphics[height=2.15cm,width=4cm]{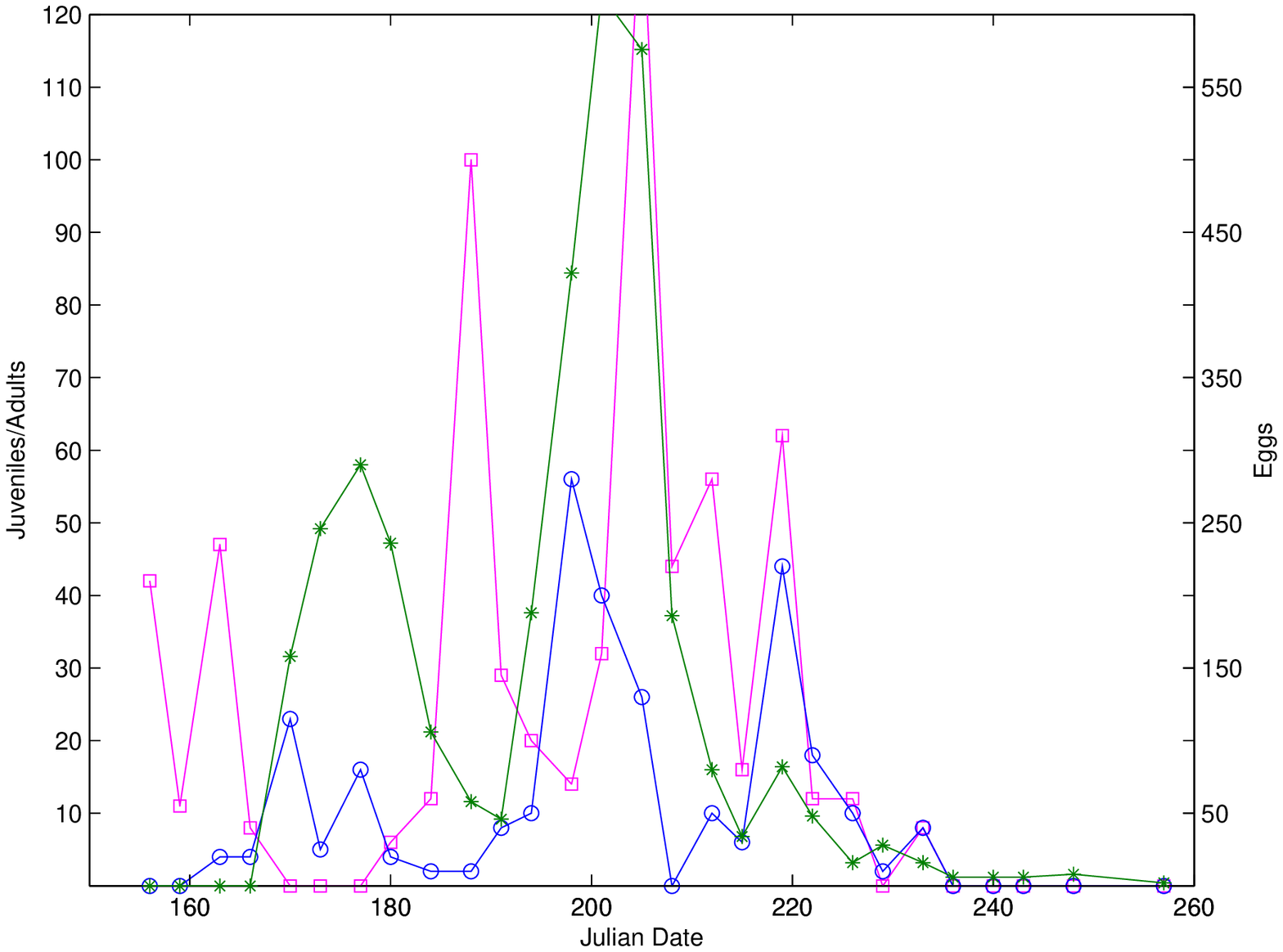} &
 \includegraphics[height=2.15cm,width=4cm]{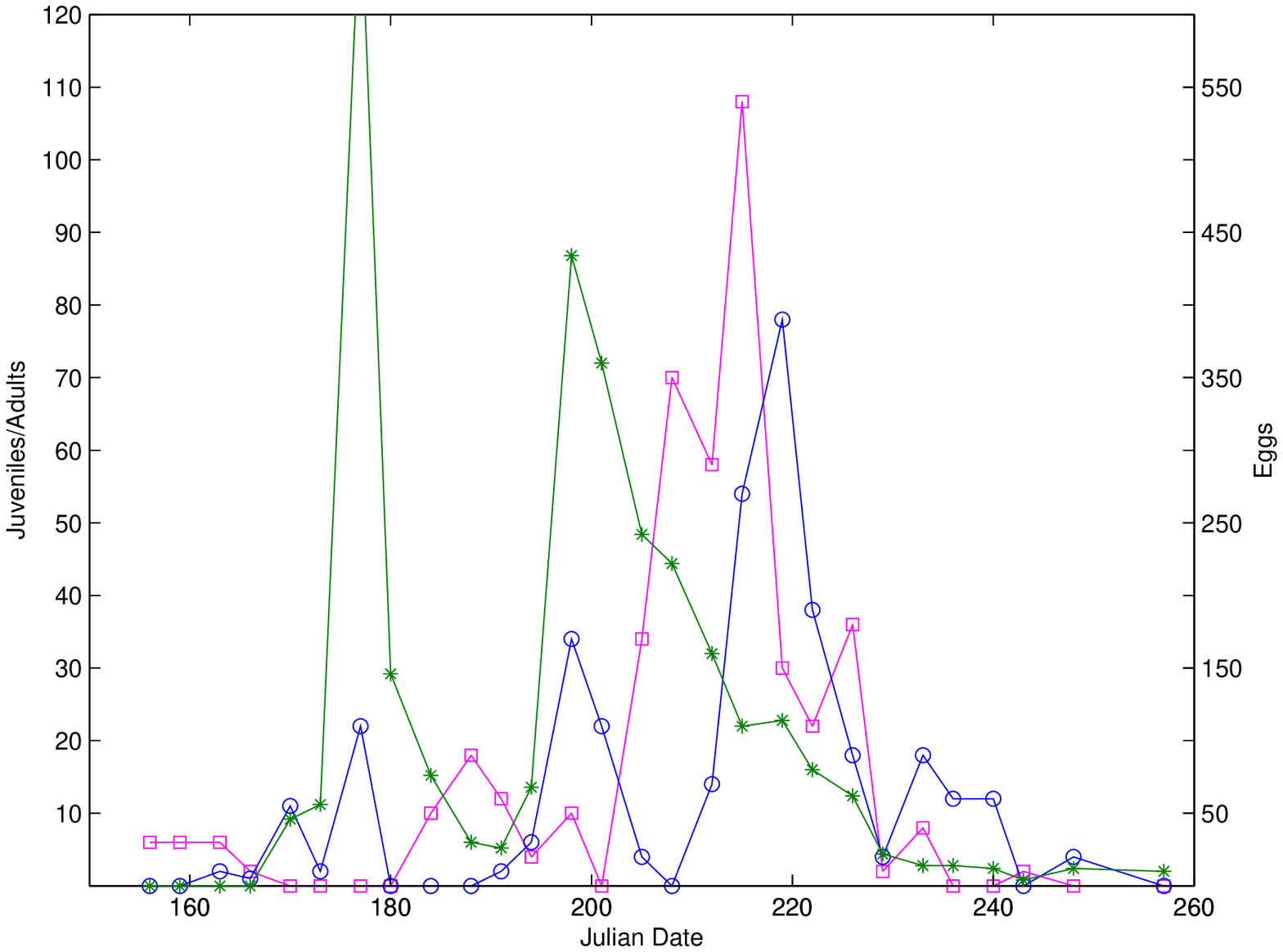} \\
 \includegraphics[height=2.15cm,width=4cm]{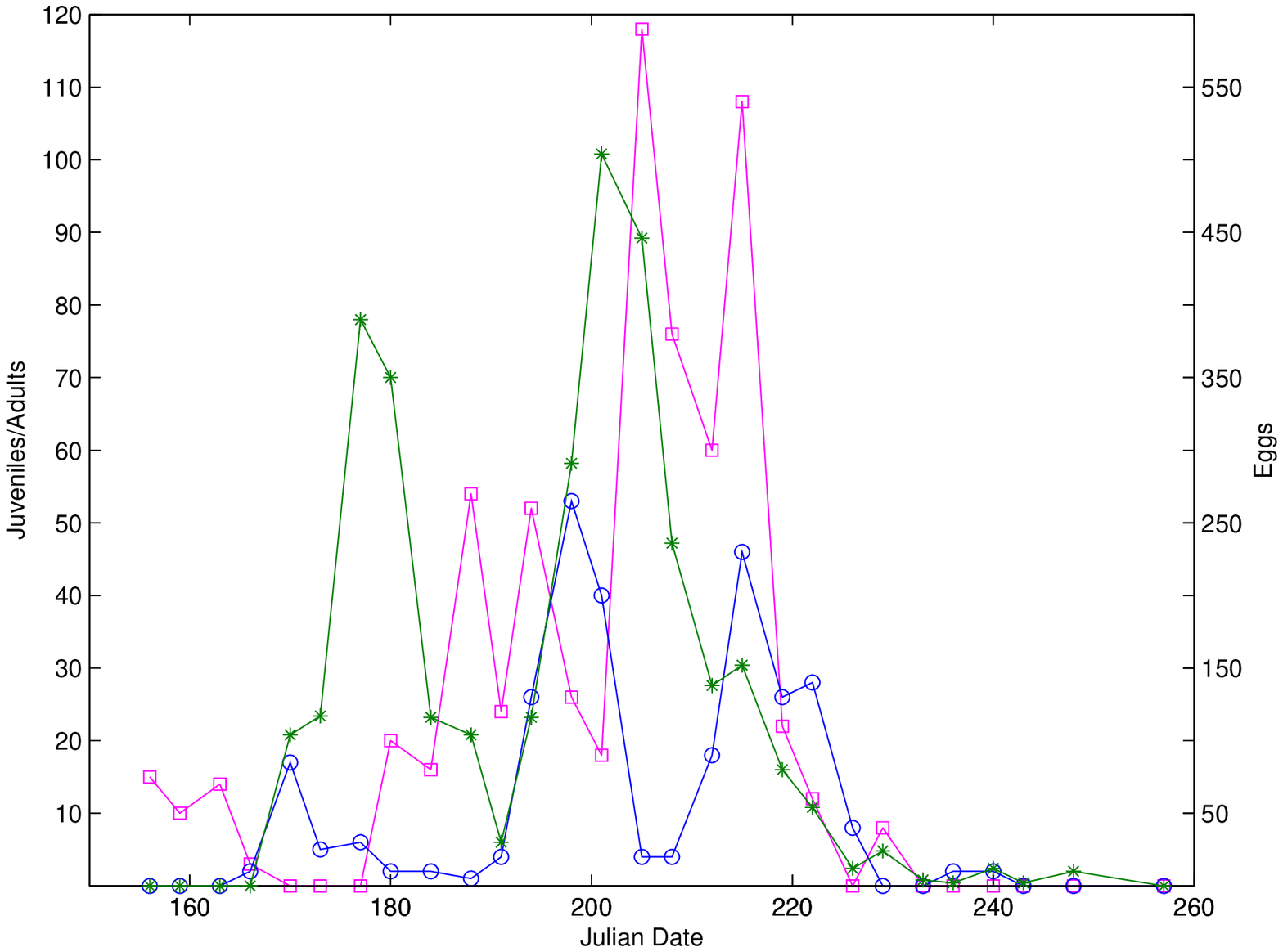}  &
 \includegraphics[height=2.15cm,width=4cm]{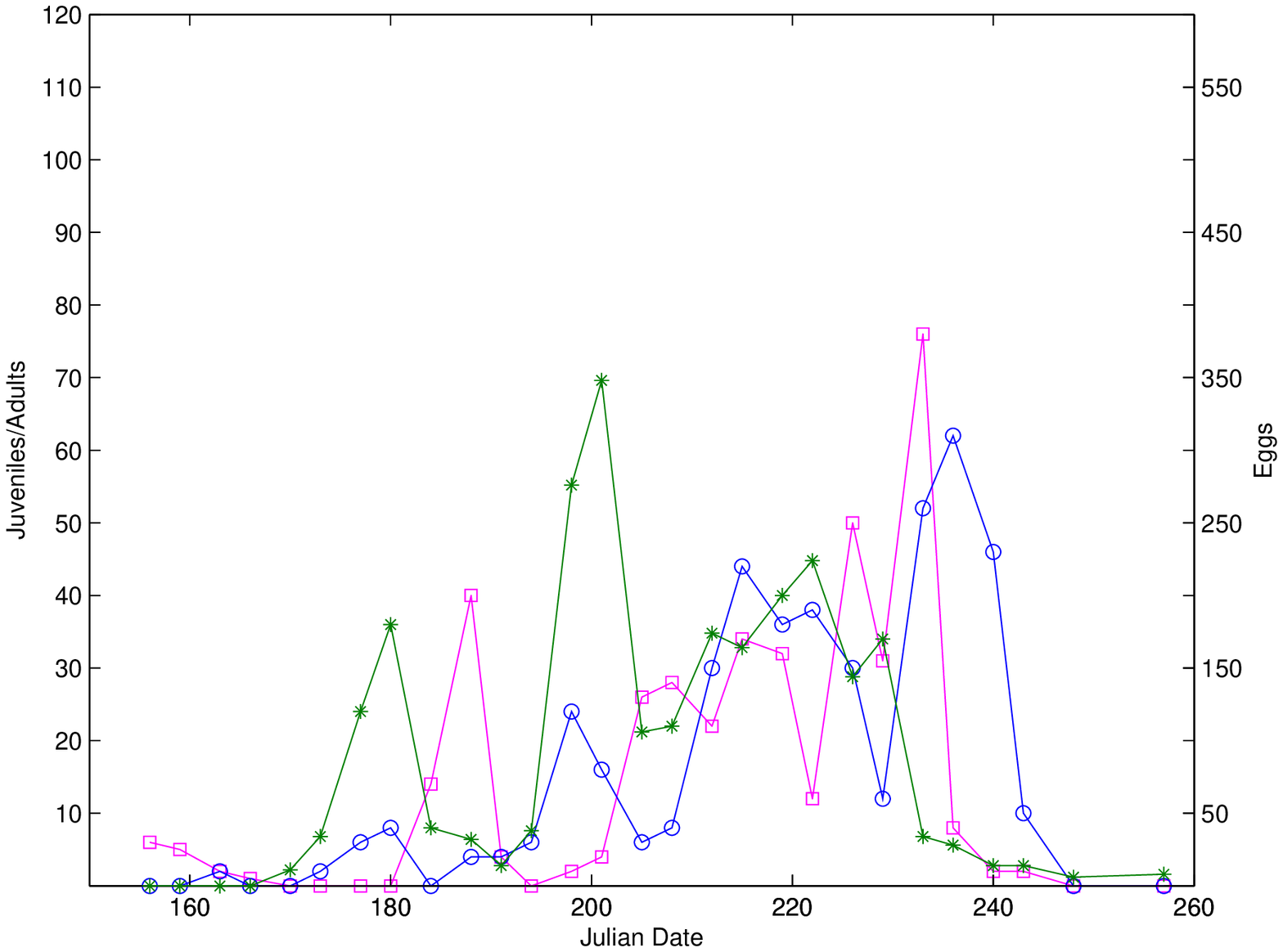} &
 \includegraphics[height=2.15cm,width=4cm]{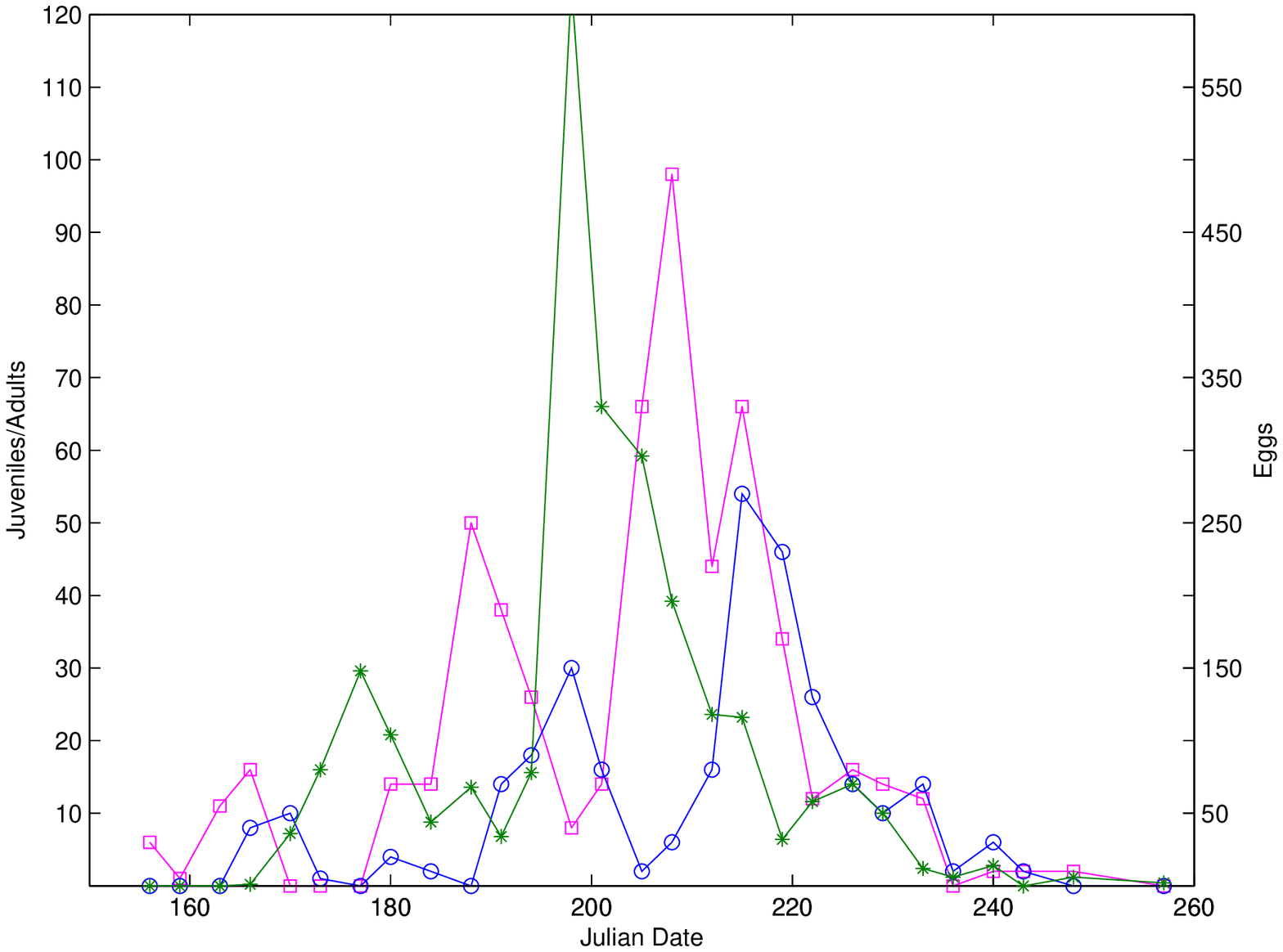} &
 \includegraphics[height=2.15cm,width=4cm]{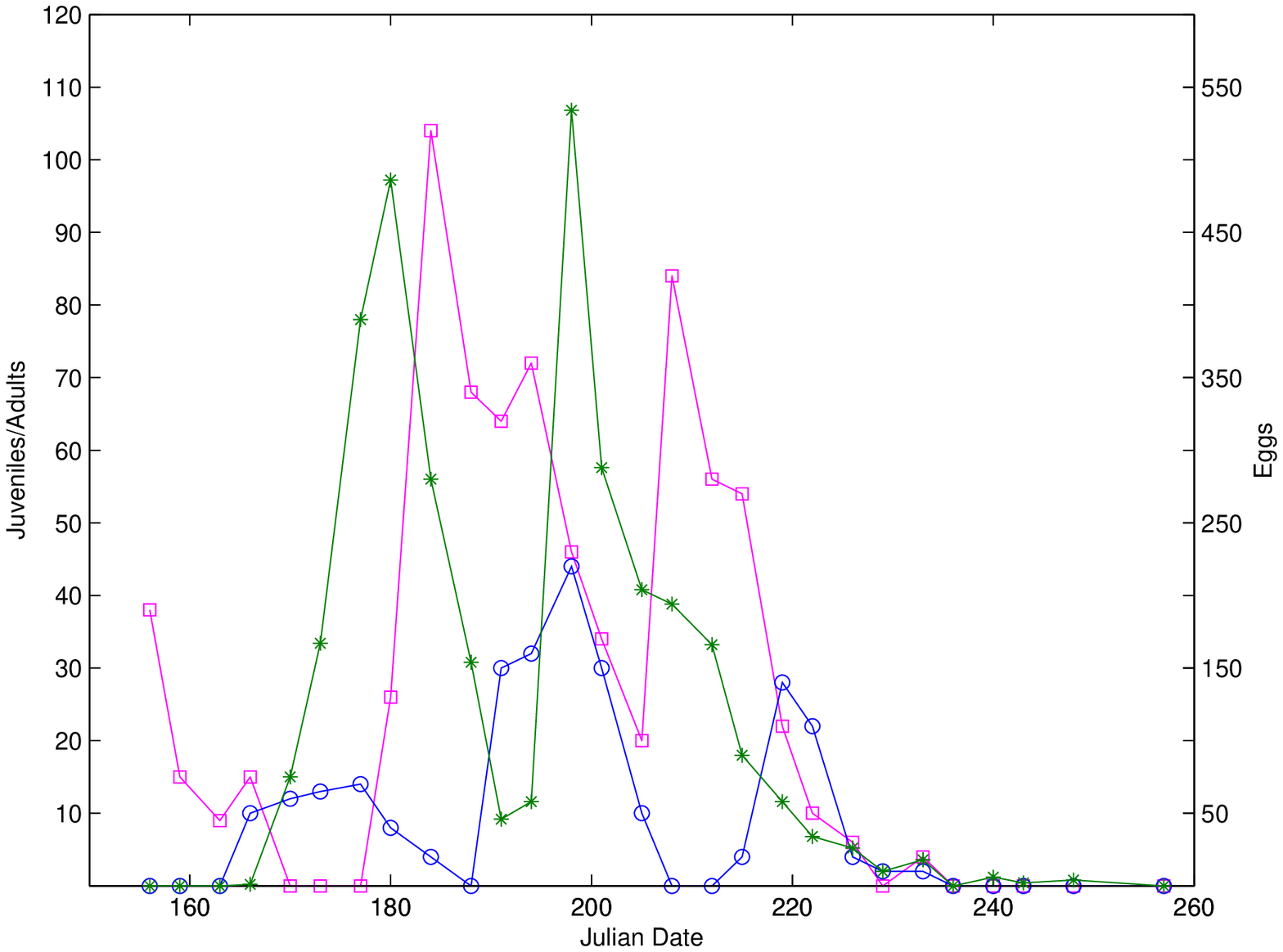}   
  \end{tabular} 
 \vspace{-4ex}
 \caption{The 24 replicates of the mite data. Fig.~\ref{Fig1} shows the life-stage averages of these data.
 (Colour coding consistent with Fig.~\ref{Fig1}.)} \label{data}
 \end{figure}  

\begin{figure}[h]
\begin{center}
\begin{tabular}{c}
 \includegraphics[height=11cm,width=9cm]{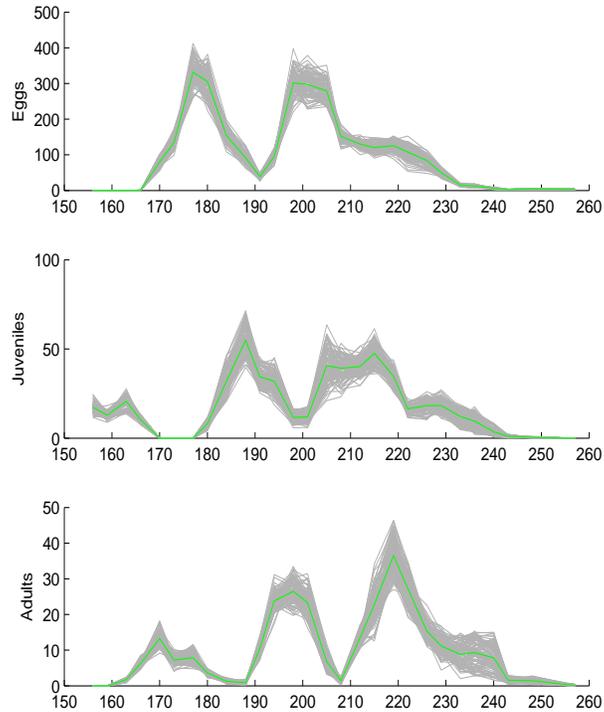} 
 \end{tabular} 
 \end{center} 
 \vspace{-8ex}
 \caption{The 100 times series (grey curves) for each life stage: top eggs; middle juveniles; bottom adults. 
 Green curves: means.} \label{random}
 \end{figure}

 The growth-rate distributions shown in Fig.~\ref{figdist} and below were computed from an enlarged data set of 100 samples, which were generated by averaging 10 randomly selected data sets of the 24 measured replicates shown in Fig.~\ref{data}.  The resulting 100 times series for each age group are exhibited in Fig.~\ref{random} (grey curves).
 \vspace{-1ex}

\section{Models}
\label{Models}
\subsection*{Simple (non-seasonal) growth models (I \& II)} 
The exponential (I) and logistic (II) models may be parametrized as
\begin{equation} N(t) = \frac{KN_0e^{rt}}{K-N_0 +N_0e^{rt} }
\label{simple}
\end{equation}
where, formally, the exponential case (I) corresponds to the 
\textit{carrying capacity} $K$ becoming arbitrarily large: $K\to \infty $.  The other two parameters $N_0$ and $r $ have the  interpretation of \textit{initial population size}  and \textit{growth rate}, respectively. 
\pagebreak 
\subsection*{Phenomenological models (III \& IV)} 
\label{F1} 
Let $X(t)$ be either $E(t)$, $J(t)$, or $A(t)$. Then the seasonal cycle in the data may be modelled by 
an exponential (III) or logistic (IV) rise (for $t\le S$), followed by a simple exponential decline ($t> S$):   
\begin{subequations}
\begin{equation} X_S(t) = \left\{ \begin{array}{ll} \frac{KX_0e^{rt}}{K-X_0 +X_0e^{rt} }  
, & t\le S \\
\frac{KX_0e^{rS}}{K-X_0 +X_0e^{rS} }  e^{-\mu (t-S)} , & t> S \end{array} \right. ,
\label{cycle}
\end{equation} 
where, as above, the linear case (III) corresponds to the limit $K\to \infty $.  
The five parameters $X_0,K,r,S,\mu $ have the interpretation: $X_0$ initial size of the age group; $K$ carrying capacity; $r$ growth rate; $S$ \textit{seasonal parameter}: time during the season at which mites stop laying same-season eggs; $\mu $ \textit{per-capita death rate}.  

The purely seasonal curve $X_S(t)$ is modulated by generational, non-seasonal, oscillations:
\[ X(t) = X_S(t) \left[1+ I\sin \left(2\pi ( t - \phi _0)/P\right)\right] \]   
This adds 3 more parameters: (relative) amplitude $I$, period $P $, and phase shift $\phi _0$. 

Now letting $X(t) = E(t)$ and assuming that the three age groups only differ by total phase shifts (parameters 
$\alpha _0$ and $\alpha _1$) 
and size (parameters $f$ and $g$), results in the 12-parameter model   
\begin{eqnarray}
 E(t) \hspace{-1.5ex} &=& \hspace{-1.5ex}X_S(t) \left[1+ I\sin \left(2\pi ( t - \phi _0)/P\right)\right]  \label{F1E}\\
 J(t) \hspace{-1.5ex} &=& \hspace{-1.5ex} fE(t-\alpha _0)  \hspace{-.5ex}= \hspace{-.2ex}
 f X_S(t-\alpha _0) \hspace{-.3ex} \left[1+ I\sin \left(2\pi ( t - \phi _0-\alpha _0)/P\right)\right] \qquad
   \label{F1J}\\
 A(t) \hspace{-1.5ex} &=& \hspace{-1.5ex}g E(t-\alpha _1) 
 \hspace{-.2ex}= gX_S(t-\alpha _1) \hspace{-.3ex} \left[1+ I\sin \left(2\pi ( t - \phi _0-\alpha _1)/P\right)\right]  \qquad \label{F1A}
  \end{eqnarray} 
\end{subequations} 
 \subsection*{Demographic models (V \& VI)}   
 \label{DDE} 
 The underlying  demographic model   is the 
 \textit{Sharpe-Lotka-McKendrick} partial differential equation (PDE) \cite{webb1985}.  Under the assumption of piecewise--constant demographic data (fecundity and mortality), \textit{delay}--differential--equation (DDE) models may rigorously be derived  from the full demographic model \cite{gould2007}. 
 These DDE models are reminiscent of \textit{ordinary}--differential--equation (ODE) compartment models, 
 but they have the advantage of properly accounting for the time individuals spend in the various life stages.
 Here eggs take $a_0$ days to hatch and the emerging juveniles take $\delta $ days to turn into reproducing adults.  
 The model we used reads
 \begin{subequations}
\begin{eqnarray}
\dot{E} (t)\hspace{-1.5ex} &=& \hspace{-1.5ex}e^{a_0\mu _0}\beta (t) A(t) -  \beta (t-a_0) A(t-a_0) -\mu _0E(t)  \label{dE}\\
\dot{J} (t)\hspace{-1.5ex} &=&\hspace{-1.5ex}f\beta (t-a_0) A(t-a_0)\hspace{-0.2ex} - \hspace{-0.2ex}fe^{-\delta \mu _1} \beta(t-a_1)A(t-a_1)
\hspace{-0.7ex}-  \hspace{-0.7ex}\mu _1J(t)   \qquad \label{dJ}\\
\dot{A} (t) \hspace{-1.5ex}&=&\hspace{-1.5ex}fe^{-\delta \mu _1} \beta(t-a_1)A(t-a_1)- \mu _2A(t)  \label{dA}
\end{eqnarray}  
where the $\dot{X} (t) = \frac{dX}{dt} $, $a_1=a_0+\delta $, 
\begin{equation} \beta (t) = \beta _me^{-a_0\mu _0}e^{-\nu M(t)} \left\{ \begin{array}{cl} 1, & t\le S \\
 0 , & t>S\end{array} \right\} \, \textrm{and} \,M(t) = J(t) + A(t). \label{dF} \end{equation}  
 \end{subequations} 
The parameters are: $a_0$ and $a_1=a_0+\delta $ --  \textit{hatching} and \textit{maturation age}; 
$\mu _0$, $\mu_1 $, and $\mu _2$ -- \textit{per-capita death rate} for eggs, juveniles and adults, 
respectively; $f$ -- \textit{egg-to-juvenile attrition rate}; $\beta _m$ -- (maximal) \textit{fecundity}; $\nu $ -- 
\textit{nonlinearity coefficient} parametrizing the reduction in fecundity due to overcrowding\footnote{According to the 
biology of the species under consideration (European red mite \textit{Panonychus ulmi}), this is the dominant density effect; increased mortality 
due to overcrowding is believed to be a secondary effect. Note that logistic growth model (\ref{simple}) may also be  
 interpreted in this way by writing it as an ODE: 
$\frac{d}{dt} N(t) = \beta _m f(N(t)) N(t) -\mu N(t)$, where the density dependence is given by 
 $f(N) = 1-\nu N $ (which may be viewed as an expansion of the exponential expression 
 $e^{-\nu N}$ appearing in the demographic model). Then $r=\beta _m -\mu $ and $K = r/\beta _m\nu$.}.  
The linear model (V) arises by setting $\nu = 0$. 
 
 To obtain \textrm{unique} solutions , the system has to be complemented by initial conditions, which 
in the case of DDEs, take the form of initial \textrm{functions}, $E_h(t),N_h(t),A_h(t)$ ($0\le t \le a_1 $), 
called  \textit{histories}.  Strictly speaking, prescribing whole \textrm{functions} amounts to prescribing   
an \textrm{infinite} number of parameters.  However, we  found that a \textrm{two-dimensional} parameterization
of the histories is sufficient in this modelling exercise; we chose the width of the initial age distribution and the total number of
beginning-of-season eggs.   

The \textit{Euler-Lotka equation} for this model takes the form
\begin{equation}
\lambda  = fe^{-[a_0\mu_0 + \delta \mu _1]} \beta _me^{-a_1\lambda } - \mu_2 ,
     \label{Euler-Lotka}
     \end{equation} 
i.e. the \textit{growth rate} (Lotka's $r$) is the largest real root $r=\lambda $ of (\ref{Euler-Lotka}). 
\newpage 
\section{Parameter estimation (model fitting)} \label{Fit}
\begin{figure}[h]
\hspace{-15ex}
\begin{tabular}{|c||c|c|cc|}
\hline  
   &   \multicolumn{3}{c}{Time Window} &\\
 \hline 
    Model & 30 days & 50 days  & 70 days & \\
    \hline \hline 
 \multirow{6}{*}{\centering I} &  \multirow{6}{*}{\includegraphics[height=3cm,width=5cm]{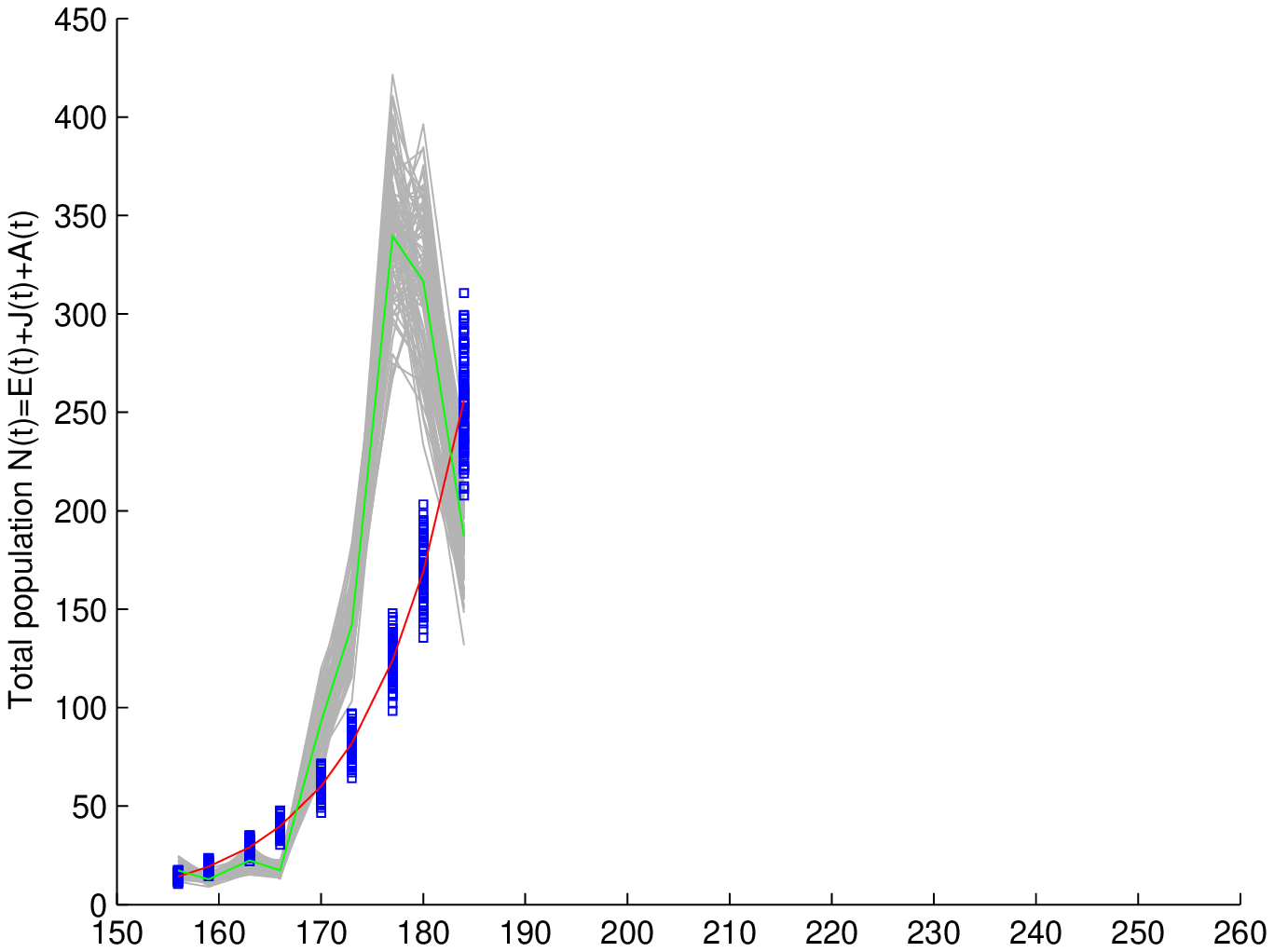} }&
  \multirow{6}{*}{ \includegraphics[height=3cm,width=5cm]{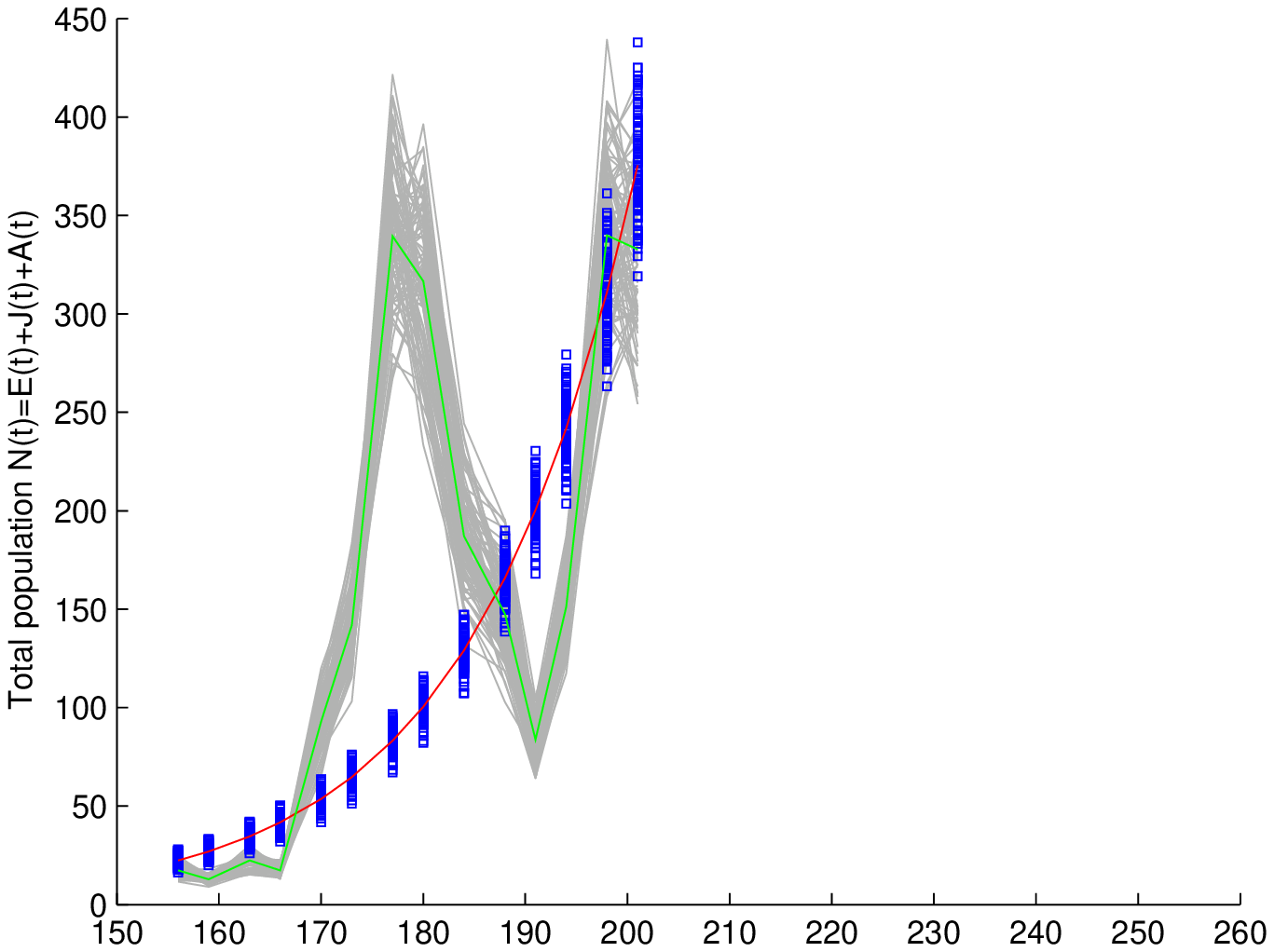}} &
  \multirow{6}{*}{\includegraphics[height=3cm,width=5cm]{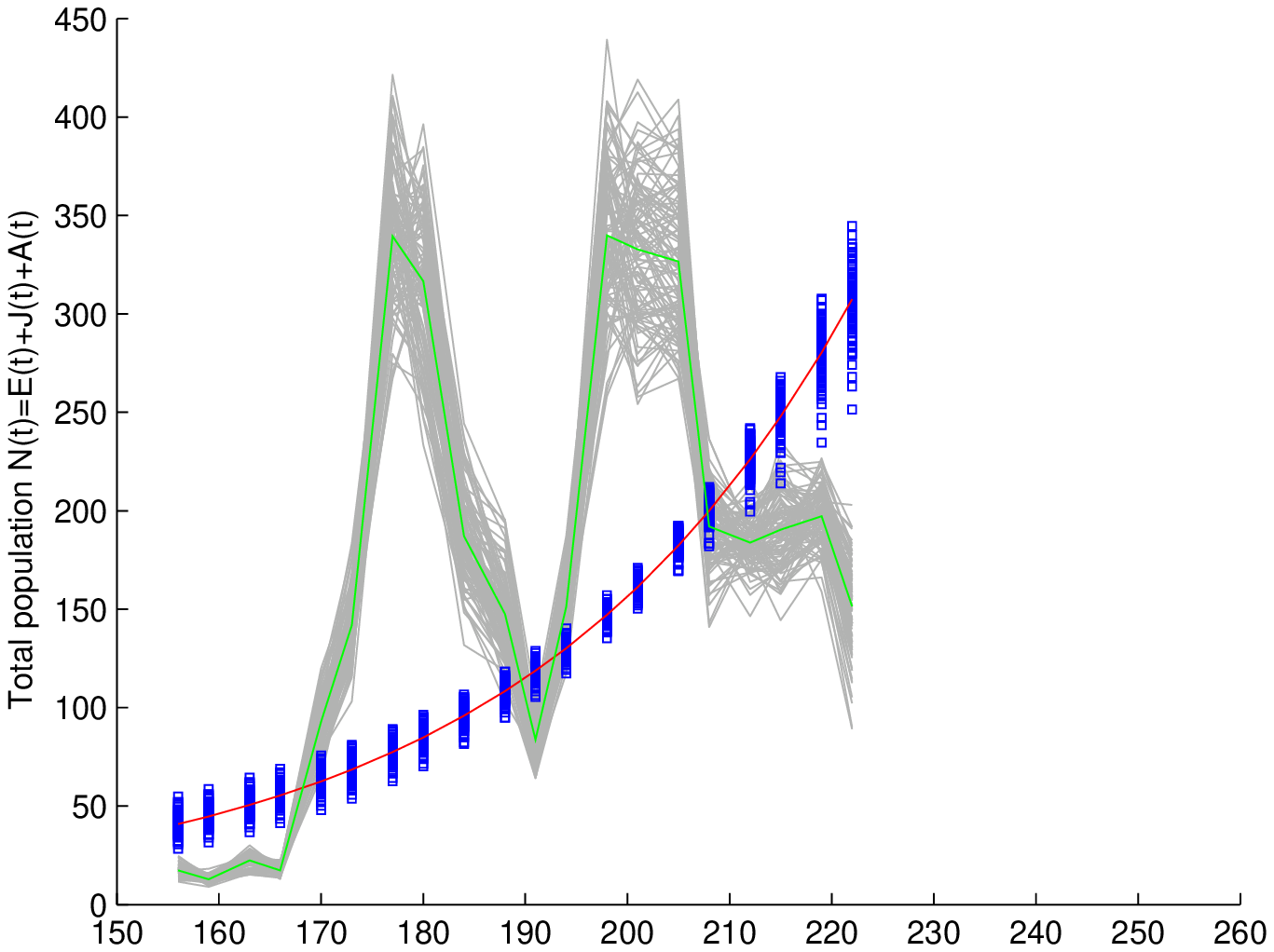}}  &\\
  & & & & \\
  & & & & \\
  & & & & \\
  & & & & \\
  & & & & \\
 \hline 
\multirow{6}{*}{\centering II} &\multirow{6}{*}{\includegraphics[height=3cm,width=5cm]{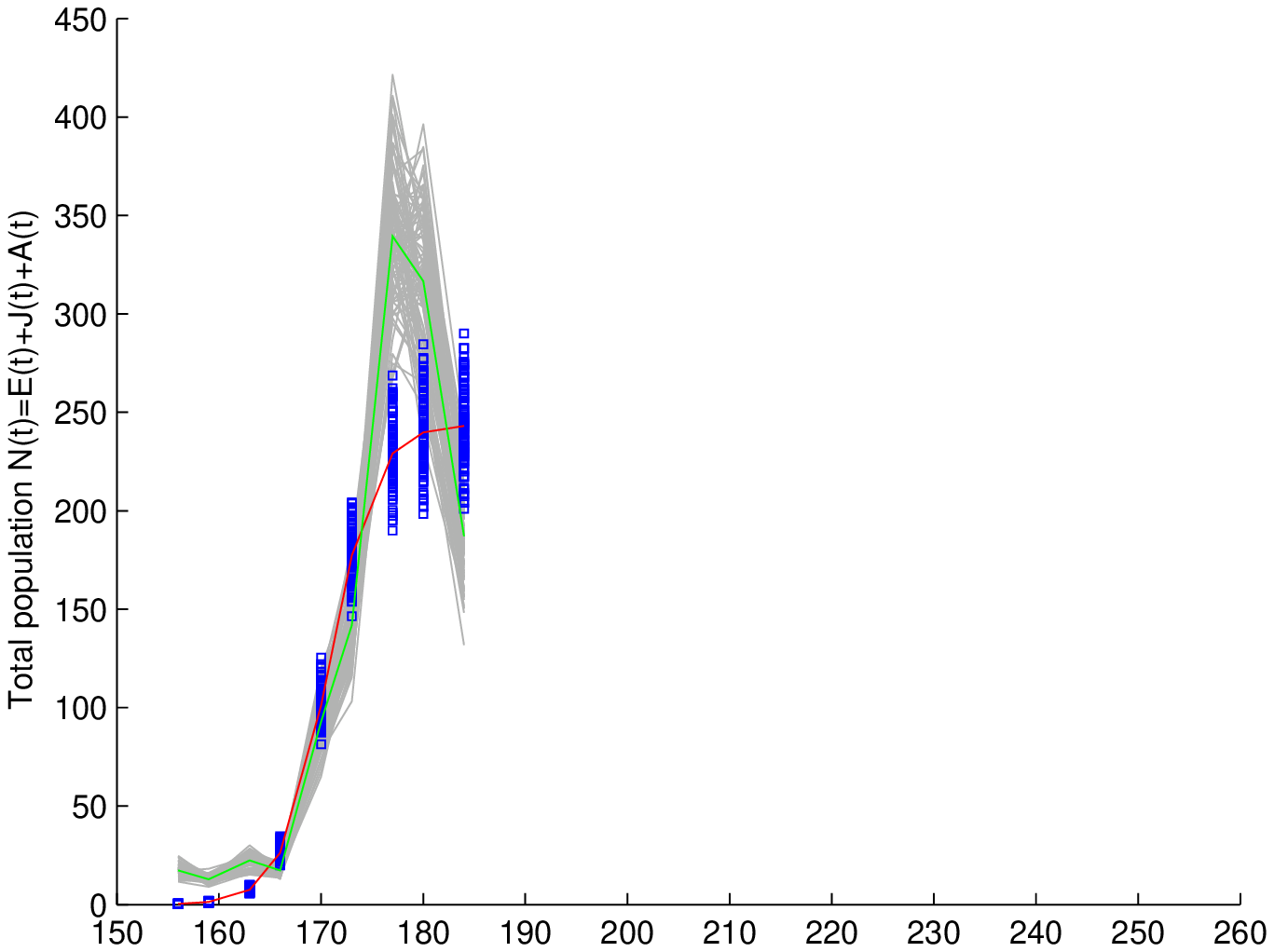}} &
  \multirow{6}{*}{\includegraphics[height=3cm,width=5cm]{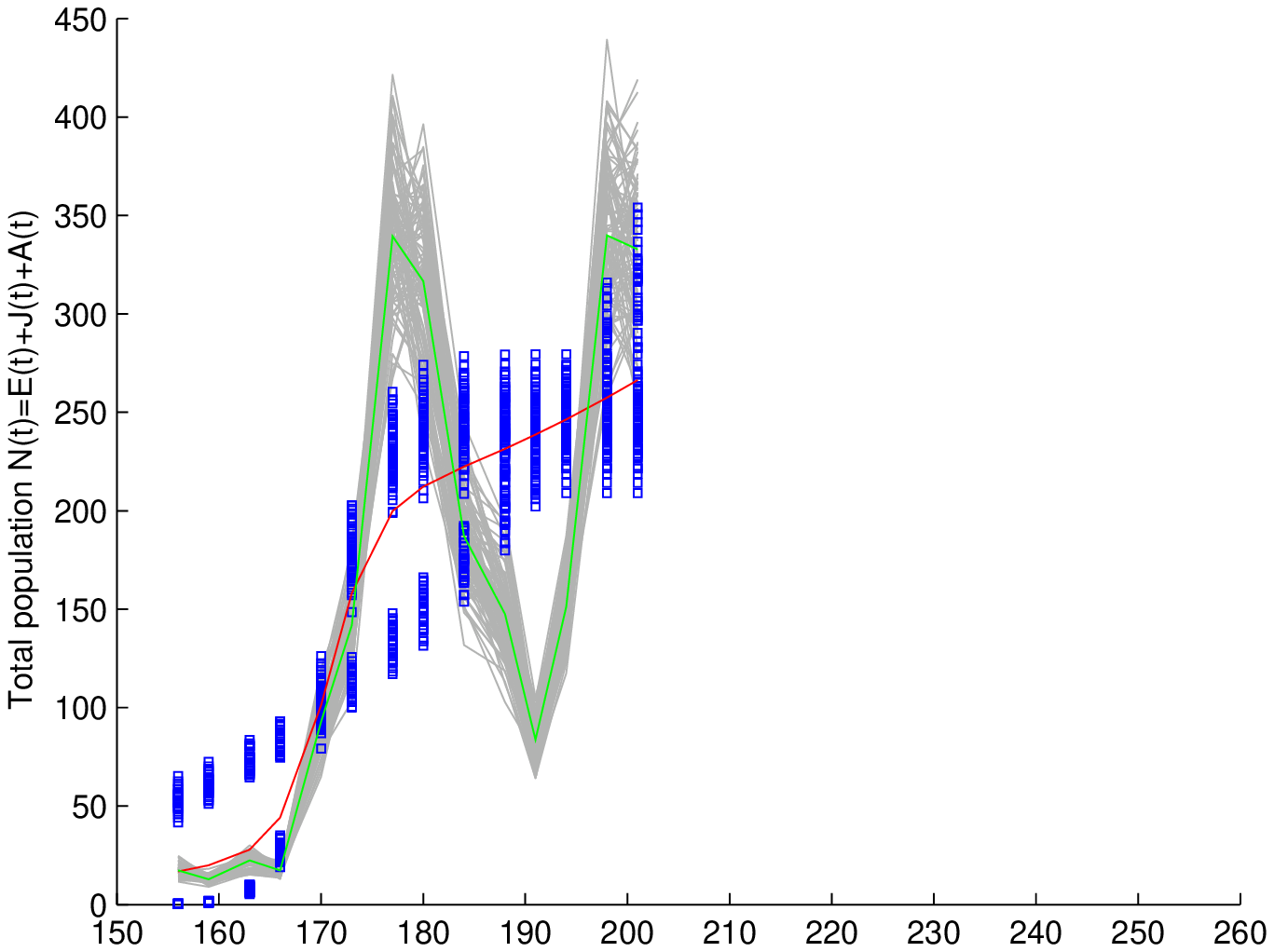} }&
 \multirow{6}{*}{\includegraphics[height=3cm,width=5cm]{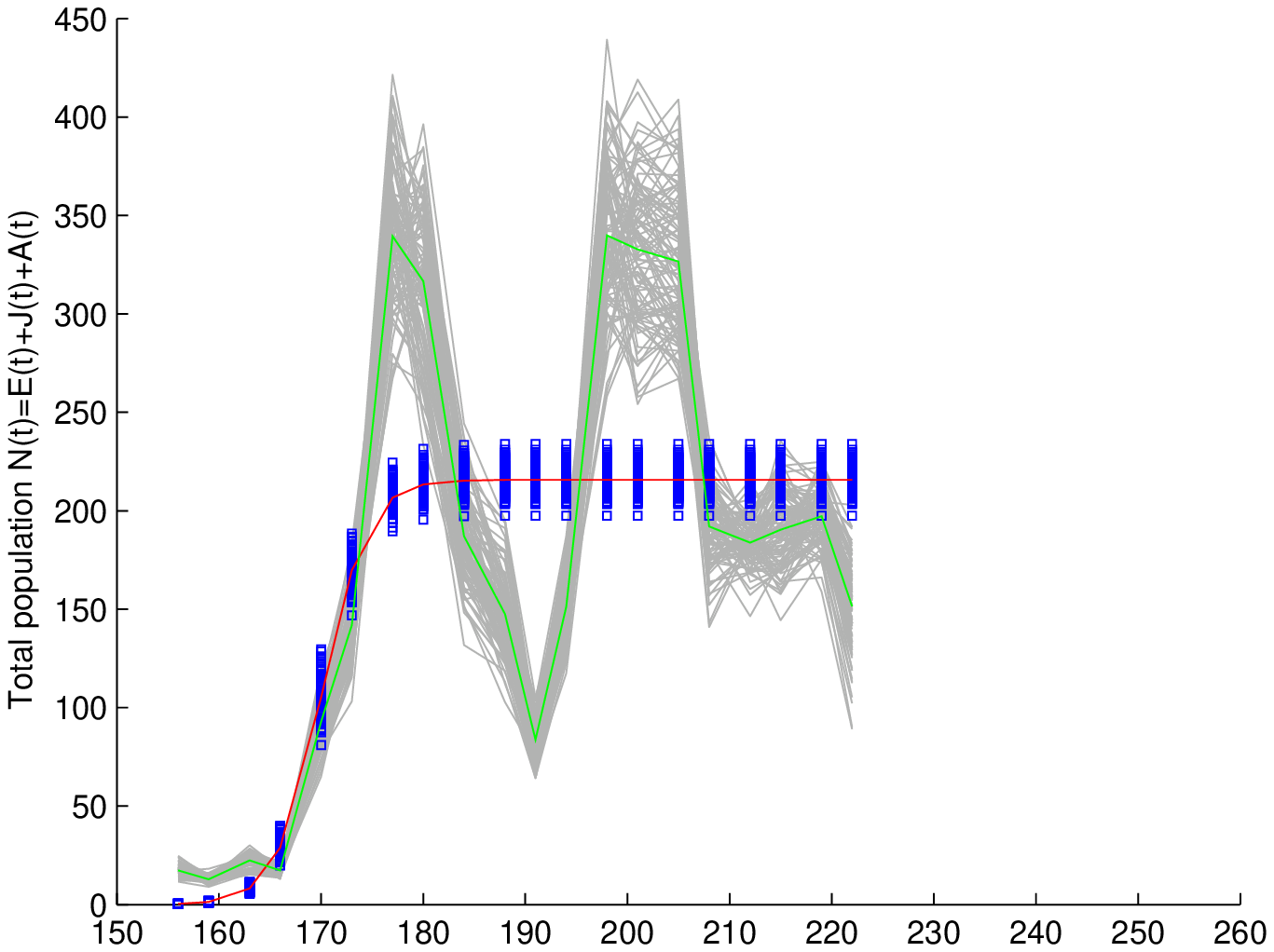} } & \\
  & & & & \\
  & & & & \\
  & & & & \\
  & & & & \\
  & & & & \\
 \hline  
  \end{tabular} 
 \vspace{-1.8ex}
 \caption{Fits of simple (non--seasonal) growth models I \& II. Each panel shows the  total population 
 $N(t) = E(t) + J(t) + A(t)$ for the 100 data sets (grey) and the values of the fitted models at the data points (blue).
 Green curve: mean of the 100 data sets; red curve: mean of fitted time series.} \label{fitsSimple}
 \end{figure}

\begin{figure}
\hspace{-15ex}
\begin{tabular}{|c||c|c|cc|}
\hline  
   &   \multicolumn{3}{c}{Time Window} &\\
 \hline 
    Model & 50 days & 70 days  & full season & \\
    \hline \hline 
 \multirow{9}{*}{\centering III} &  \multirow{3}{*}{\includegraphics[height=4cm,width=5cm]{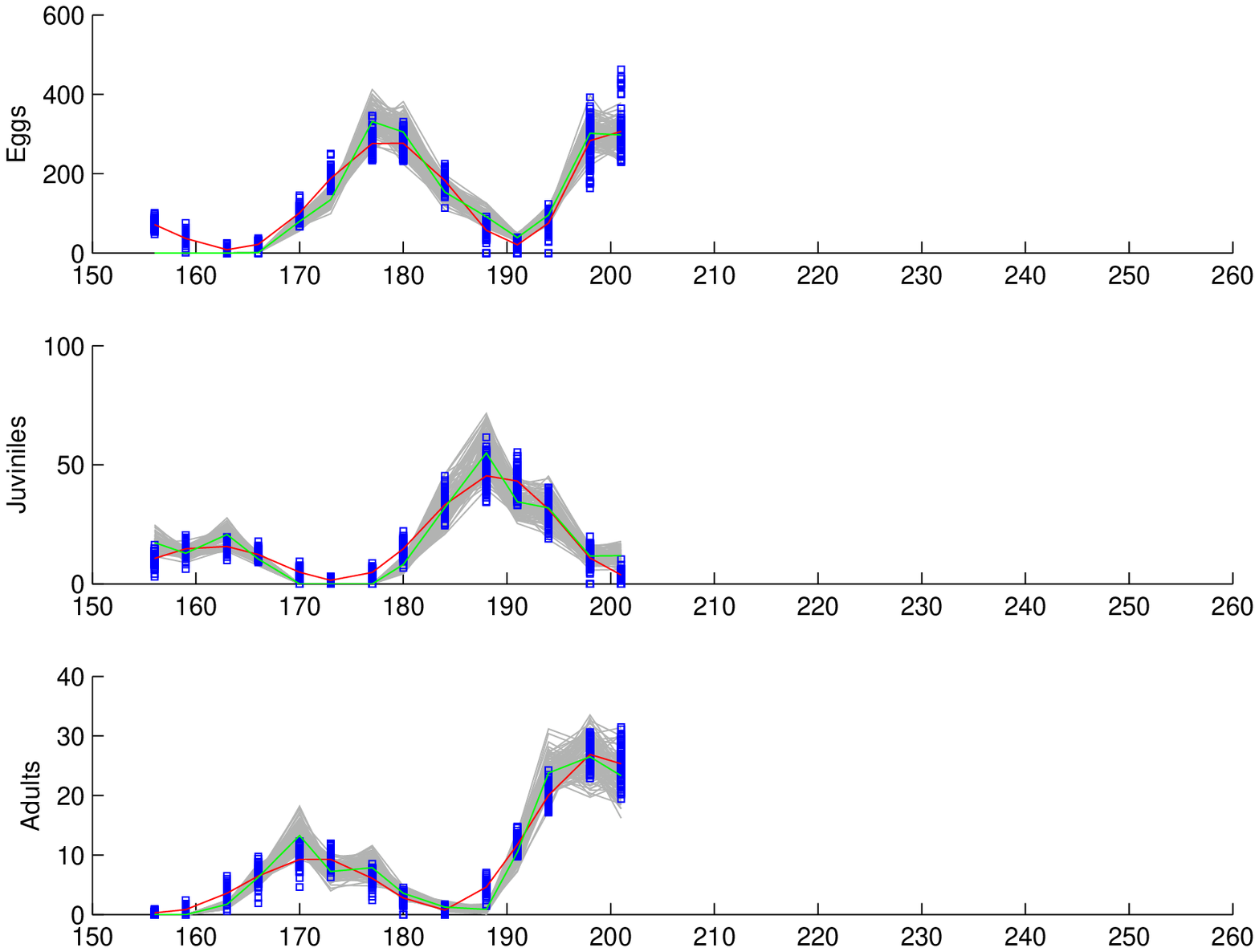} }&
  \multirow{9}{*}{ \includegraphics[height=4cm,width=5cm]{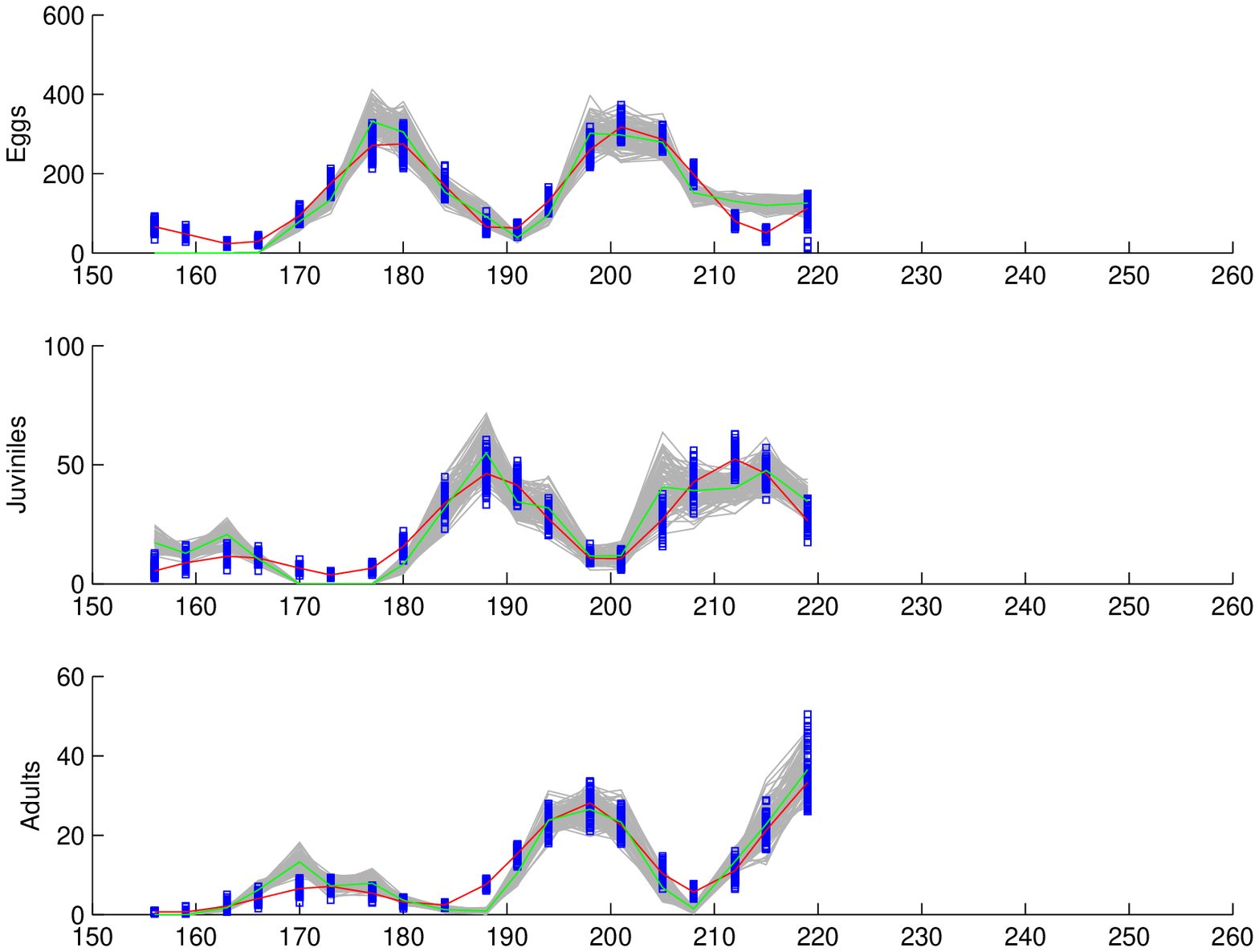}} &
  \multirow{9}{*}{\includegraphics[height=4cm,width=5cm]{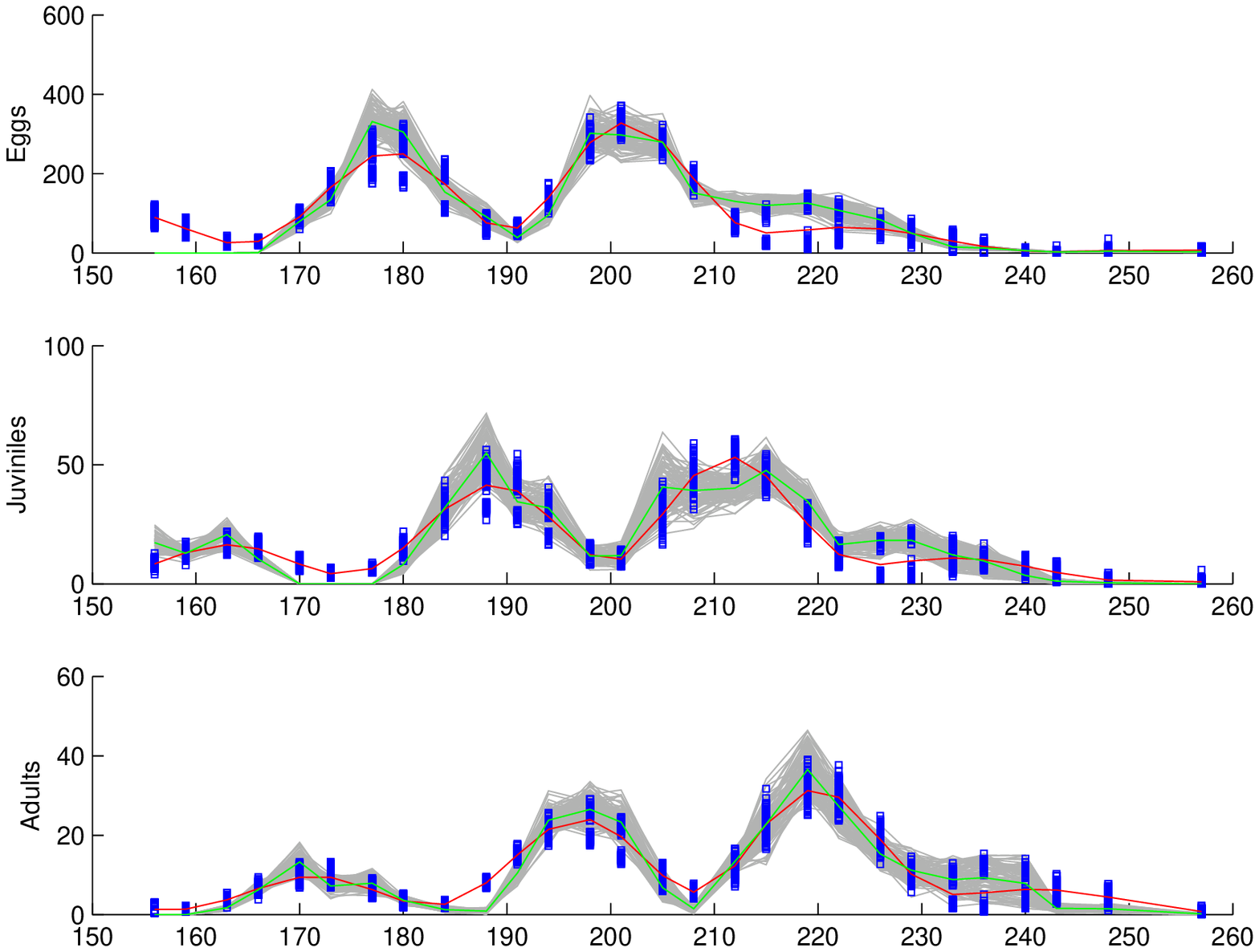}}  &\\
  & & & & \\
  & & & & \\
  & & & & \\
  & & & & \\
  & & & & \\
  & & & & \\
  & & & &\\
  & & & &\\
 \hline 
\multirow{9}{*}{\centering IV} &\multirow{9}{*}{\includegraphics[height=4cm,width=5cm]{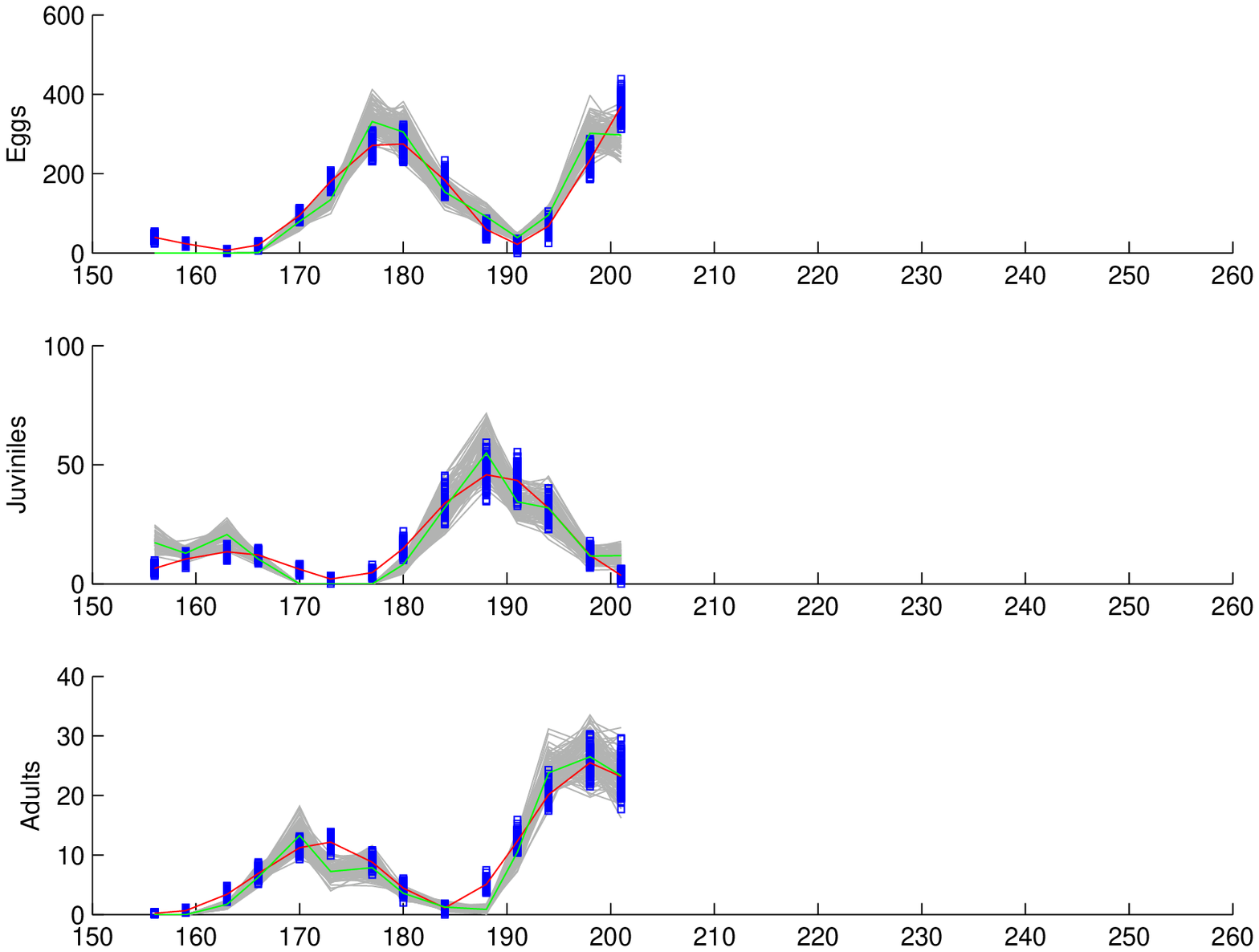}} &
  \multirow{9}{*}{\includegraphics[height=4cm,width=5cm]{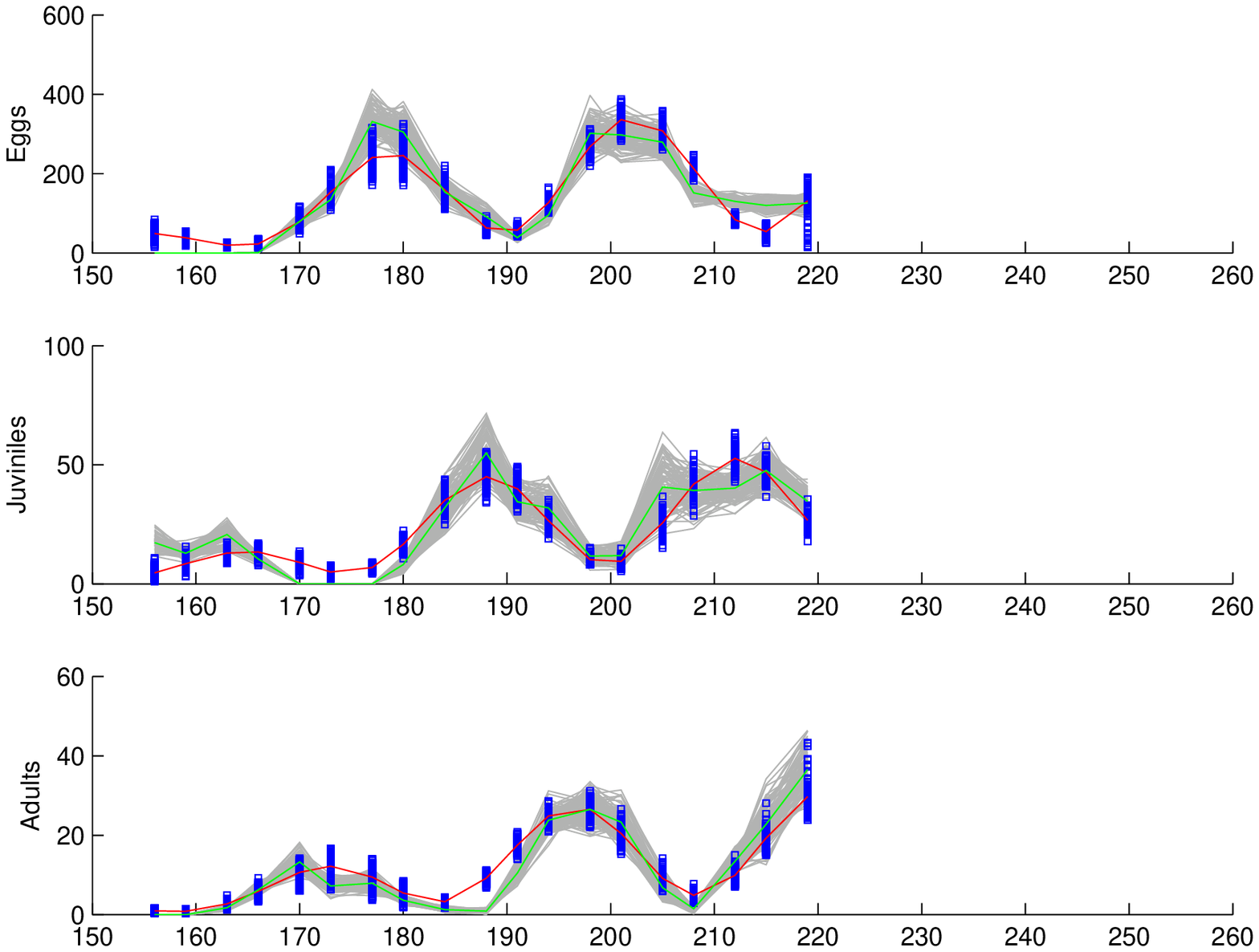} }&
 \multirow{9}{*}{\includegraphics[height=4cm,width=5cm]{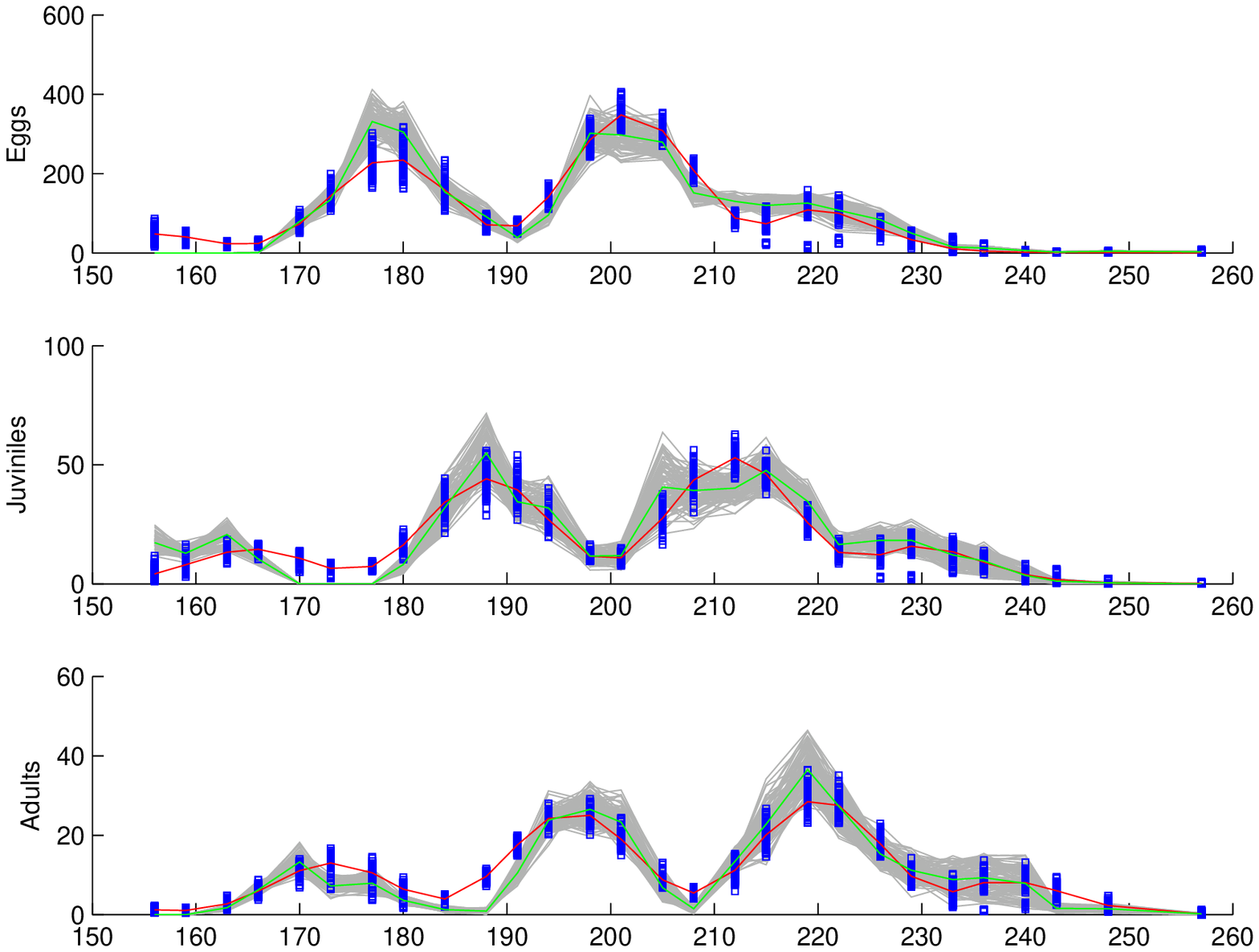} } & \\
  & & & & \\
  & & & & \\
  & & & & \\
  & & & & \\
  & & & & \\
  & & & & \\
  & & & &\\
  & & & &\\
  \hline 
 \multirow{9}{*}{\centering V} &  \multirow{3}{*}{\includegraphics[height=4cm,width=5cm]{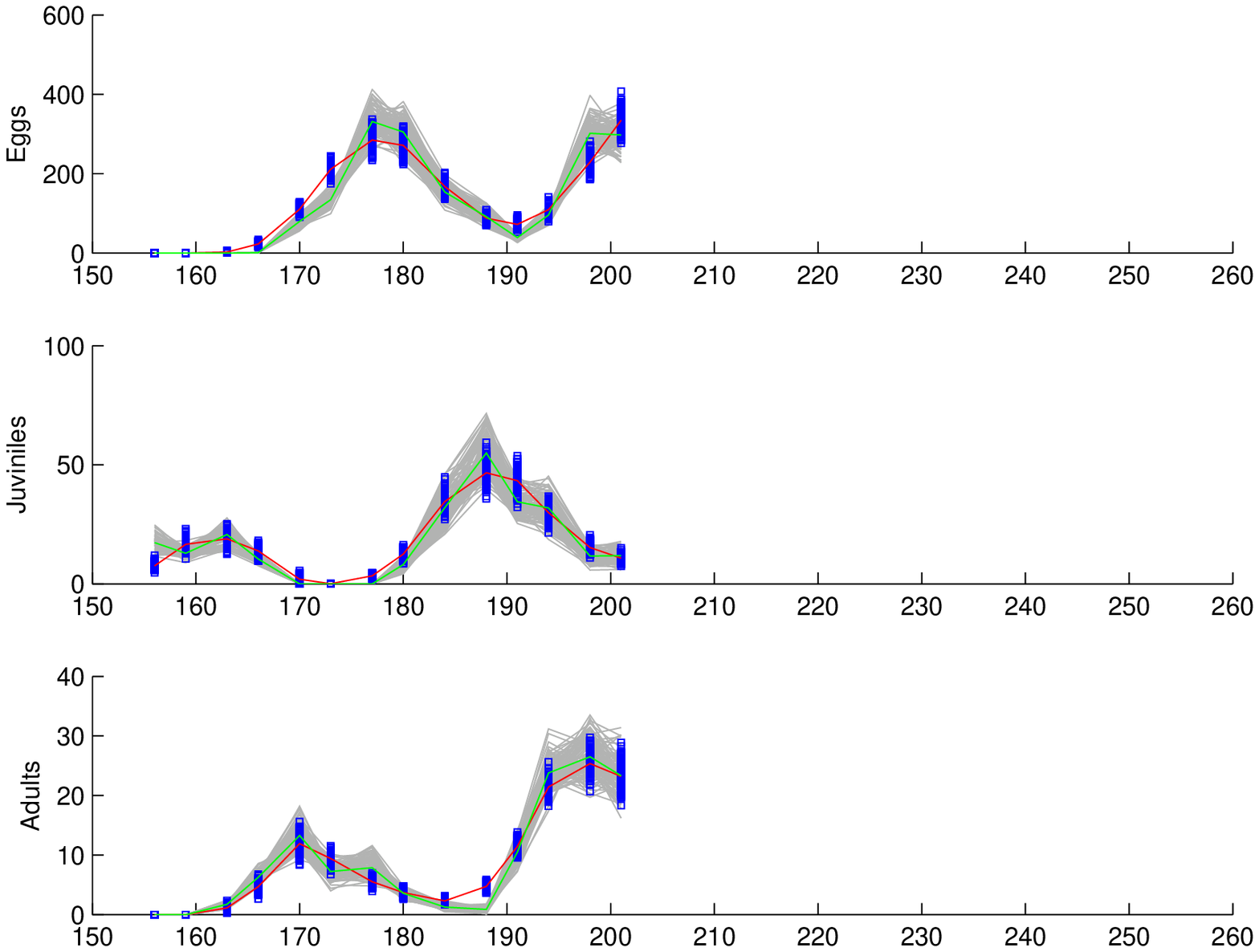} }&
  \multirow{9}{*}{ \includegraphics[height=4cm,width=5cm]{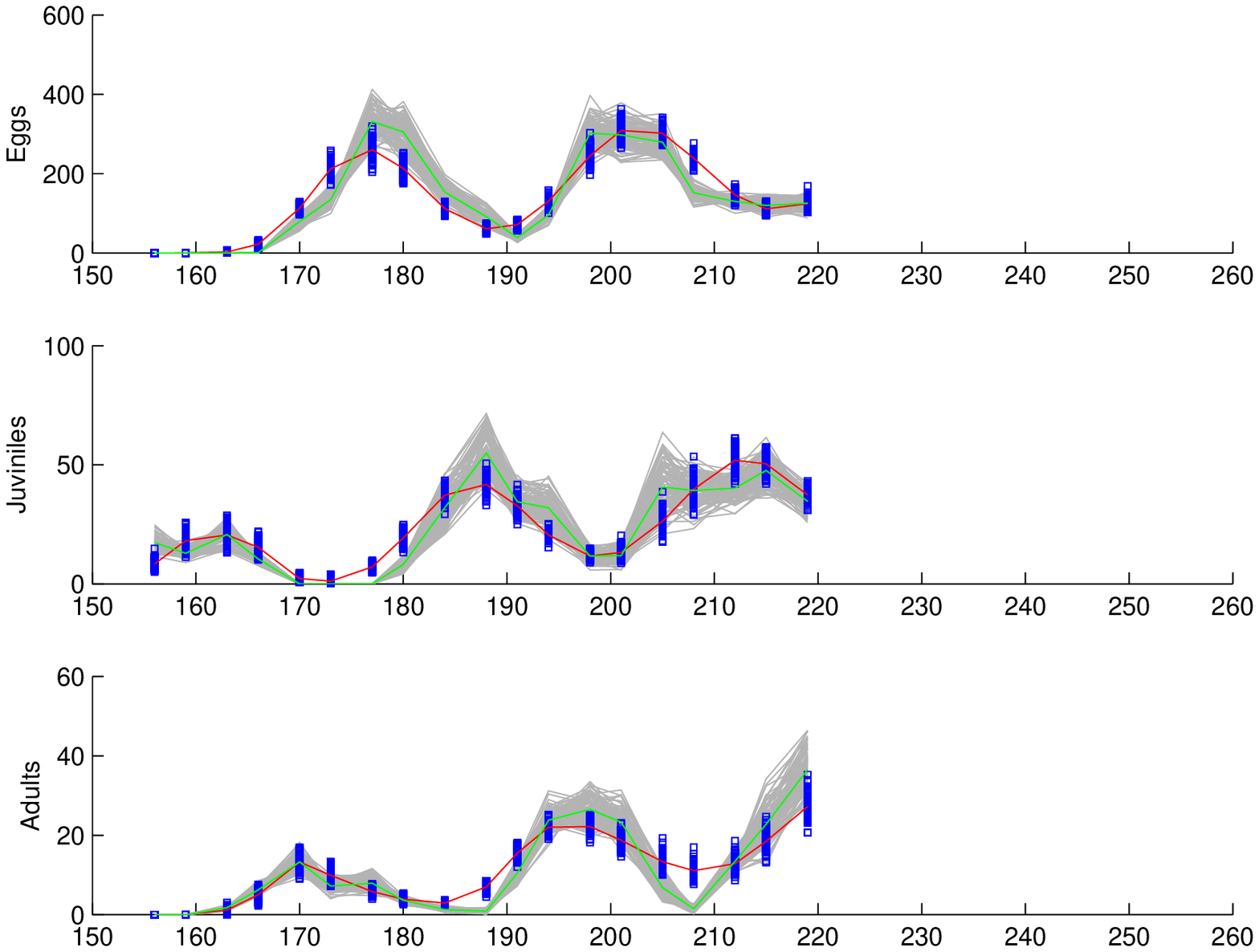}} &
  \multirow{9}{*}{\includegraphics[height=4cm,width=5cm]{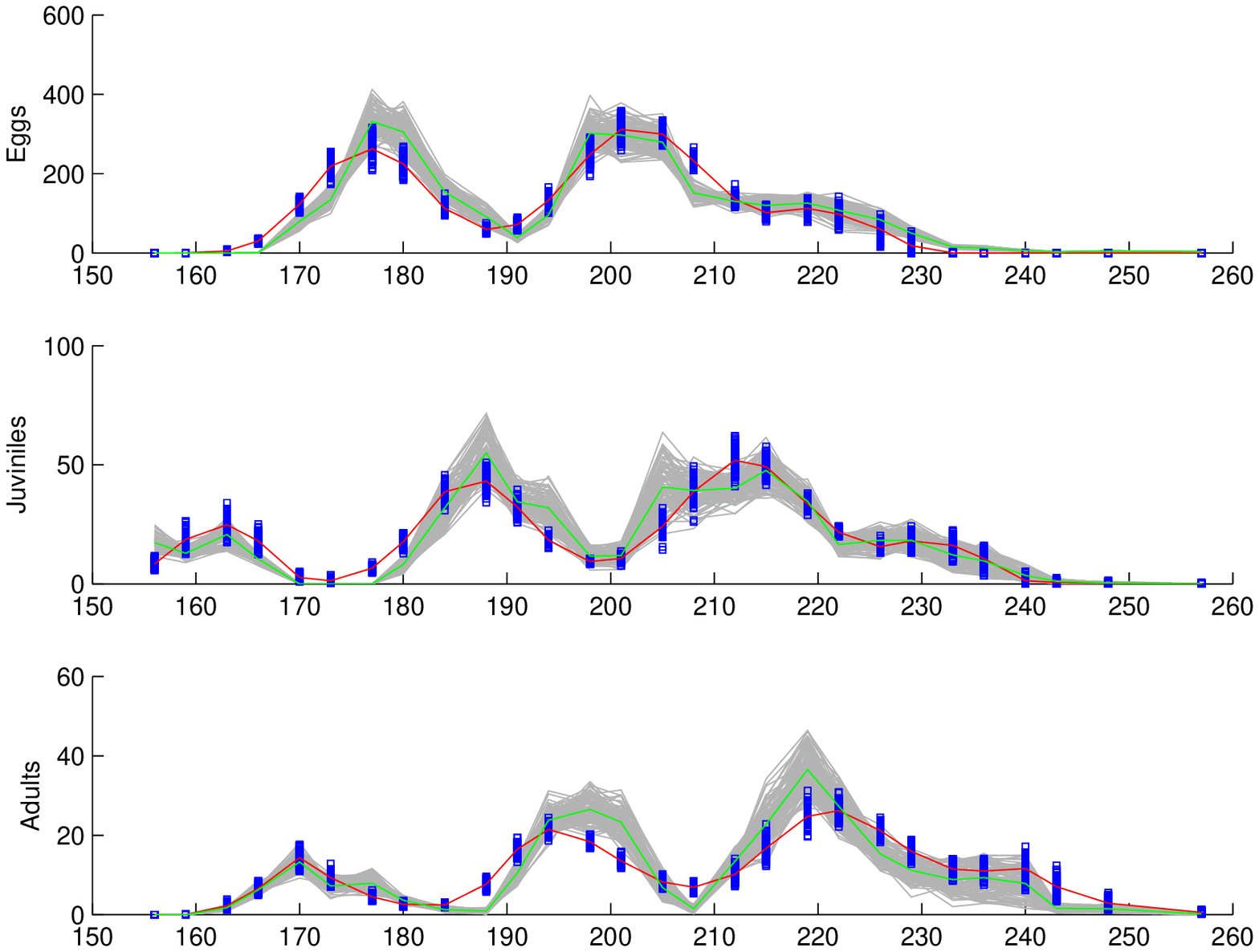}}  &\\
   & & & & \\
  & & & & \\
  & & & & \\
  & & & & \\
  & & & & \\
  & & & & \\
  & & & &\\
  & & & &\\
 \hline 
\multirow{9}{*}{\centering VI} &\multirow{9}{*}{\includegraphics[height=4cm,width=5cm]{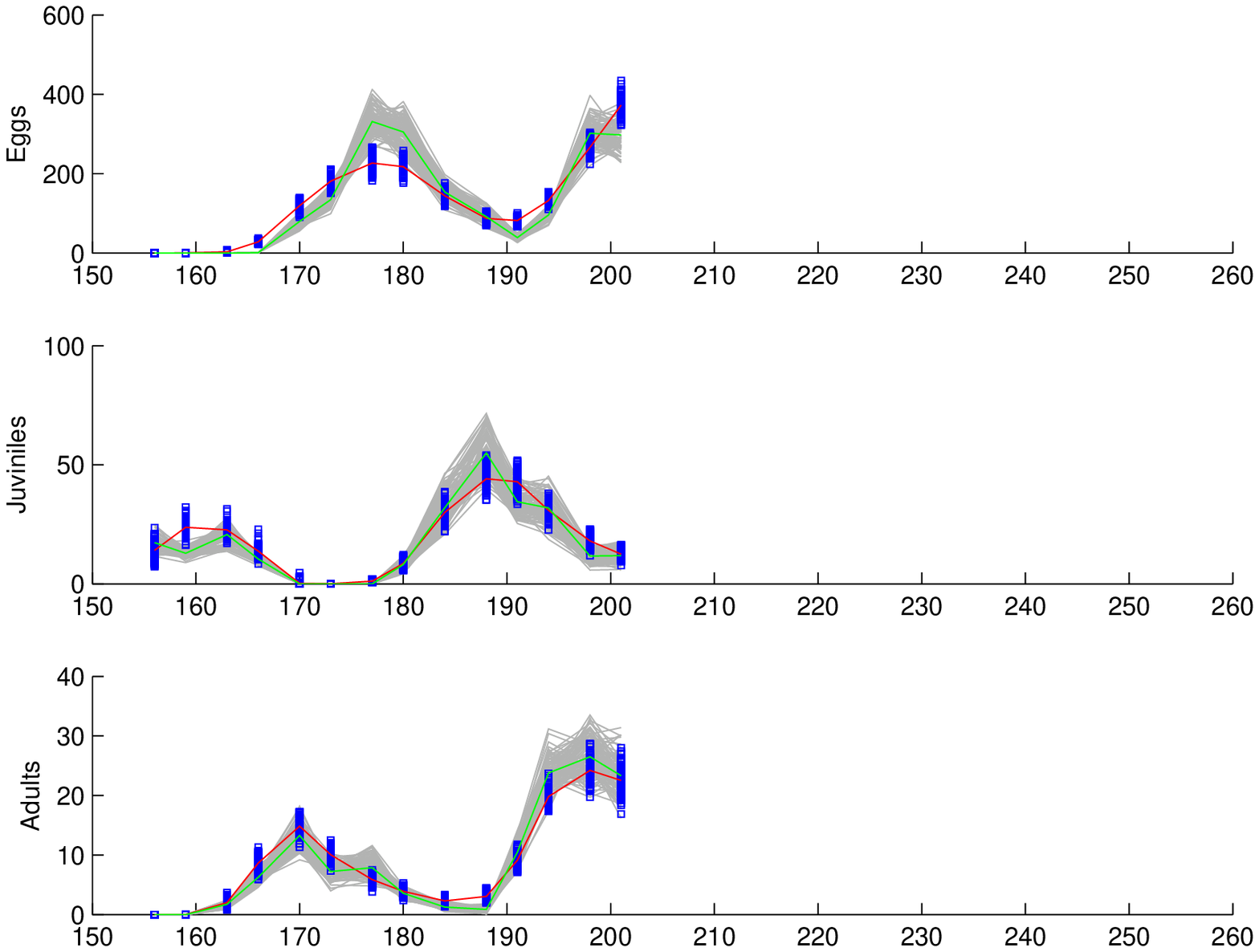}} &
  \multirow{9}{*}{\includegraphics[height=4cm,width=5cm]{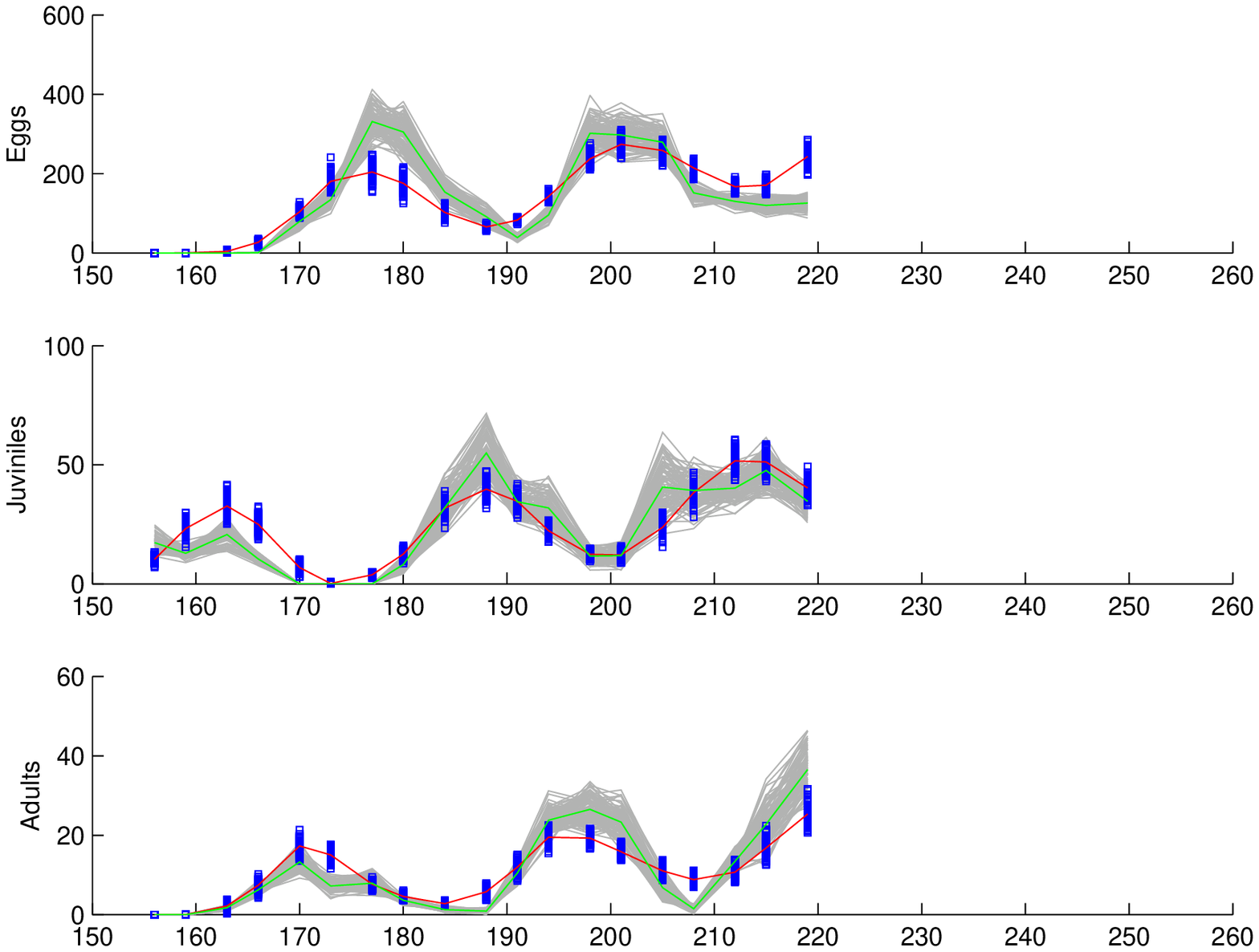} }&
 \multirow{9}{*}{\includegraphics[height=4cm,width=5cm]{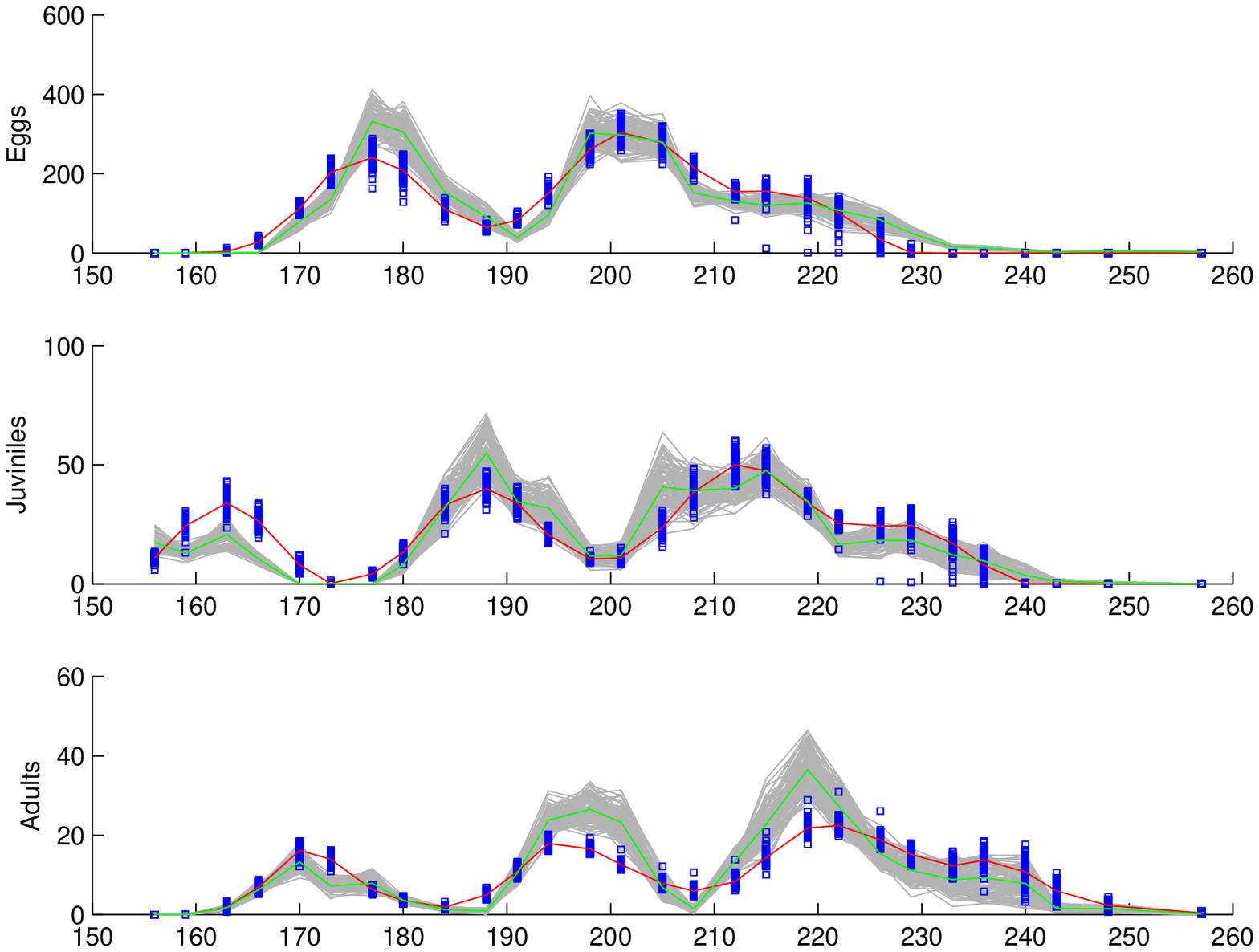} } & \\
 & & & & \\
  & & & & \\
  & & & & \\
  & & & & \\
  & & & & \\
  & & & & \\
  & & & &\\
  & & & &\\
 \hline  
  \end{tabular} 
 \vspace{-1.8ex}
 \caption{Fits of seasonal models: each panel shows the 100 data sets (grey) 
 for eggs (top), juveniles (middle), and adults (bottom), as well as the values of the fitted models at the data points (blue).
 Green curve: mean of the 100 data sets; red curve: mean of fitted time series.} \label{fitsDemo}
 \end{figure}

Model evaluation and fitting was performed in Matlab using standard Matlab routines such as \verb~dde23~, 
\verb~lsqcurvefit~, and \verb~nlinfit~.  The results are visualized in Figures \ref{fitsSimple} (models I \& II) and \ref{fitsDemo} (models (III--VI). 
\vspace{-1ex}
\mbox{} \vspace{-5ex}

\parbox[t]{12cm}{ \mbox{} \\  \mbox{} \\ \mbox{} \\ \mbox{} \\  \mbox{} \\ \mbox{} \\ \mbox{} \\  \mbox{} \\ \mbox{} \\ \mbox{} \\ }

\section{Growth-rate distributions with error bars}
\begin{figure}[h]
\mbox{}\hspace{-13ex}
\begin{tabular}{c}
 \includegraphics[height=12cm,width=8cm]{pl9cK.eps} 
\includegraphics[height=12cm,width=8cm]{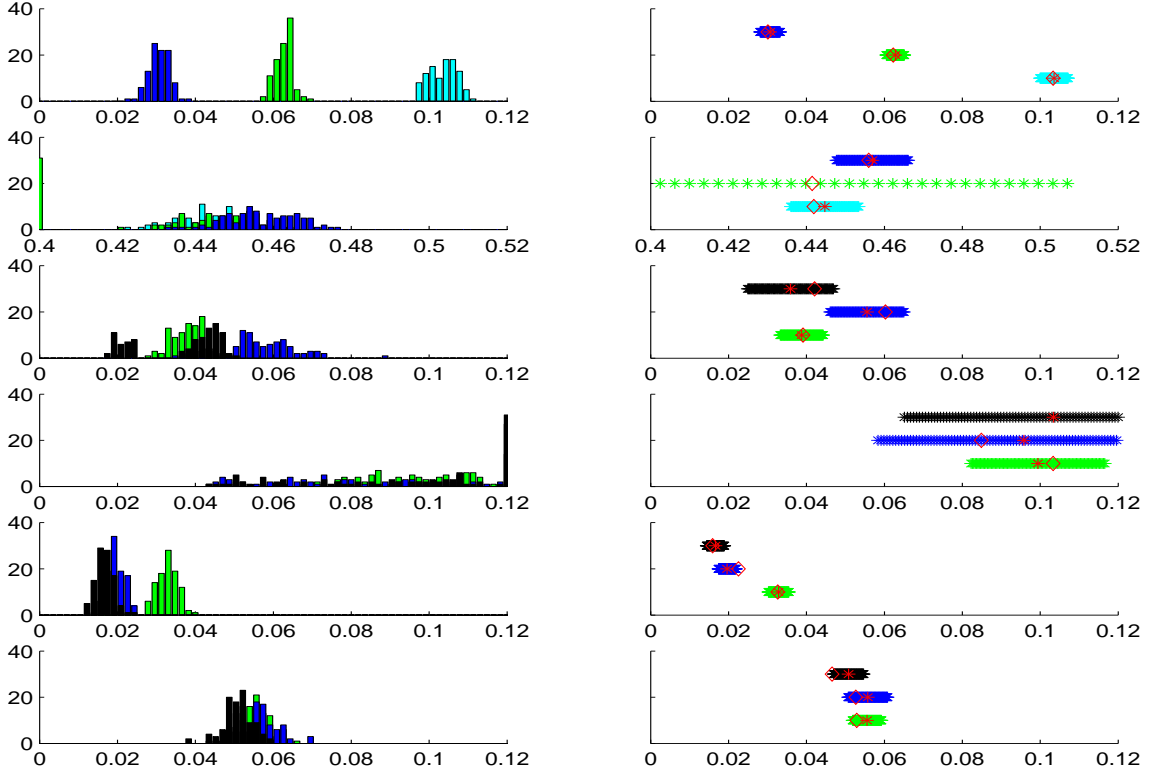} 
 \end{tabular} 
 \vspace{-8ex}
 \caption{Growth-rate distributions for models I-VI (rows 1-6). 
 Left: Fig.~\ref{figdist}. Right: Means (red stars) with error bars (colour coding consistent with left panel); red diamonds: growth-rate values of Table \ref{tab} 
derived from the averaged data displayed in Fig.~\ref{Fig1}.
} \label{figdist1}
 \end{figure} 
These are presented in Fig.~\ref{figdist1}. 
For the models in rows I, V and VI 
we observe that (to a good approximation) 
\begin{equation}
 \left(\parbox{2.9cm}{growth rate of {\it averaged} data set} \right) \,=\, \left(\parbox{4.1cm}{{\it average} of growth rates of individual data sets}\right). 
 \label{av}
 \end{equation} 

\section{Re-weighting age groups 1: ad-hoc method} \label{Weight}
The weights $w_1,w_2,w_3$ of the heuristic re-wheighting method described in the main body of the paper
were computed as 
\begin{equation}
 w_1 = \frac{\int _0^{T/2} A(t)dt }{\int _0^{T/2} E(t)dt}  , \quad w_2 =\frac{\int _0^{T/2} A(t)dt }{\int _0^{T/2} J(t)dt} ,
 \quad w_3 = 1 . 
\label{weights} 
\end{equation}  
where $T$ denotes the length of the season (101 days).  
\vspace{-1ex}
\begin{figure}[h]
\hspace{-15ex}
\begin{tabular}{|c||c|c|cc|}
\hline  
   &   \multicolumn{3}{c}{Time Window} &\\
 \hline 
    Model & 30 days & 50 days  & 70 days & \\
    \hline \hline 
 \multirow{6}{*}{\centering I} &  \multirow{6}{*}{\includegraphics[height=3cm,width=5cm]{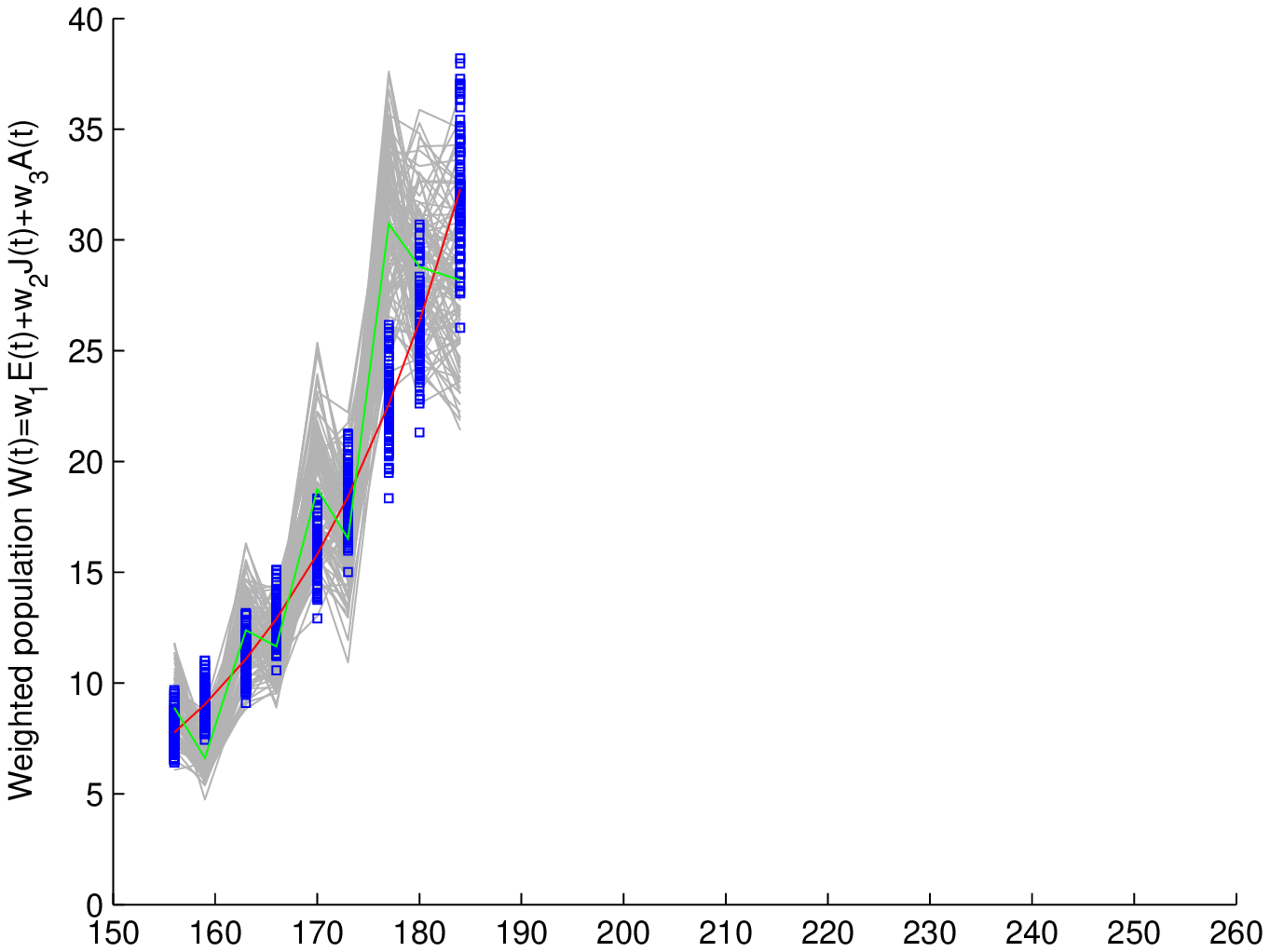} }&
  \multirow{6}{*}{ \includegraphics[height=3cm,width=5cm]{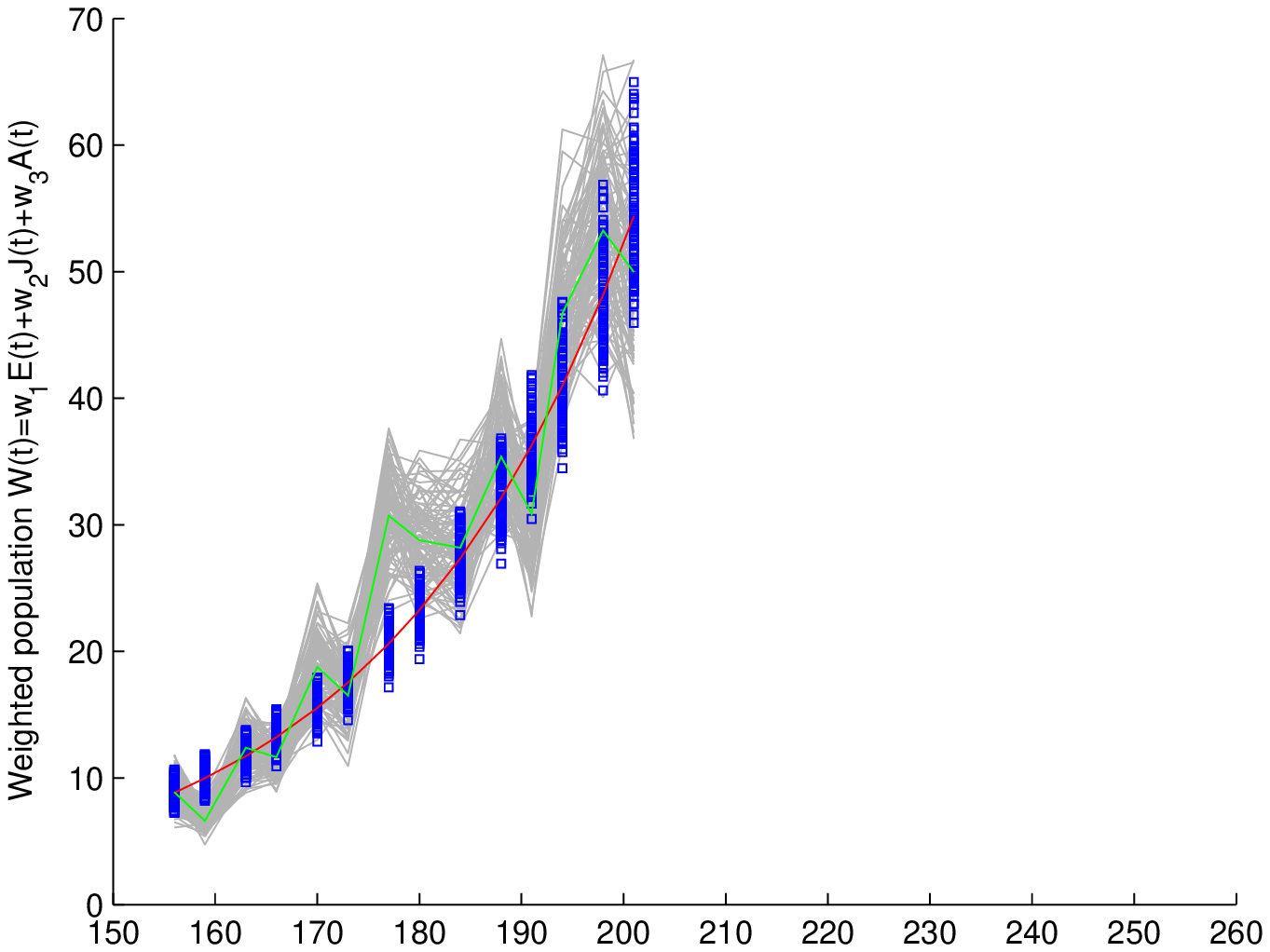}} &
  \multirow{6}{*}{\includegraphics[height=3cm,width=5cm]{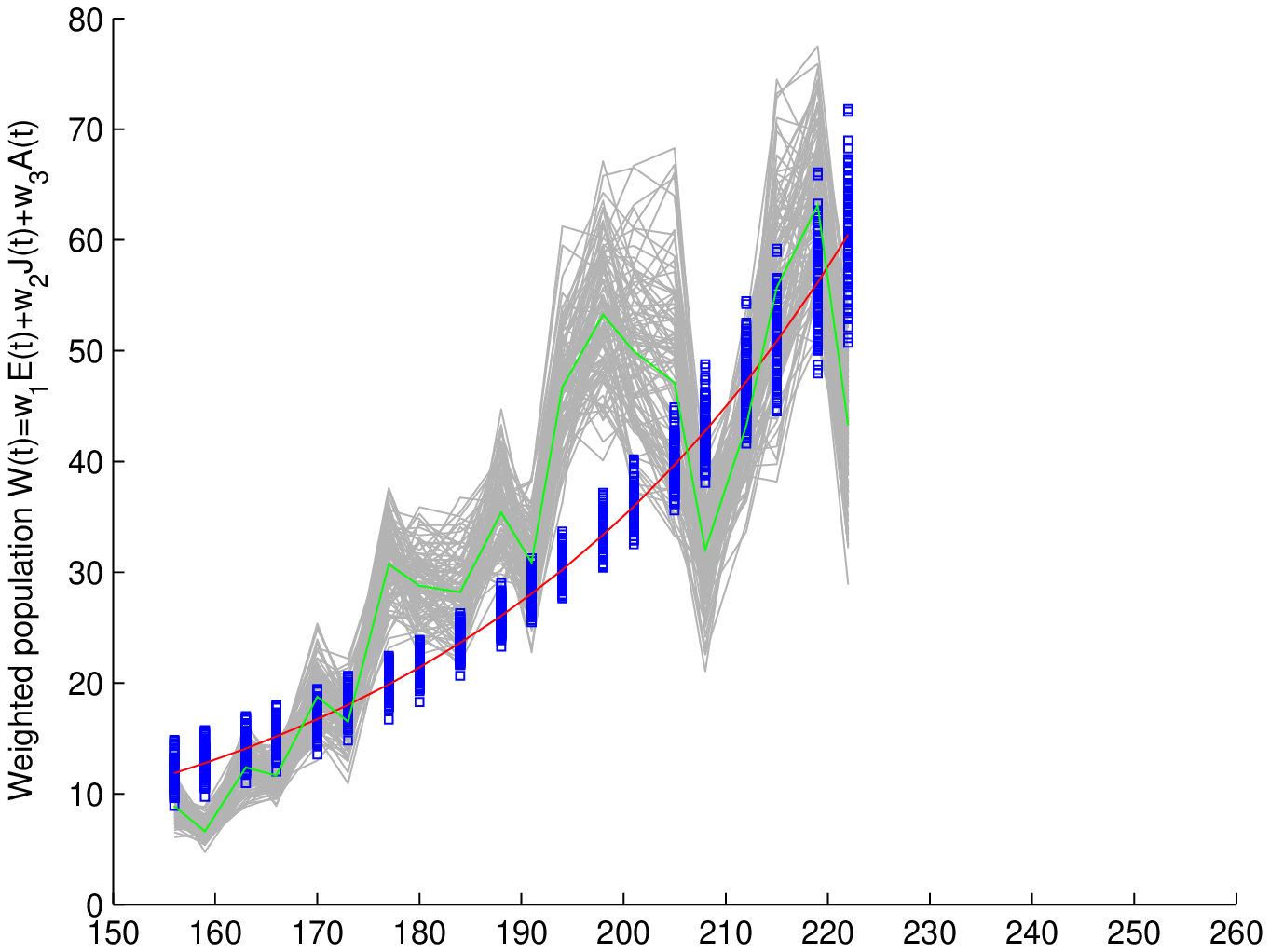}}  &\\
  & & & & \\
  & & & & \\
  & & & & \\
  & & & & \\
  & & & & \\
 \hline 
\multirow{6}{*}{\centering II} &\multirow{6}{*}{\includegraphics[height=3cm,width=5cm]{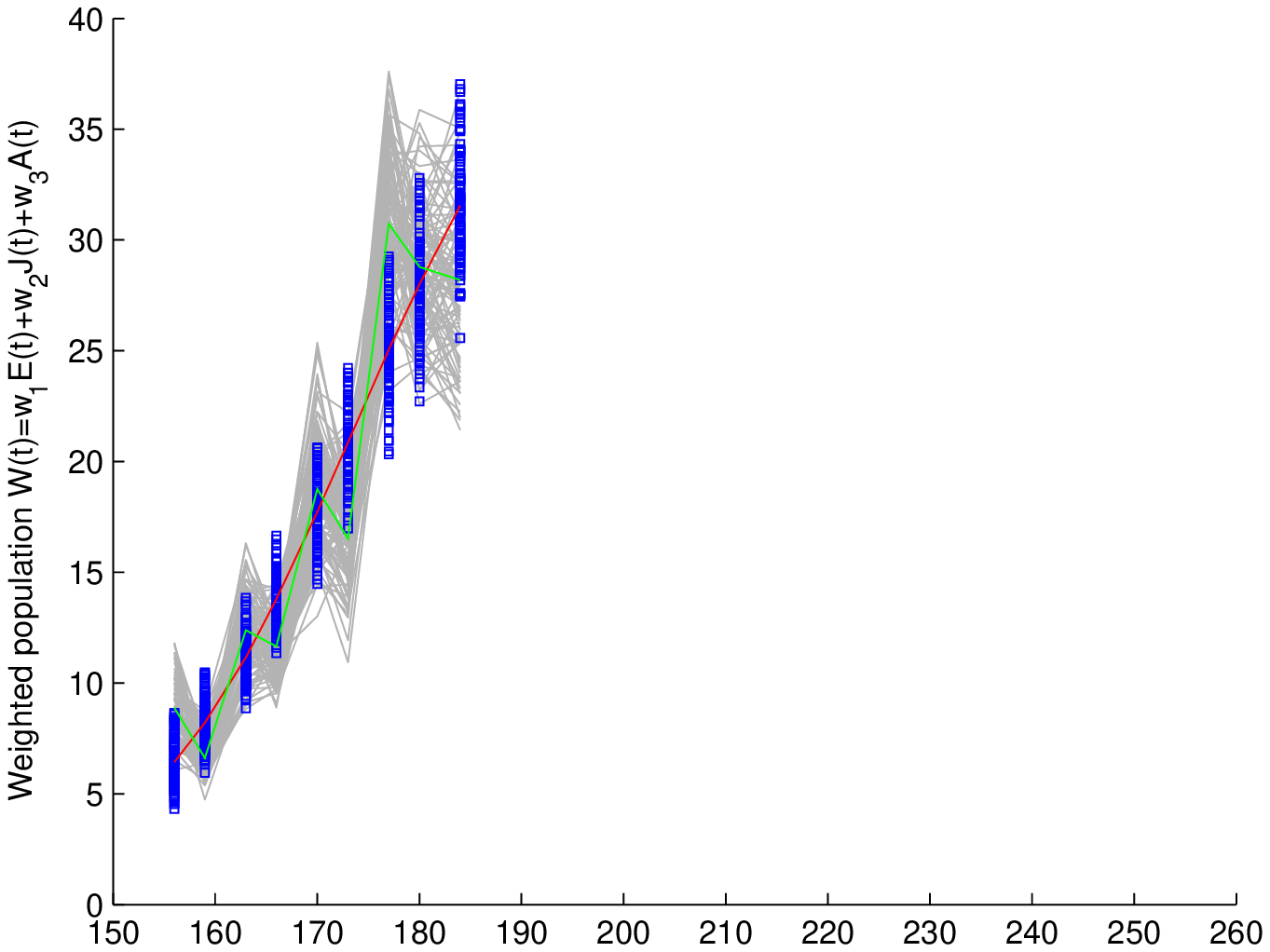}} &
  \multirow{6}{*}{\includegraphics[height=3cm,width=5cm]{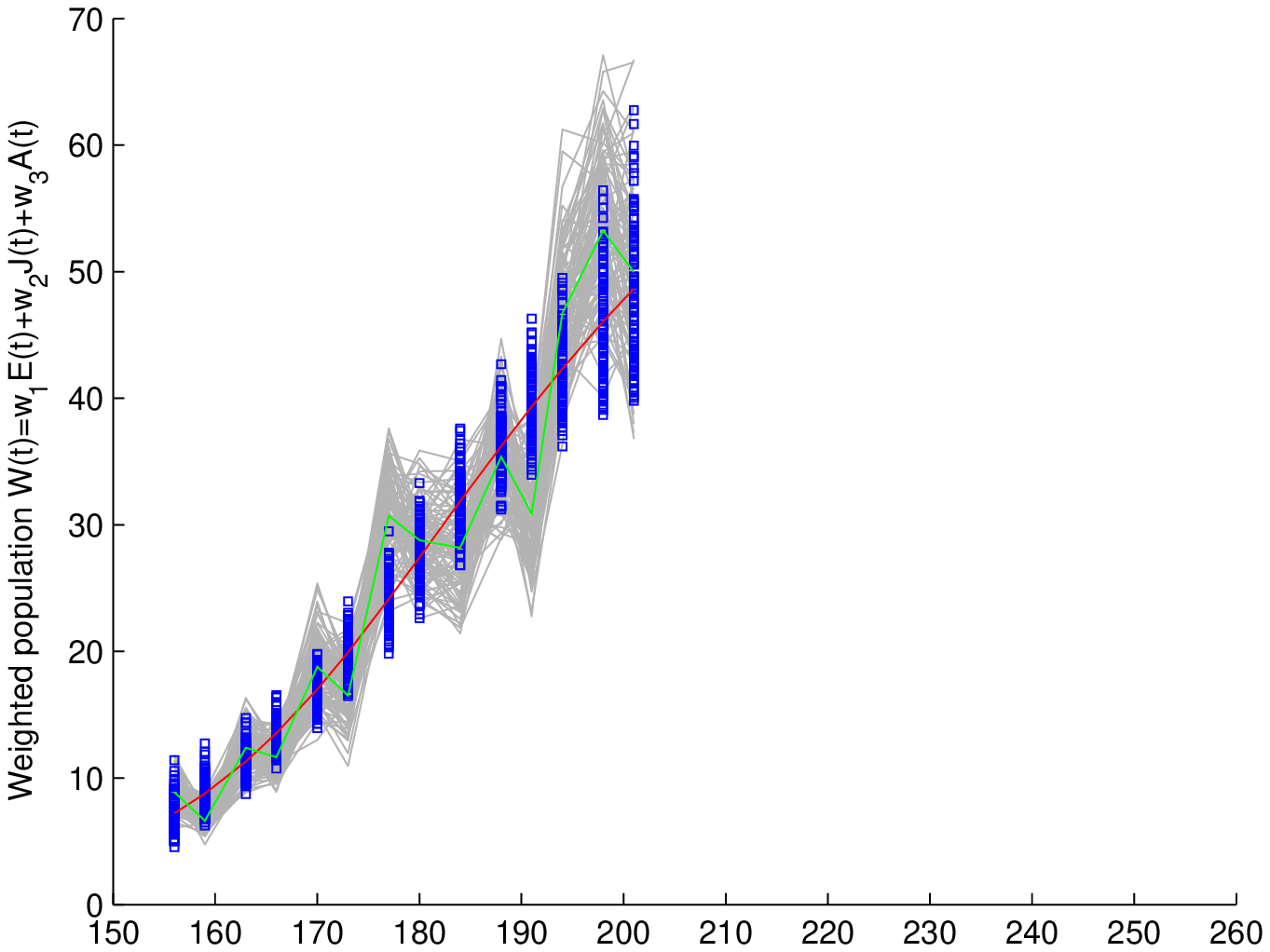} }&
 \multirow{6}{*}{\includegraphics[height=3cm,width=5cm]{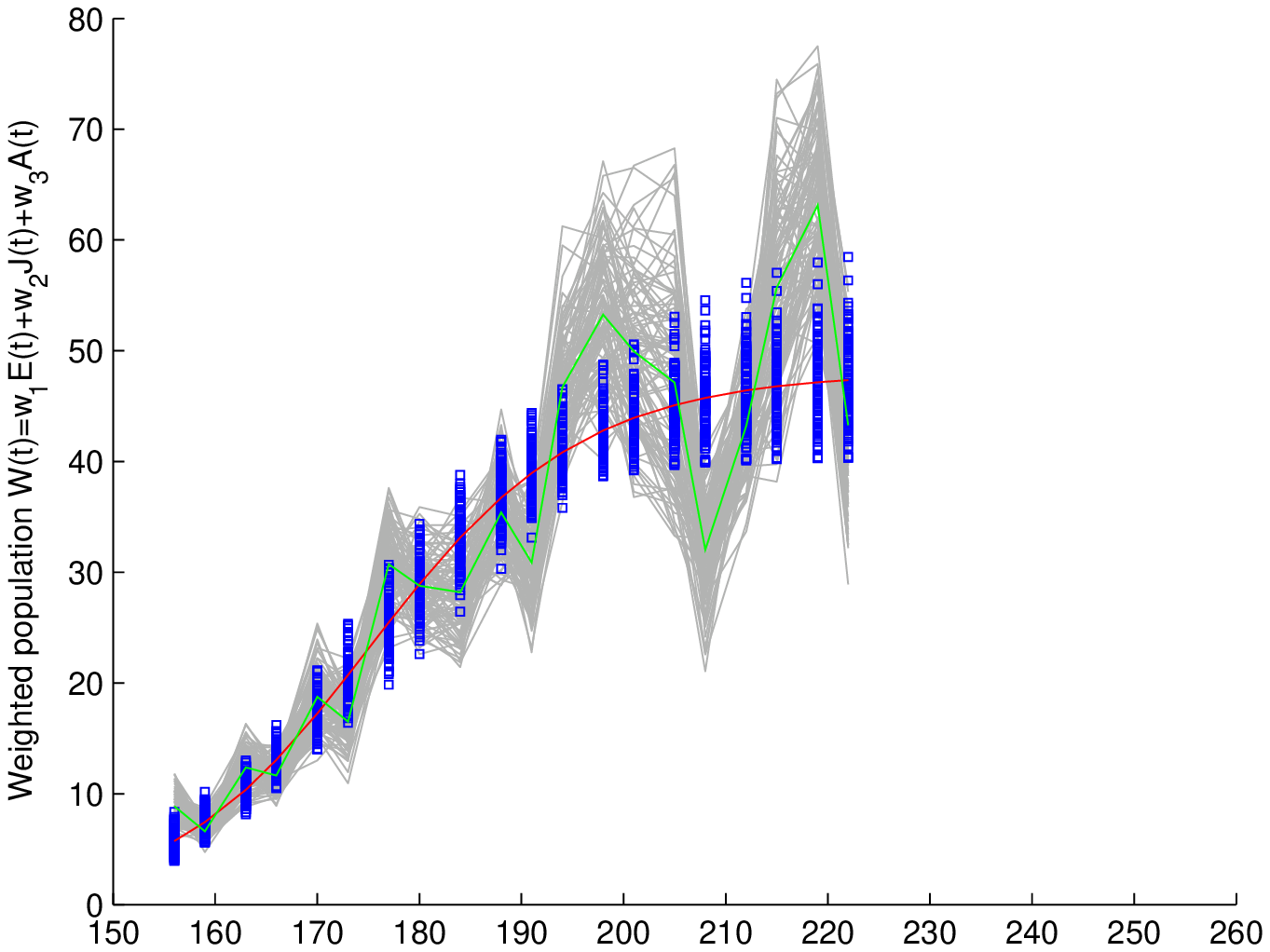} } & \\
  & & & & \\
  & & & & \\
  & & & & \\
  & & & & \\
  & & & & \\
 \hline  
  \end{tabular} 
 \vspace{-1.8ex}
 \caption{Fits of models I \& II to re-weighted data $W(t) = w_1E(t) + w_2J(t) + w_3A(t)$. 
 Plots analogous to \ref{fitsSimple}. The reduction  of the generational fluctuations is evident
 (see main text).} \label{fitsW}
 \end{figure} 
 
\vspace{-1.5ex}
\begin{figure}[h]
\hspace{-13ex}\begin{tabular}{cc}
 \includegraphics[height=3.8cm,width=7.8cm]{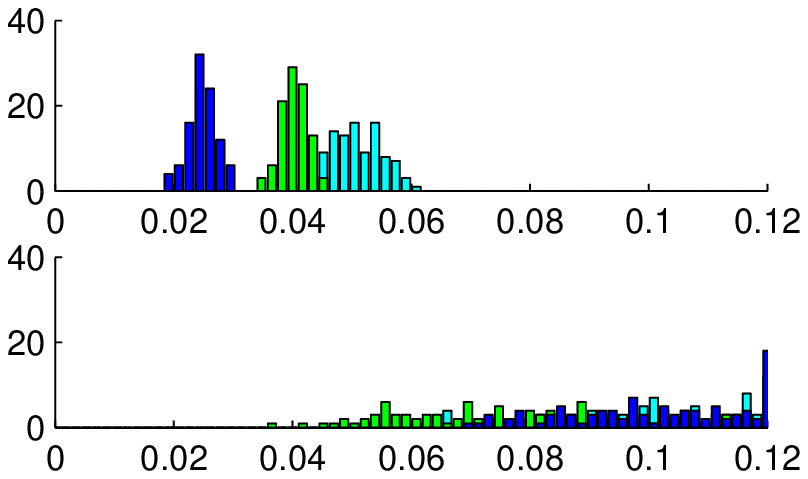} &
 \includegraphics[height=3.8cm,width=7.8cm]{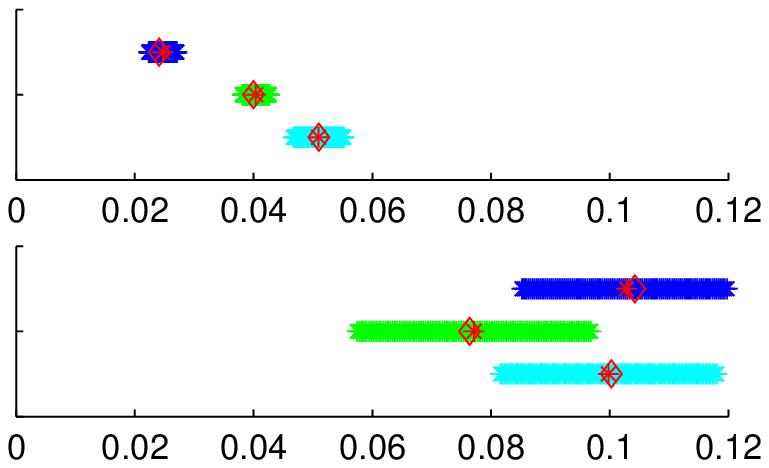} 
 \end{tabular} 
 \vspace{-4ex}
 \caption{Plot analogous to the first two rows of Fig.~\ref{figdist1}, but scale in row 2  changed! The dependence on the time window is substantially
 reduced and identity (\ref{av}) is satisfied to a high degree.} \label{figdistW}
 \end{figure} 
 The growth-rate distributions with errors bars are shown on Fig.~\ref{figdistW}.

 \section{Re-weighting age groups 2: reproductive value} 
\begin{figure}[b]
\hspace{-17ex}
\begin{tabular}{|c||c|c|cc|}
\hline  
   &   \multicolumn{3}{c}{Time Window} &  \\
 \hline 
    Model & 30 days & 50 days  & 70 days &  \\
    \hline \hline 
 \multirow{6}{*}{\centering \parbox{1cm}{\begin{center}I \\(+V)\end{center} }} &  \multirow{6}{*}{\includegraphics[height=3cm,width=5cm]{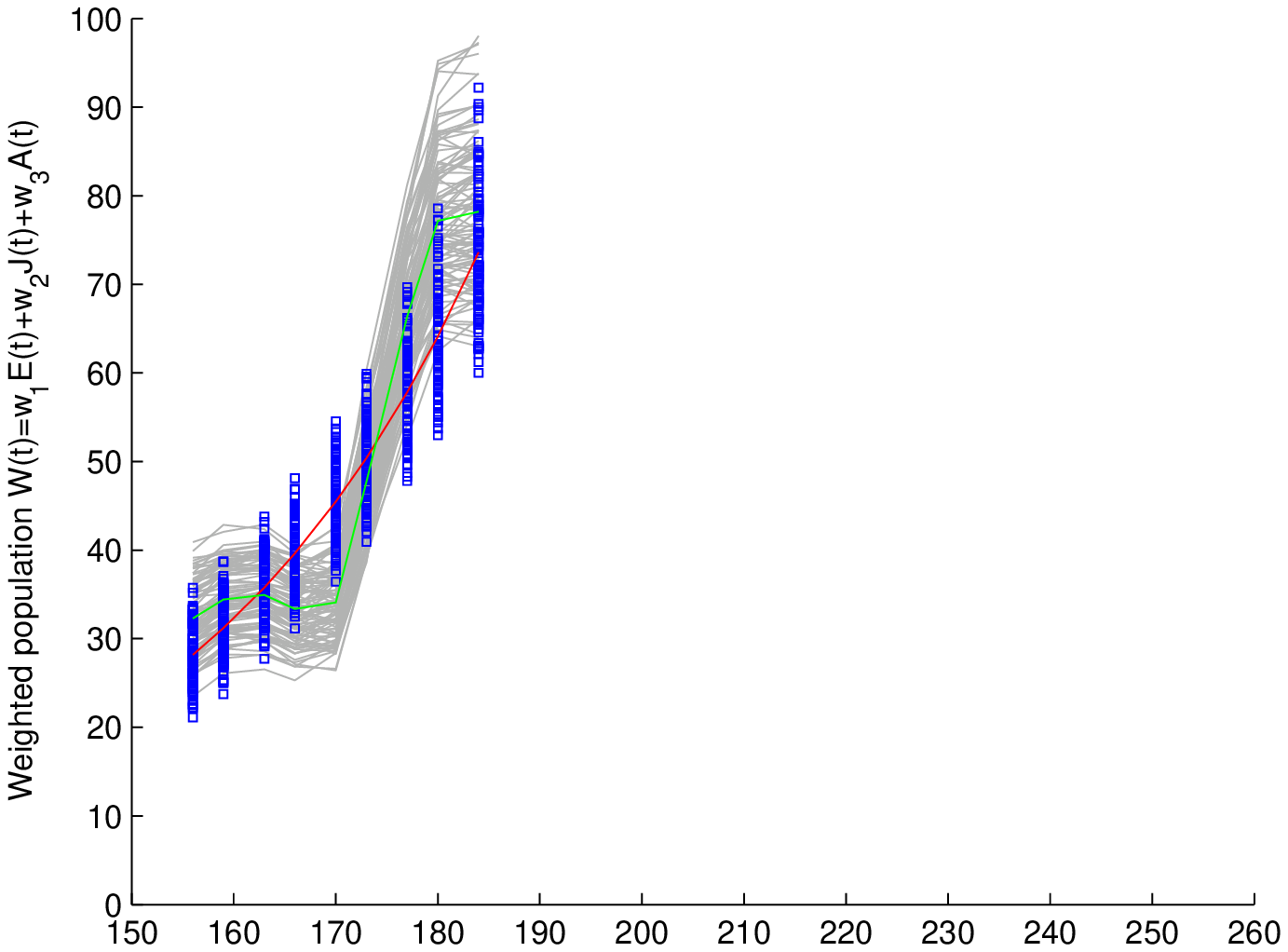} }&
  \multirow{6}{*}{ \includegraphics[height=3cm,width=5cm]{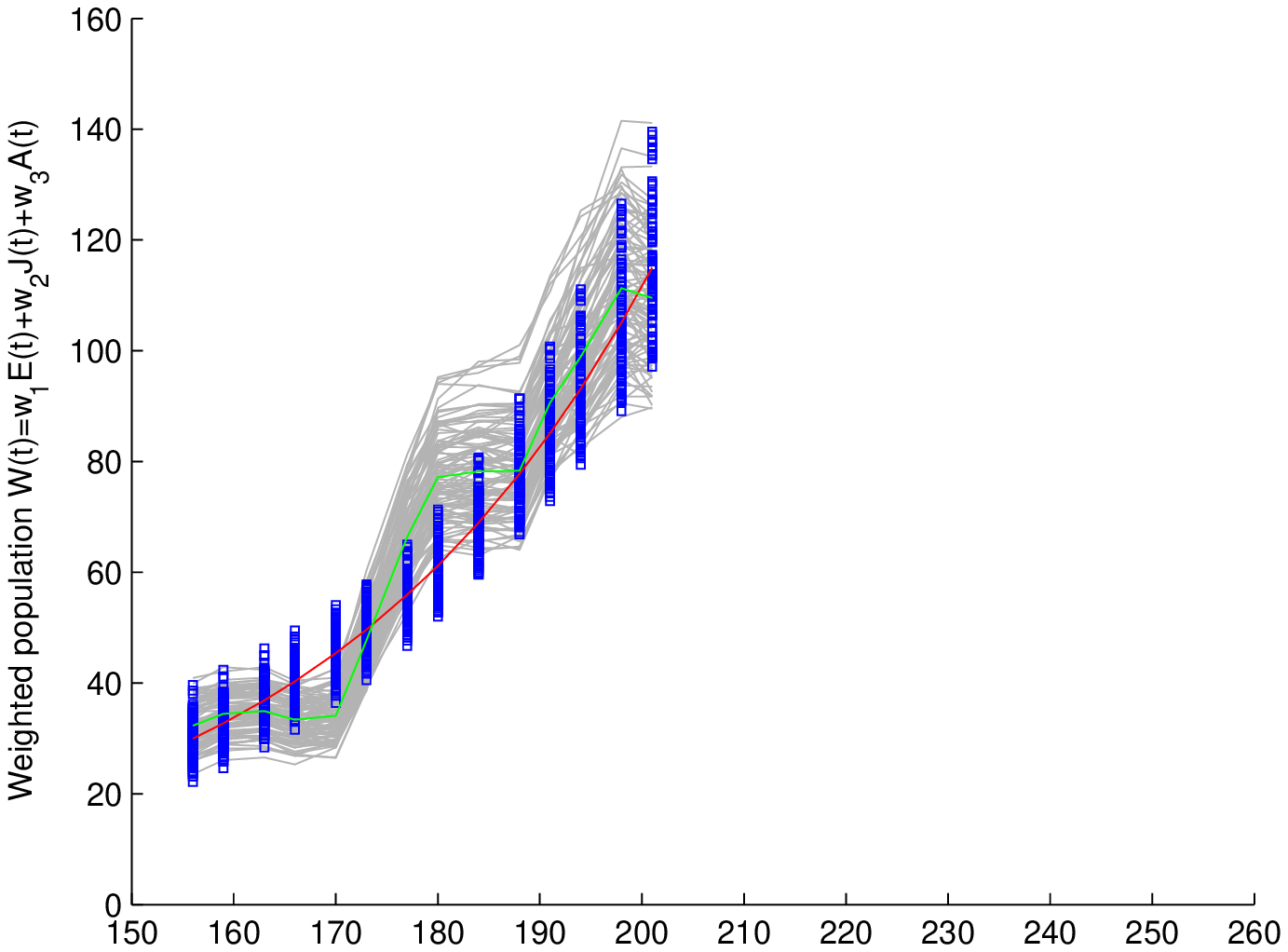}} &
  \multirow{6}{*}{\includegraphics[height=3cm,width=5cm]{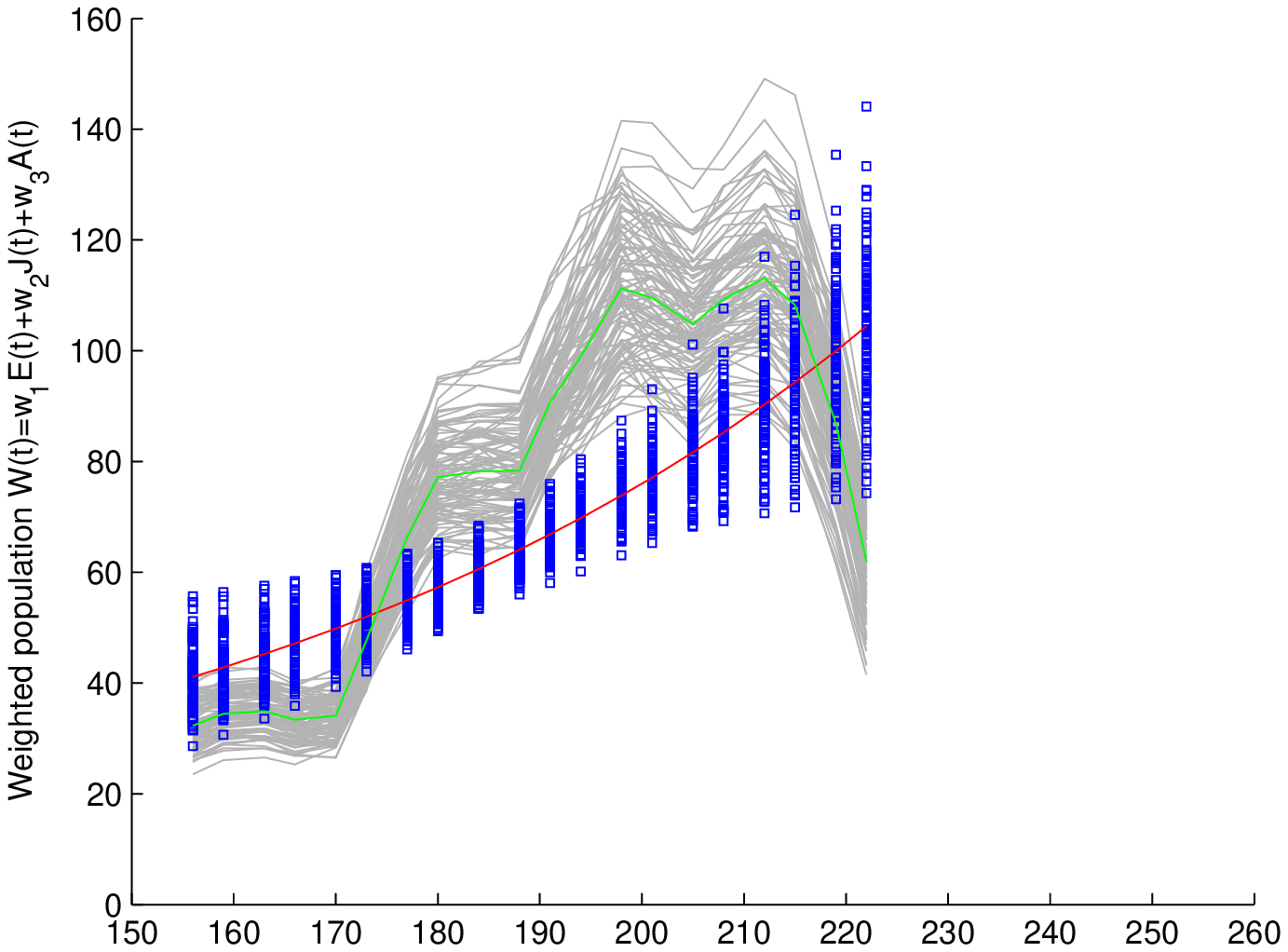}}   &\\
  & & & & \\
  & & & & \\
  & & & & \\
  & & & & \\
  & & & & \\
 \hline 
\multirow{6}{*}{\centering \parbox{1cm}{\begin{center}II \\(+V)\end{center} }} &\multirow{6}{*}{\includegraphics[height=3cm,width=5cm]{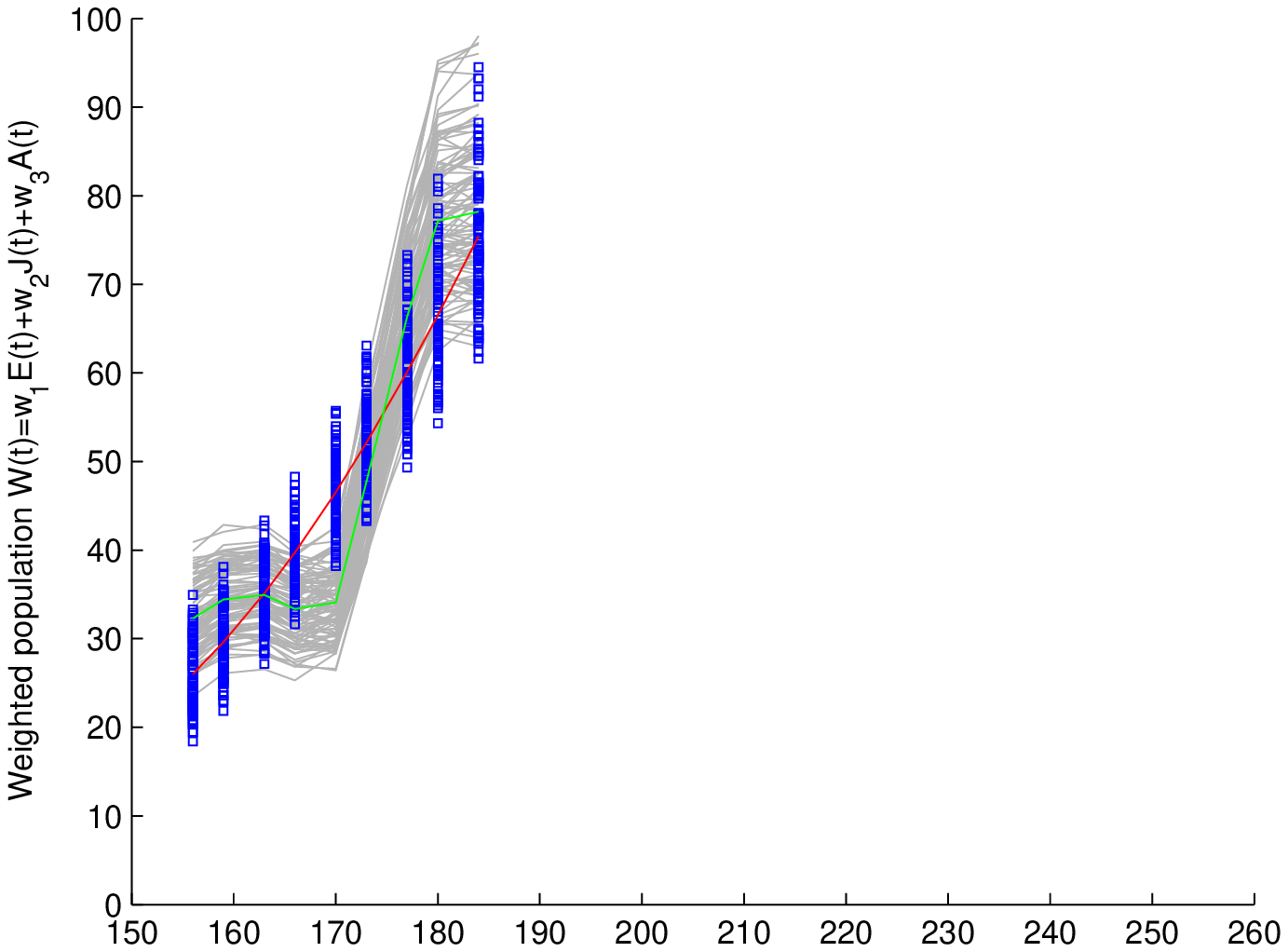}} &
  \multirow{6}{*}{\includegraphics[height=3cm,width=5cm]{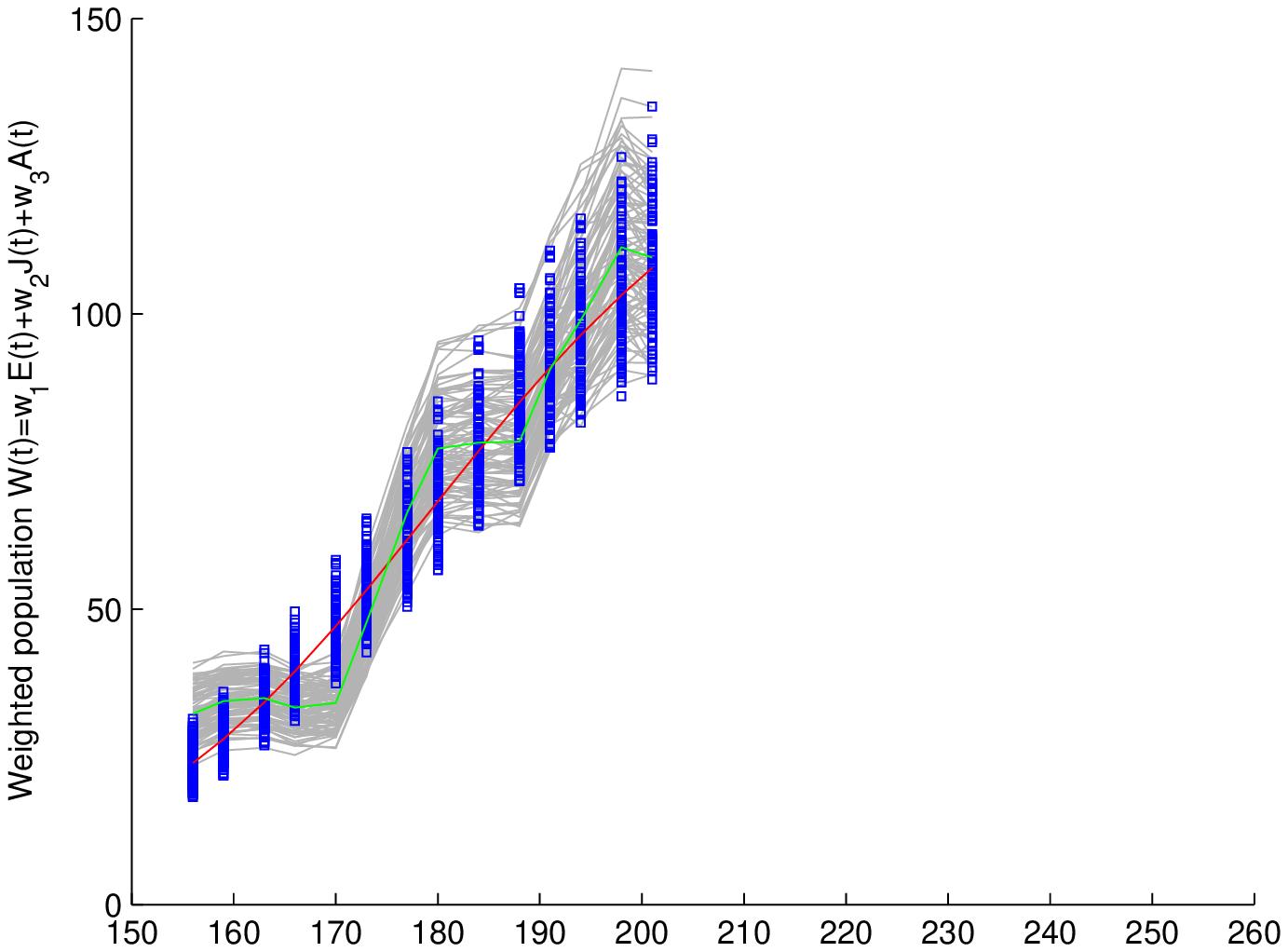} }&
 \multirow{6}{*}{\includegraphics[height=3cm,width=5cm]{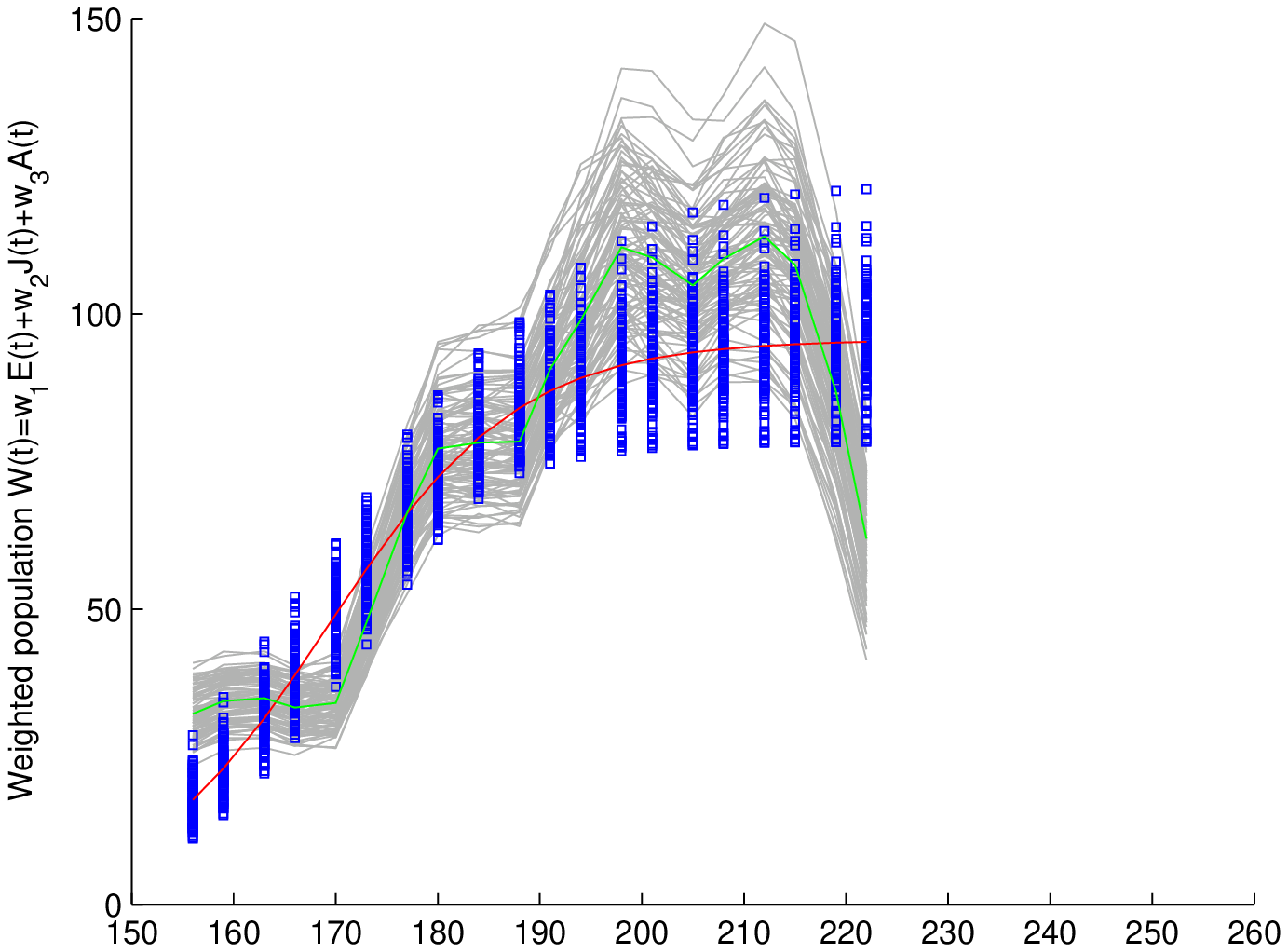} }  &  \\
  & & & & \\
  & & & & \\
  & & & & \\
  & & & & \\
  & & & & \\
 \hline  
 \multirow{6}{*}{\centering \parbox{1cm}{\begin{center}I \\(+VI)\end{center} }} &  \multirow{6}{*}{\includegraphics[height=3cm,width=5cm]{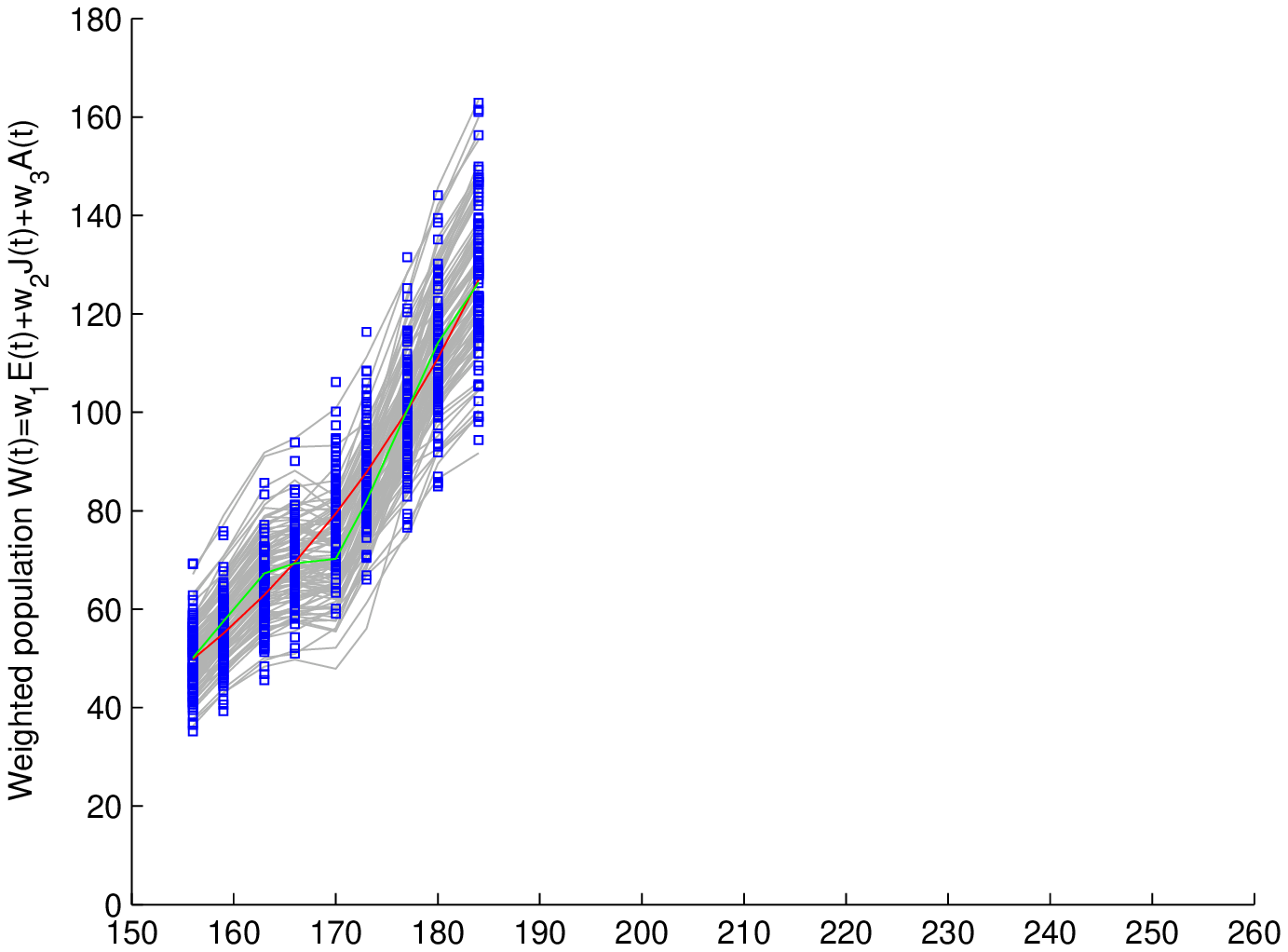} }&
  \multirow{6}{*}{ \includegraphics[height=3cm,width=5cm]{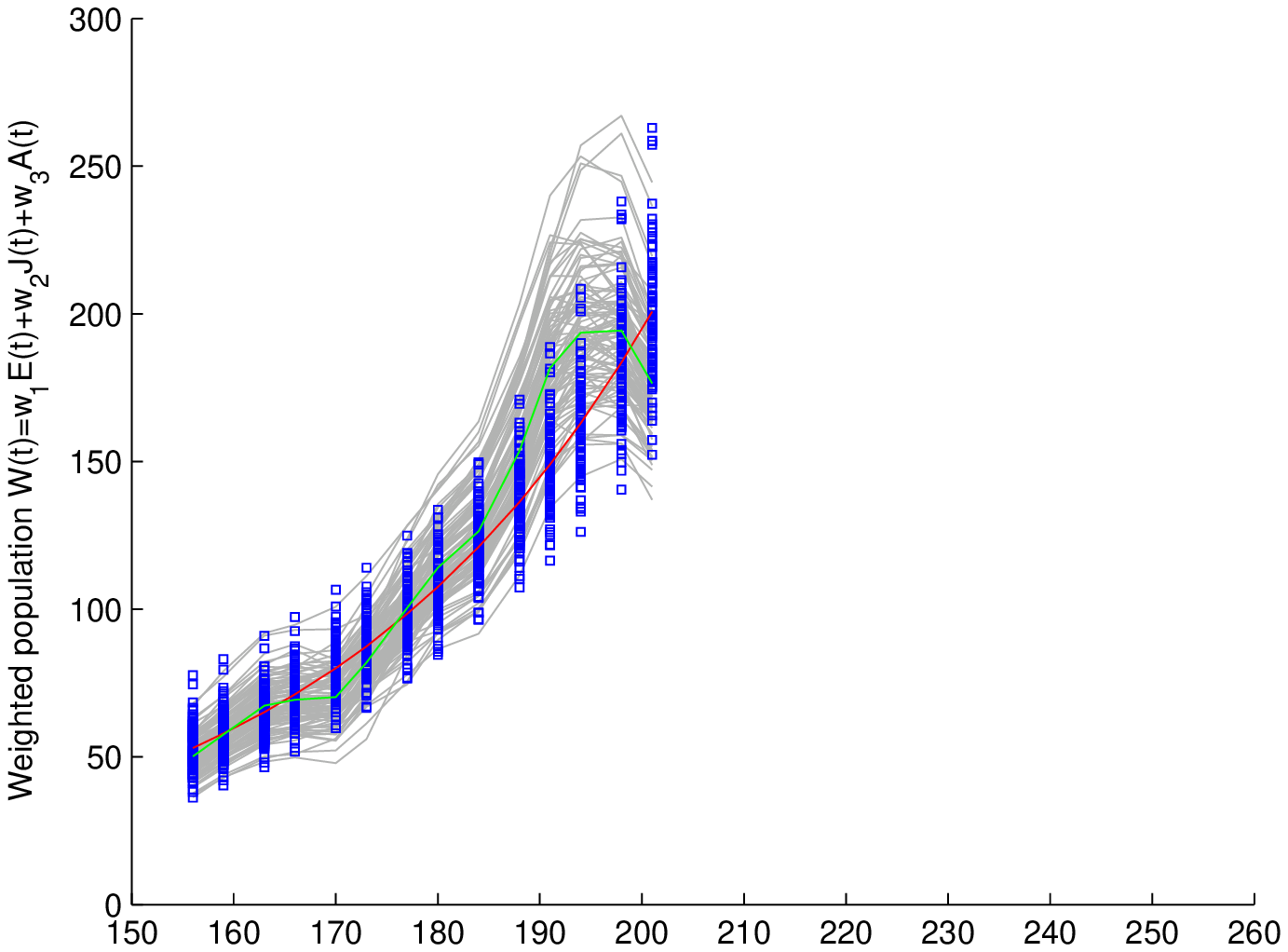}} &
  \multirow{6}{*}{\includegraphics[height=3cm,width=5cm]{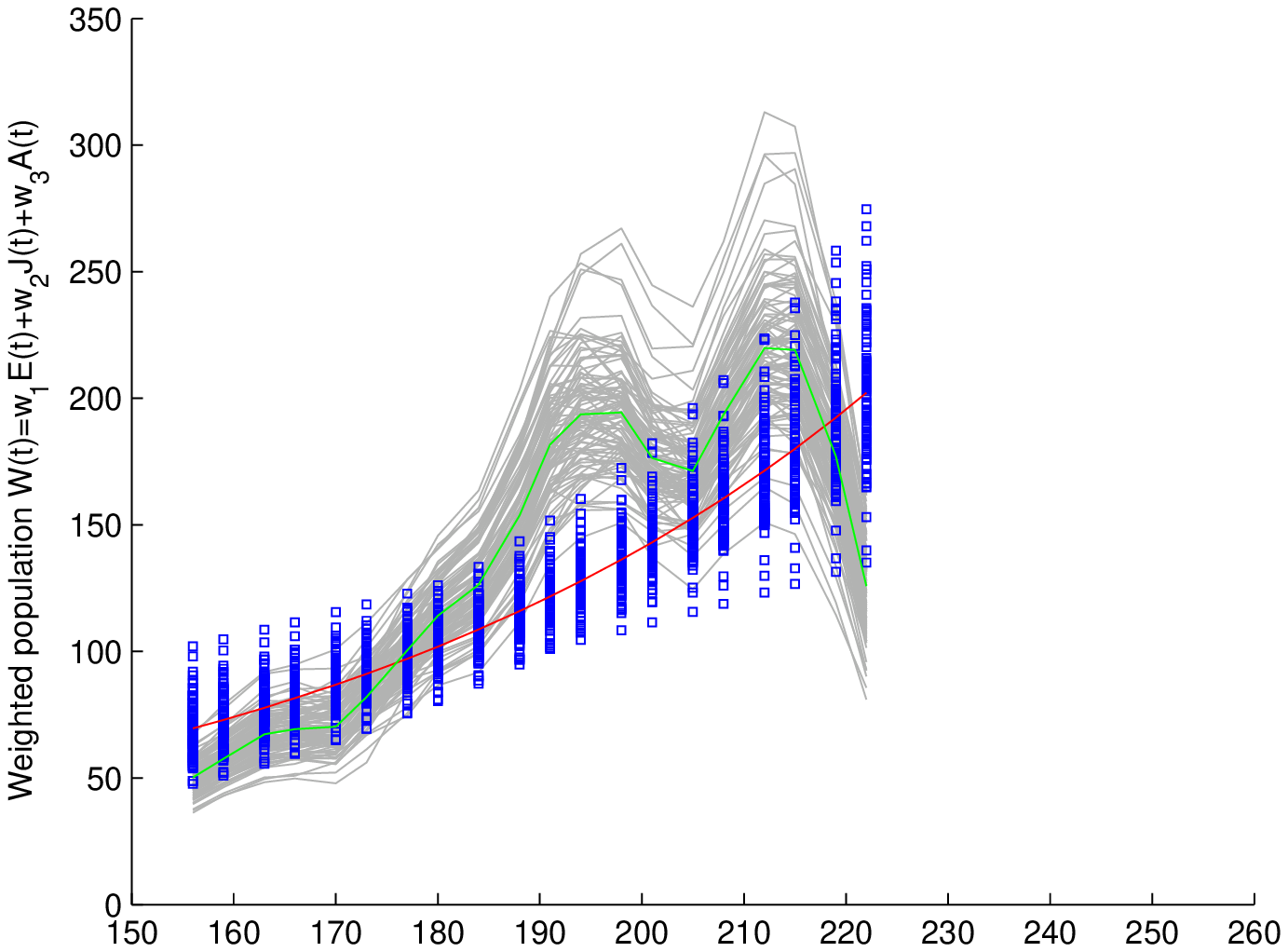}} & \\
  & & & & \\
  & & & & \\
  & & & & \\
  & & & & \\
  & & & & \\
 \hline 
\multirow{6}{*}{\centering \parbox{1cm}{\begin{center}II \\(+VI)\end{center} }} &\multirow{6}{*}{\includegraphics[height=3cm,width=5cm]{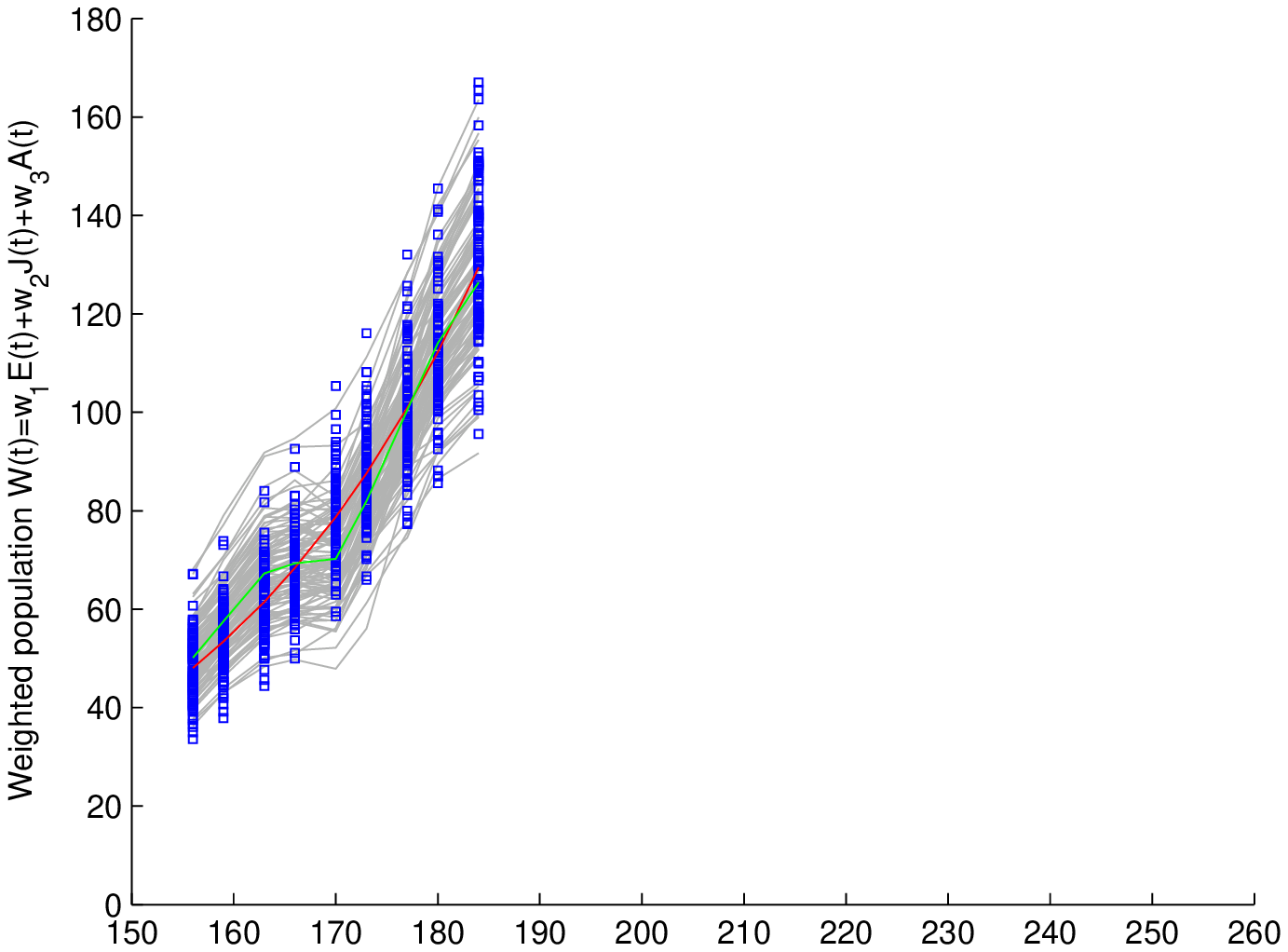}} &
  \multirow{6}{*}{\includegraphics[height=3cm,width=5cm]{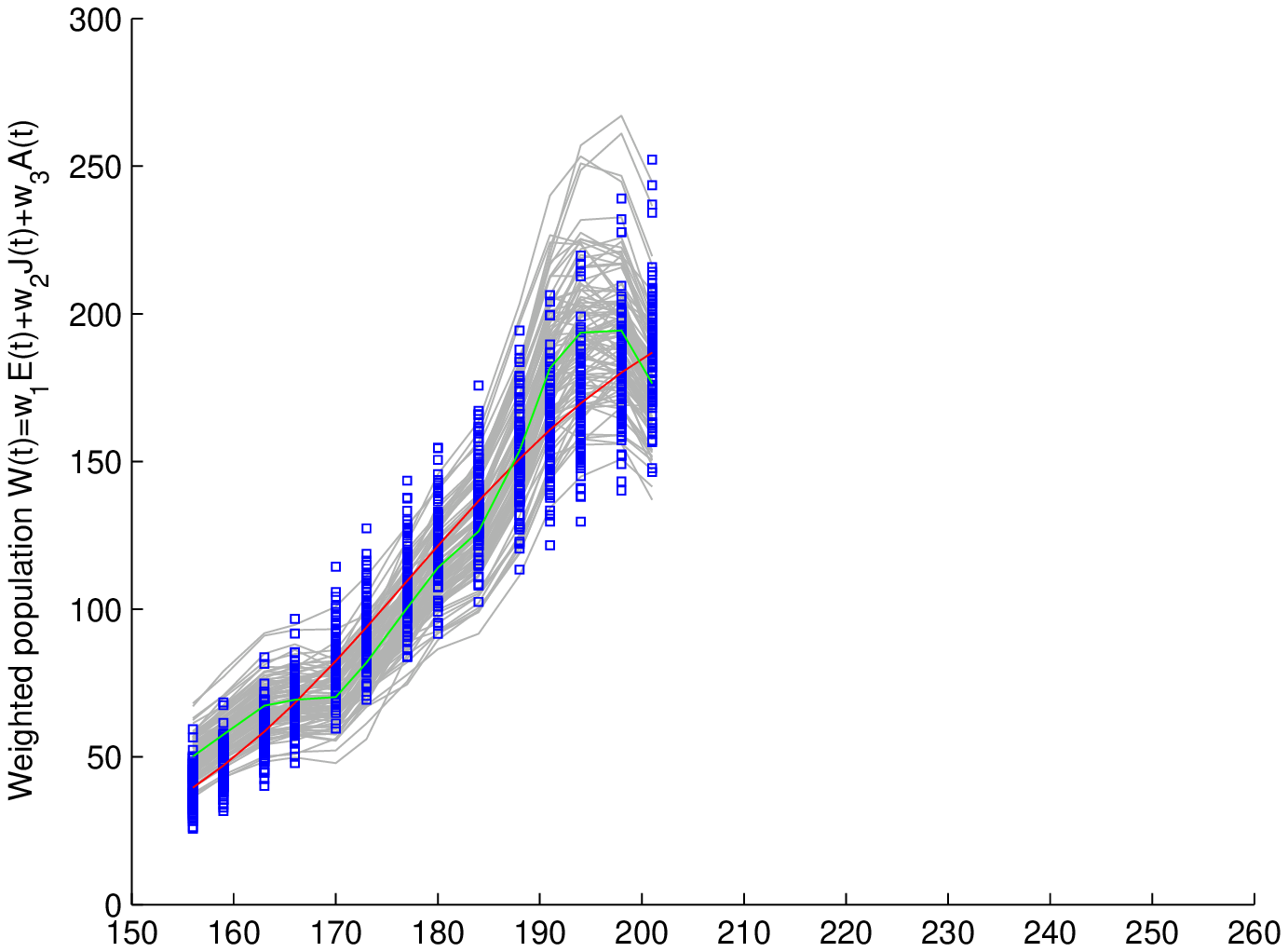} }&
 \multirow{6}{*}{\includegraphics[height=3cm,width=5cm]{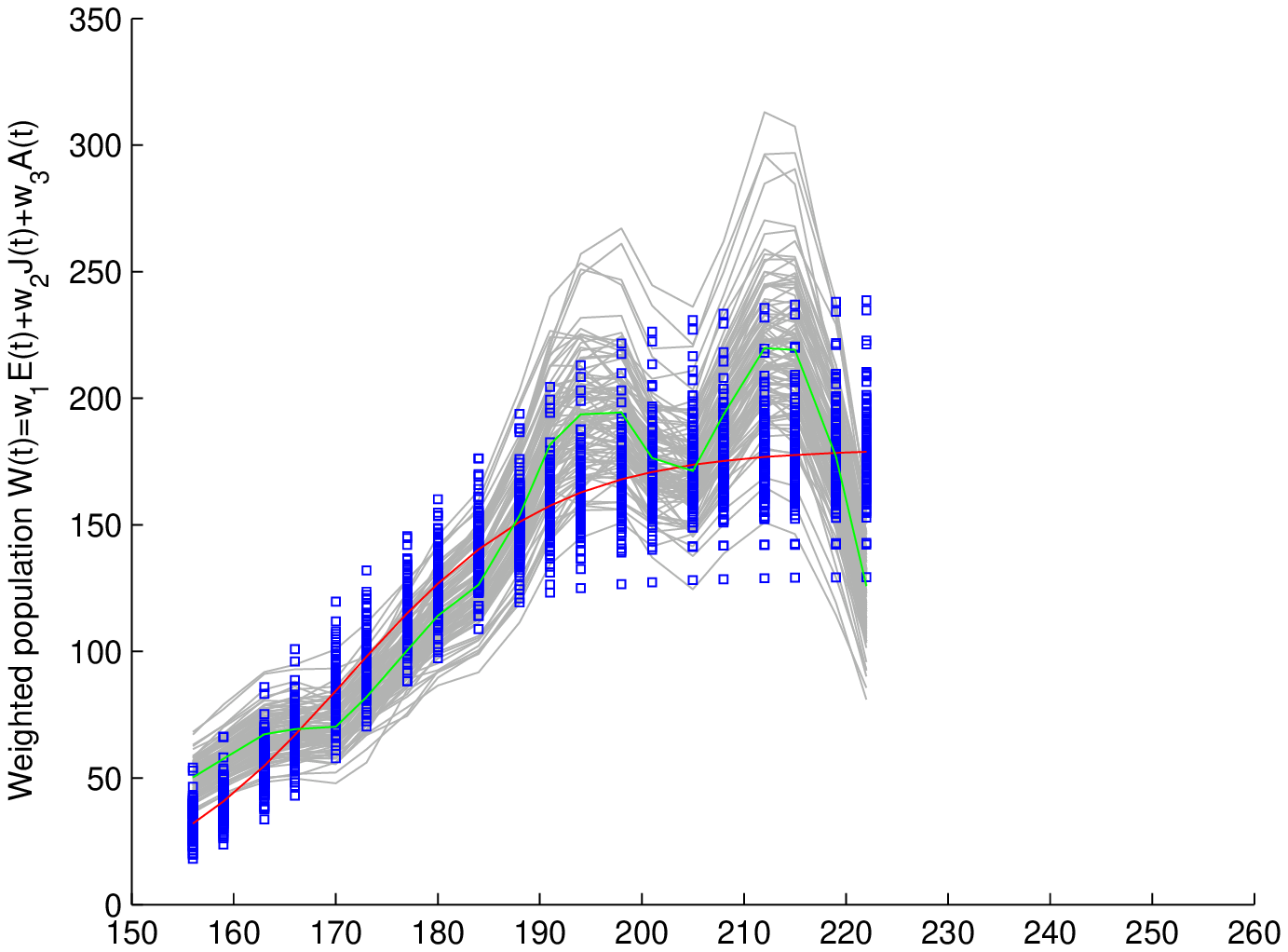} } & \\
  & & & & \\
  & & & & \\
  & & & & \\
  & & & & \\
  & & & & \\
 \hline  
  \end{tabular} 
 \vspace{-1.8ex}
 \caption{Fits of models I \& II to total reproductive value $V(t)$.  
 Rows 1 \& 2 and 3 \& 4 are analogous to Fig.~\ref{fitsW}, respectively. Rows 1 \& 2: $V(t)$ is computed from the (best-fit) parameter values 
 for the \textit{linear} demographic model (V). Rows 3 \& 4: $V$ is computed from the (best-fit) parameter values 
 for the \textit{nonlinear} linear demographic model (VI).} \label{fitsV}
 \end{figure}

As described in the main body of the paper, ``naive" aggregate population measures such as the total population may not follow an exponential growth law. 
Rather, one needs to take an appropriately \textit{weighted} average (the weight function being given by the (age-dependent) 
reproductive value $v(a)$)  of the various age groups to obtain an exponentially growing quantity (the total reproductive value $V(t)$). 
Unfortunately, computing $v(a)$ is just as (or even more) involved as determining the growth rate $r$; so the exponential growth law for $V(t)$ 
does not seem to be of much practical value for finding $r$.  

However, this property \textit{can} be used ``after the fact" as another consistency or robustness check for the 
 demographic models. The idea is this: 
after fitting the models to the data, one can use the (best-fit) model parameters to calculate $V(t)$.
In a perfect world, $V(t)$ would be an exponential function $e^{rt}$ with $r$ derived from the Lotka-Euler equation
(\ref{Euler-Lotka}).  So fitting 
the simple exponential model (I) to the $V(t)$ time series would just reproduce  $r$. Of course, the real world is messier than etherial theory and things are not quite as perfect; see Figures \ref{fitsV} and \ref{figdist2} below, where this 
procedure was carried out for the model parameters derived from the linear \textit{and} nonlinear models (V \& VI) and the 
resulting time series $V(t)$ was fitted to the exponential \textit{and} logistic growth models (I \& II).    

\vspace{-2ex}
\begin{figure}[h]
\mbox{}\hspace{-13ex}
\begin{tabular}{c}
 \includegraphics[height=12cm,width=8cm]{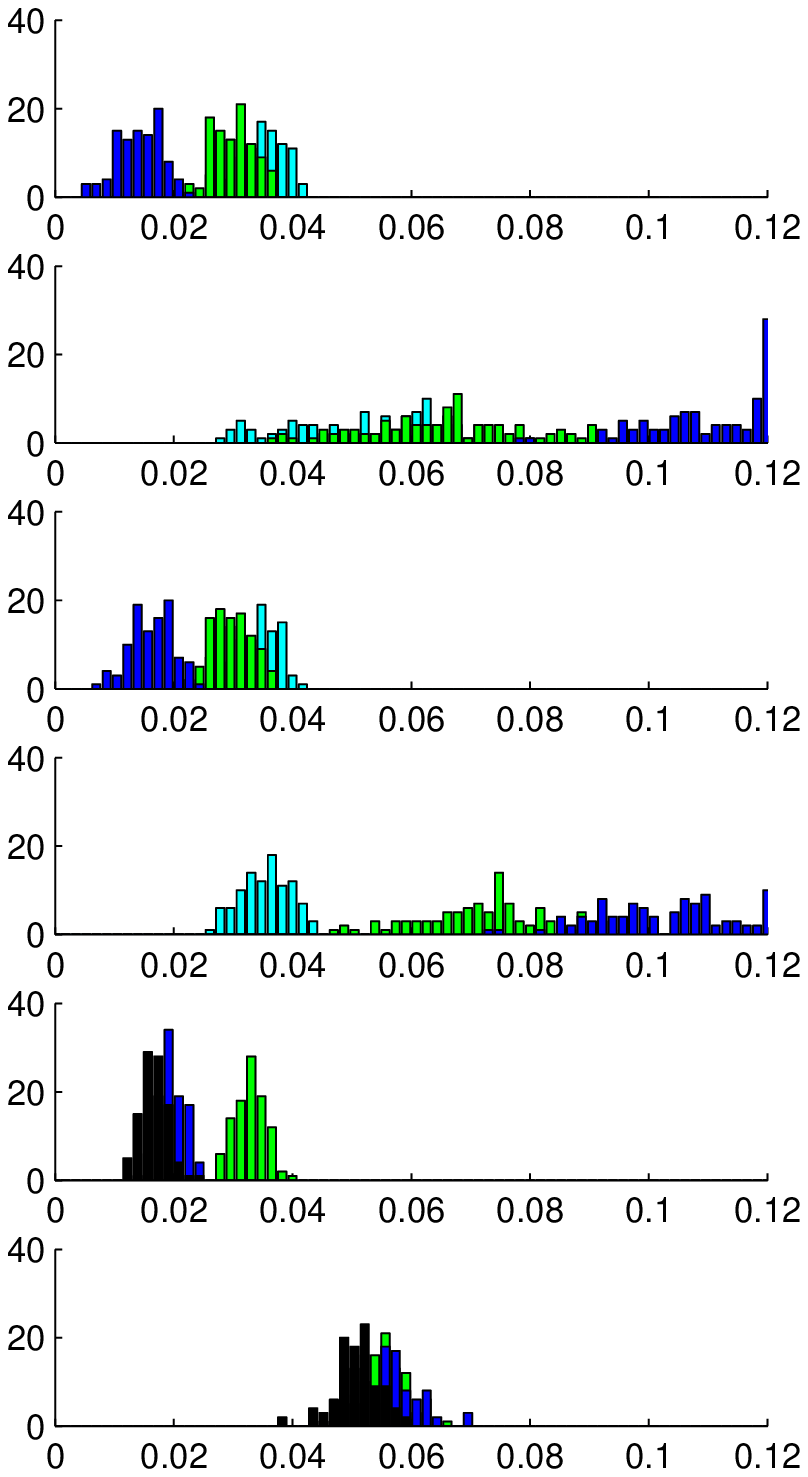} 
 \includegraphics[height=12cm,width=8cm]{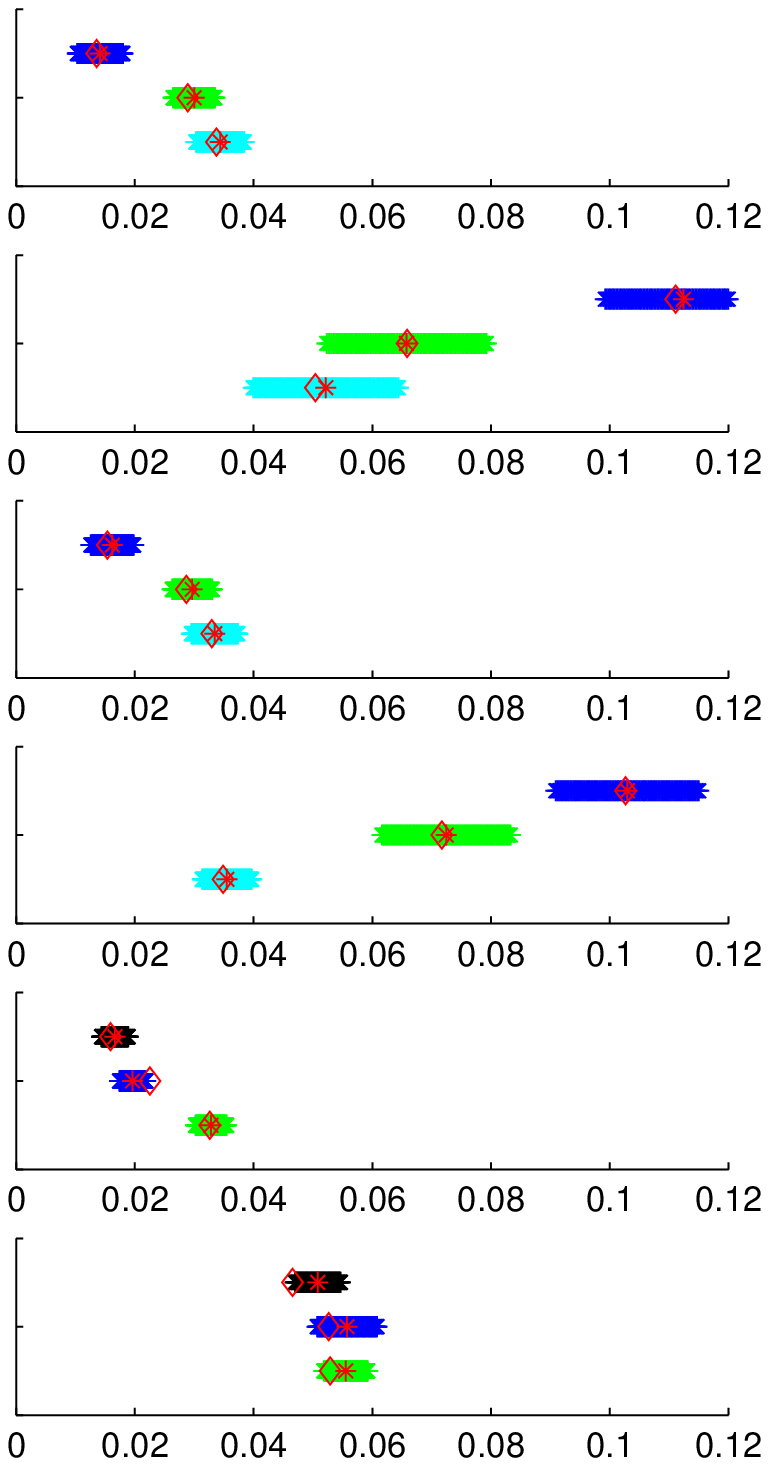} 
 \end{tabular} 
 \vspace{-8ex}
 \caption{Growth-rate distributions corresponding to Fig.~\ref{fitsV} (rows 1-4).  Rows 5 and 6 show 
 the distributions 
 derived directly from models V and VI (identical 
 with rows 5 and 6 of Fig.~\ref{figdist1}); shown for comparison.} \label{figdist2}
 \end{figure} 

\end{document}